\documentclass[paper]{agujournal2019}
\usepackage{url}
\usepackage{lineno}
\usepackage{color}
\usepackage{amsmath}
\draftfalse

\newcommand{\solphys}{{Solar Phys. }}
\newcommand{\planss} {{Planatary Space Science }}  
\newcommand{\ssr}{   {Space Sci. Rev. }}

\newcommand{\jgr}{   {J. Geophys. Res.}}
\newcommand{\grl}{   {Geophys. Res. Lett.}}
\newcommand{\apj}{   {Astrophys. J.}}
\newcommand{\apjl}{   {Astrophys. J. Lett.}}
\newcommand{\nat}{   {Nature}}

\newcommand{\mb}[1]{\mathbf{#1}}
\newcommand{\sbra}[1]{\left( #1 \right)}
\newcommand{\mbra}[1]{\left[ #1 \right]}
\newcommand{\lbra}[1]{\left\{ #1 \right\}}
\newcommand{\p}{\prime}

\graphicspath{{./}{figures/}}

\journalname{JGR: Space Physics}

\begin{document}


\title{Stability of the magnetotail current sheet with normal magnetic field and field-aligned plasma flows}

\authors{Chen Shi \affil{1}, Anton Artemyev \affil{1}, Marco Velli \affil{1}, Anna Tenerani \affil{2}} 
\affiliation{1}{Earth, Planetary, and Space Sciences, University of California, Los Angeles, Los Angeles, CA 90095, USA}
\affiliation{2}{Department of Physics, The University of Texas at Austin, Austin, TX 78712, USA}

\correspondingauthor{Chen Shi}{cshi1993@ucla.edu}

\begin{keypoints}
\item {1D magnetotail current sheet model with a finite $B_z$ and plasma flows is developed}
\item {Strong stabilizing effect of $B_z$ is confirmed for current sheets with field aligned flow}
\item {Field-aligned plasma flows cannot overtake the $B_z$ stabilizing effect}
\end{keypoints}

\begin{abstract}
One of the most important problems of magnetotail dynamics is the substorm onset and the related instability of the magneotail current sheet. Since the simplest 2D current sheet configuration with monotonic $B_z$ was proven to be stable to the tearing mode, the focus of the instability investigation moved to more specific configurations, e.g. kinetic current sheets with strong transient ion currents and current sheets with non-monotonic $B_z$ (local $B_z$ minima or/and peaks). Stability of the latter current sheet configuration has been studied both within kinetic and fluid approaches, whereas the investigation of the transient ion effects were limited to kinetic models only. This paper aims to provide detailed analysis of stability of a multi-fluid current sheet configuration that mimics current sheets with transient ions. Using the system with two field-aligned ion flows that mimic the effect of pressure non-gyrotropy, we construct 1D current sheet with a finite $B_z$. This model describes well recent findings of very thin intense magnetotail current sheets. The stability analysis of this two-ion model confirms the stabilizing effect of finite $B_z$ and shows that the most stable current sheet is the one with exactly counter-streaming ion flows and zero net flow. Such field-aligned flows may substitute the contribution of the pressure tensor nongyrotropy to the stress balance, but cannot overtake the stabilizing effect of $B_z$. Obtained results are discussed in the context of magnetotail dynamical models and spacecraft observations.
\end{abstract}

\section{Introduction}\label{sec:intro}
\begin{figure*}
\centering
\includegraphics[width=1\textwidth]{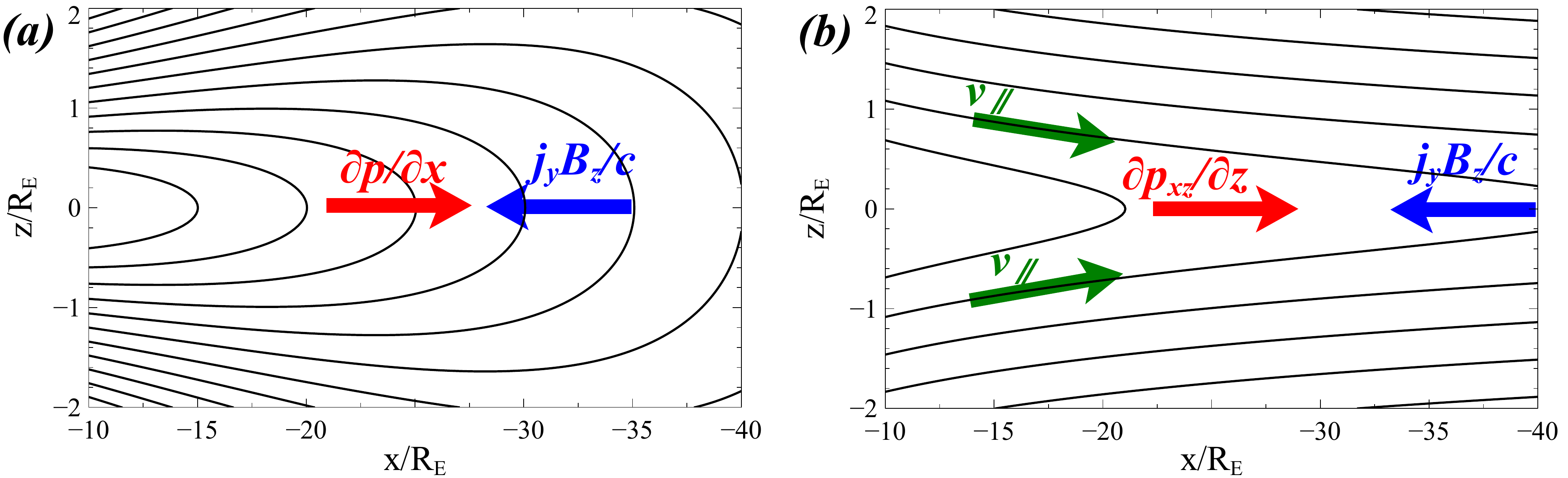}
\caption{Schematic of the magnetotail current sheet configuration: (a) 2D current sheet with the $j_yB_z/c=\partial p/\partial x$ stress balance, (b) 1D current sheet with $j_yB_z/c=\partial p_{xz}/\partial z$  stress balance.
\label{fig1}}
\end{figure*}

The problem of  current sheet stability is key for most theories and models of magnetospheric dynamics, because such stability determines magnetic reconnection onset \cite{Yamada10:reconnection, book:Gonzalez&Parker} and triggers magnetospheric substorms \cite{Baker96,Angelopoulos13,Sitnov19}. A short review of investigations of the magnetotail current sheet instability should start from the work by \cite{LPV66, Coppi66}, suggesting that magnetic reconnection results from the tearing mode driven by electron currents. \citeA{Schindler73} and \citeA{Schindler74} showed that in realistic magnetotail configuration the normal magnetic field component $B_z$ (see sketch in Figure \ref{fig1}) destroys the resonant electron interaction with tearing mode (at least for realistic $B_z$ magnitudes), so only ions can drive tearing. The inclusion of $B_z\ne 0$ into self-consistent current sheet models require either consideration of a specific class of exact 2D solutions (see, e.g., \citeA{Kan73} and the most recent generalizations in \citeA{YL05,Vasko13:pop}) or weakly-2D (with $\partial^2/\partial x^2\ll\partial^2/\partial z^2$) solutions (see, e.g., \cite{Schindler72} and the most recent generalizations in \cite{SB02,Birn04,Artemyev16:pop:cs}). Further investigation of {\it electron stabilizing effect} due to $B_z\ne 0$ \cite{Galeev76, bookGaleev85:vol2}  reduces the parametric space for instability, whereas the consideration of such weakly-2D current sheet configuration (where $j_yB_z/c$ tension force is balanced by the plasma pressure gradient $\partial p/\partial x$, see sketch in Figure \ref{fig1}) demonstrated that such current sheets are stable \cite{LP82,Pellat91,Quest96}, with a possible exception of unrealistically stretched field line configurations with extremely small $B_z$ \cite{Goldstein82}. To resolve the contradiction between well observed magnetic reconnection in the Earth's magnetotail (e.g., \citeA{Sergeev95, Nagai&Machida98, Petrukovich98, Angelopoulos08}) and theoretical current sheet stability, other models have considered the possibility of the reduction of {\it electron stabilizing effect} \cite{Kuznetsova91, Zelenyi08JASTP} and comprehensive and more general current sheet configurations \cite{Pritchett95,Sitnov10,Sitnov13}. The important role in development of such new current sheet models has been played by series of MHD \cite{Birn98:cs,Birn04MHD} and kinetic (particle-in-cell) \cite{Pritchett91,Pritchett&Coroniti94,Pritchett&Coroniti95} simulations of the thin current sheet formation in the magetotail. These simulations show formation of non-monotonic $B_z(x)$ profile with a local $B_z$ minimum and inverse $\partial B_z/\partial x<0$ gradient. Theoretically the formation of a $B_z$ minimum in the near-Earth plasma sheet may attribute to the steady earthward convection \cite{Hau89}. Further investigation of a current sheet with $\partial B_z/\partial x<0$ has shown that such current sheet is tearing unstable \cite{Sitnov14, Bessho&Bhattacharjee14, Pritchett15:pop, Merkin15,Birn18}. There is also indirect observational evidence for the formation of such current sheet configurations in the near-Earth magnetotail \cite{Sergeev18:grl, Angelopoulos20}. All such theoretical models of current sheet instability utilize the same class of 2D current sheets based on the stress balance $j_yB_z/c=\partial p/\partial x$ (for review of current sheet models of this class, see \cite{bookSchindler06,Baumjohann07} and references therein). However, as will be discussed below, modern spacecraft observations suggest that magnetotail current sheets may not always belong to this class. 

A finite $B_z$ field in the middle and distant magnetotail current sheet (where magnetic field line curvature does not contribute to the cross-tail pressure balance) requires the plasma pressure gradient $\partial p/\partial x$ to balance the tension force $j_yB_z/c$, whereas the equatorial plasma pressure $p$ must equal the lobe magnetic field pressure $B_{lobe}^2/8\pi$ \cite{Baumjohann90:pressure, Petrukovich99} that is fitted by a series of empirical models \cite{Nakai91, Shukhtina04}. These models give $B_{lobe} \approx 25{\rm nT}\cdot(-x/20R_E)^{-1}$, and the corresponding $j_y\approx (cB_{lobe}/4\pi B_z)\cdot(\partial B_{lobe}/\partial x)$ is limited to $\sim 1-2$ nA/m$^2$ for radial distances $x\sim -20 R_E$ and realistically small $B_z\sim 1$nT. As such current density amplitudes are much smaller than what is typically observed $5-10$ nA/m$^2$ \cite{Runov06,Petrukovich15:ssr,Vasko15:jgr:cs,Lu19:jgr:cs}, the stress balance $j_yB_z/c=\partial p/\partial x $ cannot universally describe the magnetotail equilibrium. If we write down the current sheet thickness $L_z\approx cB_{lobe}/4\pi j_y$ and length $L_x \approx (\partial \ln p/\partial x)^{-1}$, this stress balance gives $L_x = L_z B_{lobe}/2B_z$, whereas there are observations of current sheets with $L_x$ much exceeding this limit \cite{Artemyev11:grl, Artemyev15:grl:2dCS}. Empirical reconstructions of the magnetotail configuration during the growth phase of a substorm (i.e., before magnetic reconnection) also demonstrates the existence of very long current sheets, likely with $L_x \gg L_z B_{lobe}/2B_z$ \cite{Sitnov19:jgr, Sitnov21}. Thus, statistical observations at the middle (and distant) magnetotail of intense currents $j_y\geq 5-10$ nA/m$^2$, indirect $L_x$ estimates and empirical magnetotail reconstructions suggest that 2D current sheet models with $j_yB_z/c=\partial p/\partial x$ may not be suitable for investigation of the magnetotail current sheet stability (at least for a significant sample of observed configurations). An exploration of models alternative to the class with $j_yB_z/c=\partial p/\partial x$ is therefore necessary. This argumentation is not valid for the near-Earth magnetotail where the curvature force can contribute to the cross-sheet (vertical) pressure balance. Investigation of stability of the near-Earth current sheet requires analysis of a fully 2D magnetotail configurations (e.g., \citeA{Goldstein82,bookSchindler06}).

Plasma anisotropy (if any) may significantly reduce the tension force $(1-\Lambda)j_yB_z/c=\partial p/\partial x$ with $\Lambda=4\pi(p_\parallel-p_\perp)/B^2$ \cite{Rich72, Cowley78}, i.e. current sheets at the limit of fire-hose instability ($\Lambda\to 1$) may be almost 1D \cite{FP76, Cowley&Pellat79}. In particular, two-dimensional MHD equilibrium models have shown that both firehose pressure anisotropy and field-aligned flows may yield more stretched magnetotail configurations \cite{Hau93, Hau96:pop}. However, ions, mostly contributing to the thermal pressure, are mainly isotropic in the magnetotail current sheet \cite{Wang13:ions, Artemyev19:jgr:ions}, whereas parallel anisotropic electrons \cite{Walsh11,Artemyev14:jgr} do not contribute more than $\Lambda\sim 0.1-0.3$ for the absolute majority of current sheets \cite{Artemyev20:jgr:electrons}. More subtle (less measurable and more complex to be described) with respect to anisotropy, another kinetic effect, pressure non-gyrotropy ($p_{xz}\ne0$), allows balancing of 1D current sheet magnetic forces $j_yB_z/c=\partial p_{xz}/\partial z$ \cite{Eastwood72, Burkhart92TCS, Pritchett92, Maha96, Mingalev07, Mingalev18}. This effect has been included into the series of kinetic current sheet models (e.g., \citeA{Sitnov03, Sitnov06, Sitnov&Merkin16, Zelenyi11PPR} and references therein) that can describe many properties of observed current sheets \cite{Artemyev&Zelenyi13}. Although direct spacecraft measurements of ion non-gyrotropy in the magnetotail current sheet are quite challenging (see discussion in \citeA{Zhou09, Artemyev10:jgr, Artemyev19:jgr:ions}), such a non-gyrotropy seems to be a prospective solution of the dilemma why magnetotail current sheets are much longer than 2D stress balance limit $L_x = L_z B_{lobe}/2B_z$. Therefore, investigations of stability of 1D current sheets (i.e., current sheet that mimics a $p_{xz}\ne 0$ effect) should reveal if they are more unstable to tearing mode than the very stable 2D current sheets. This $p_{xz}\ne 0$ effect is a solution for the 1D current sheet balancing alternative to the balance by the cross-sheet plasma flow. Thus, we consider both these mechanisms of 1D current sheet balance: $p_{xz}\ne0$ in absence of cross-sheet bulk flow and strong cross-sheet flow (as in the rotational discontinuity balance).

Although full PIC simulations of 2D current sheet stability (e.g., \citeA{Pritchett91,Pritchett97, Sitnov09, Sitnov11, Liu14:CS, Lu18:pop}) have provided many details on the external driver threshold needed to trigger magnetic reconnection, there are almost no investigations of the stability of a 1D nongyrotropic current sheet. To include such investigations into more general context of the tearing instability \cite{bookBiskamp00,bookBirn&Priest07}, it would be reasonable to start with a fluid resistive model that can reveal a $B_z$ role in 1D current sheet instability. To mimic the effect of $p_{xz}\ne 0$ in the stress balance of the fluid current sheet model with $B_z\ne 0$, we will adopt multi-fluid model proposed by \citeA{Steinhauer08}. In this model two counter-streaming ion flows create $v_z \partial_z v_x$ terms that balance $j_yB_z/c$ force with a zero net flow. The generalization of this model would contain imbalanced flows, that is to say a non-zero net flow. This generalization can be then reduced to a single fluid model with the field-aligned flows balancing $j_yB_z/c$ force, i.e. to the classical rotational discontinuity \cite{Hudson70} with the flow velocity equal to the Alfv\'en velocity. This paper is devoted to investigation of stability of such generalized 1D multi-fluid model.

The formation and instability of current sheets with stretched magnetic field lines is a common problem in both magnetotail and solar physics \cite{Terasawa00:AdSR, Reeves08:solar}. In the latter case magnetic reconnection is believed to explain charged particle acceleration and magnetic field energy release in eruptive flares (the so called standard CSHKP model, see \citeA{Carmichael64, Sturrock66, Hirayama74, Kopp&Pneuman76}), non-eruptive events (including, e.g., coalescence of magnetic loops, see \citeA{Sakai&deJager96}), streamers (e.g., \citeA{Riley&Luhmann12,Edmondson&Lynch17, Reville20}), and solar wind current sheets \cite{Phan06:reconnection, Gosling12}. In contrast to the magnetotail investigations, mostly focused on the stabilization of the tearing mode by $B_z$ field, the theory of tearing instability for solar applications is dominated by models of resistive tearing mode in $B_z=0$ sheets (e.g., \citeA{Dobrowolny83, Ofman91, Loureiro12,Tenerani15:apj, DelSarto16}). However, the stabilizing effect of $B_z$ has also been discussed in context of the solar physics \cite{Verneta&Somov87, Somov&Verneta93}. Therefore, our investigation of the tearing instability in 1D current sheet with $B_z\ne0$ and plasma flows may be of great interests to solar physics applications.

The stress balance in 1D current sheet model with $B_z\ne0$ is controlled by counter-streaming plasma flows (see Figure \ref{fig1}(b)), whereas imbalance of these flows results in the net plasma flow across the current sheet. For solar wind plasma such cross-sheet flow is due to current sheet (rotational discontinuity) motion relative to the solar wind \cite{Hudson70, Tsurutani&Ho99}. In the Earth's magnetotail there are a couple of mechanisms responsible for formation of such counter-streaming plasma flows. First, formation of thin current sheet is often associated with enhanced precipitations of hot plasma sheet electrons into ionosphere, and these precipitations drive the ionospheric outflow consisting of cold oxygen and hydrogen ions \cite{Keika13:outflow,Maggiolo&Kistler14, Kronberg15}. Outflow ions shape fast beams moving along magnetic field lines \cite{Sauvaud04:oxygen,Kistler05,Artemyev20:jgr:outflow} and contributing to the stress balance $p_{xz}$ \cite{Eastwood72, Eastwood74, Hill75}. Although such beams forming in the south and north hemispheres should be generally balanced (i.e., there is a stress balance without a net flow), the precise balance between them is not guaranteed, and there could be a net flow across the current sheet. Second, there are beams of energetic ions moving along plasma sheet boundary layer from the distant magnetotail and coming back to the plasma sheet after reflection from the Earth's dipole field \cite{Maha92,Maha96, Grigorenko11:SSR}. These are solar wind protons accelerated in the distant magnetotail by convection \cite{Cowley83, Maha93,Zelenyi06:beamlet} or reconnection electric fields \cite{Maha06, Grigorenko09}. Such acceleration mechanisms in combination with Speiser motion \cite{Speiser65, Lyons82} shape ion field-aligned beams that contribute significantly to the stress balance in 1D current sheet \cite{Burkhart92TCS, Pritchett92,Mingalev07}. The asymptotic solutions (for infinitely small $B_z$) of such 1D current sheets with counter-streaming plasma flows form a class of models developed in \citeA{Sitnov00} and then generalized by \citeA{Sitnov03, Sitnov06} and \cite{Zelenyi06,Zelenyi11PPR}. Models of this class describe bifurcated and embedded current density profiles with properties similar to current sheet properties in the Earth's magnetotail (see model/observation comparison in \cite{Sitnov06,Artemyev08:angeo,Zhou09}). The distinguishing feature of these models is the sufficiently long 1D current sheet with the current density magnitude significantly exceeding estimates for 2D isotropic current sheet equilibria, $c(\partial p/\partial x)/B_z$. In this study we do not specify particular mechanism of field-aligned plasma flows, and focus on stability of the 1D current sheet model with such flows.

The paper is organized as follows. In Section \ref{sec:equation} we present the equations for the zeroth-order equilibrium and the derivation of equations for the perturbed fields. In Section \ref{sec:results}, we show the results, i.e. the dispersion relation of the resistive tearing mode, obtained by numerically solving the perturbation equations derived in Section \ref{sec:equation}. In Section \ref{sec:discussion} we discuss the results in the context of Earth's magnetosphere. In Section \ref{sec:conclusion} we conclude this study.

\section{Basic equations}\label{sec:equation}
In this section, we present in detail the model of the background field and derivation of the linearized equation set which is solved numerically. The numerical results are presented in Section \ref{sec:results}.
\subsection{Background fields}\label{sec:backgroundfields}
Incompressibility is assumed throughout the paper such that the plasma density is homogeneous and unperturbed: $\rho(\mb{x},t) \equiv \rho_0$. The background magnetic field consists of a Harris-type anti-parallel component and a uniform normal component:
\begin{equation}
    \mb{B} = B_0 \tanh{\left(\frac{z}{a} \right)} \hat{e}_x + B_z \hat{e}_z
\end{equation}
The current density flows along $y$ across magnetic field lines lying in $(x,z)$ plane, whereas a field-aligned plasma flow does not change this magnetic field geometry. One can show that $\nabla \times \left( \mb{B} \cdot \nabla \mb{B} \right) \neq 0$ with this configuration and thus it is impossible to maintain equilibrium with a scalar pressure $\mb{P} = P \mb{I}$ solely. As already discussed in Section \ref{sec:intro}, one simple way to establish equilibrium is introducing the plasma flow such that the shear stress of the flow balances the tension force of the magnetic field. In one-fluid MHD model, an Alfv\'enic flow $\mb{V} \equiv \mb{B}/\sqrt{4 \pi \rho}$ is required such that $\rho \mb{V} \cdot \nabla \mb{V} \equiv  \mb{B} \cdot \nabla \mb{B}/4\pi$ and the equilibrium is achieved with a uniform total pressure $P^T = P + B^2 /8\pi $. 

\begin{figure}[htb!]
    \centering
    \includegraphics[width=\hsize]{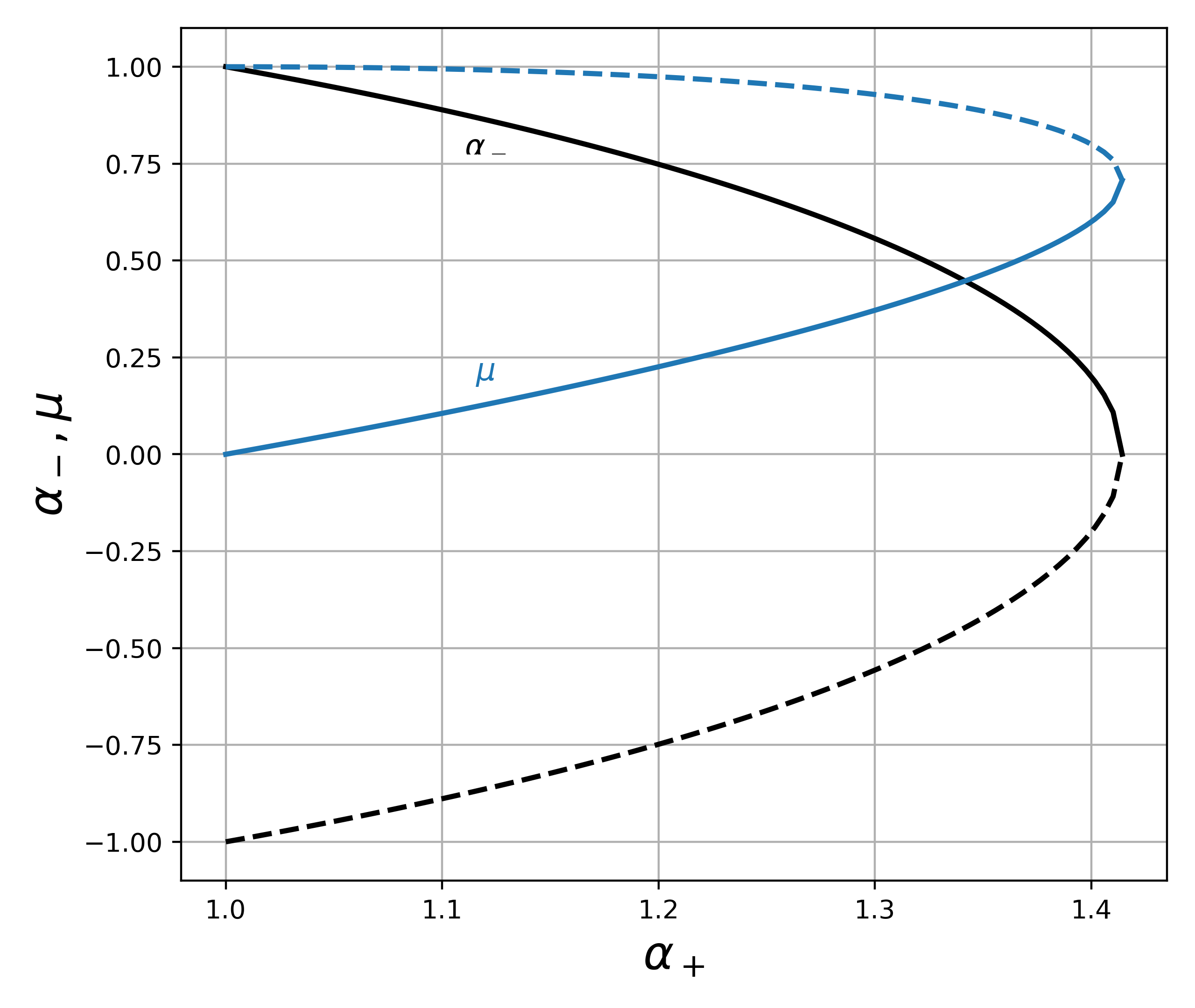}
    \caption{$\alpha_-$ (black) and $\mu = (\alpha_+-\alpha_-)/2$ (blue) as functions of $\alpha_+$. The dashed segments of the two curves correspond to the negative branch $\alpha_- = -\sqrt{2 - \alpha_+^2}$, i.e. the two ion populations flowing in the same direction. The solid segments of the two curves correspond to the positive branch $\alpha_- = \sqrt{2 - \alpha_+^2}$, i.e. the two ion populations flowing in the opposite direction.}
    \label{fig:alpha-_mu}
\end{figure}

We can consider a more generalized case where the protons shape two populations, denoted by subscripts ``+'' and ``-'' respectively such that the two corresponding momentum equations are:
 \begin{equation}\label{eq:momentum_plus_minus}
    \rho_\pm \mb{V_\pm} \cdot \nabla \mb{V_\pm} = -\nabla P_\pm + \frac{q}{m_p} \rho_\pm \sbra{\mb{E} +\frac{1}{c} \mb{V_\pm} \times \mb{B}}
\end{equation}
where $\mb{E}$ is the electric field, $q$ is the charge of proton, and $m_p$ is the proton mass. Sum up the two equations and use the relation $\mb{E} \approx - \mb{V_e} \times \mb{B}/c$ where $\mb{V_e}$ is the electron flow velocity from the massless and cold electron momentum equation
 \begin{eqnarray*}
    0= \sum_{\pm}\rho_\pm \sbra{\mb{E} +\frac{1}{c} \mb{V_e} \times \mb{B}},
\end{eqnarray*}
we get
\begin{equation}
    \rho_+ \mb{V_+} \cdot \nabla \mb{V_+} +  \rho_- \mb{V_-} \cdot \nabla \mb{V_-} =  -\nabla{\sbra{P + \frac{B^2}{8\pi}}} + \frac{1}{4 \pi} \mb{B} \cdot \nabla \mb{B}
\end{equation}
where $P = P_+ + P_-$ and we have used the approximation $\mb{J} \approx c \nabla \times \mb{B}/4\pi$ where $\mb{J}$ is the electric current density. For convenience we assume $\rho_+ = \rho_- = \rho/2$, i.e. the two ion populations are of the same density. Then we write $\mb{V_\pm} = \pm \alpha_\pm \mb{V_A}$ where $\mb{V_A} = \mb{B}/\sqrt{4 \pi \rho}$ is the magnetic field in Alfv\'en velocity unit and $\alpha_\pm$ are two constants controlling the speeds of the two ion flows. Apparently we have assumed that both of the two ion populations are streaming along the magnetic field lines. With these assumptions, we get two equilibrium criteria:
\begin{subequations}
    \begin{equation}
         \frac{d}{dz}\sbra{P + \frac{B^2}{8 \pi}} = 0
    \end{equation}
    \begin{equation}
        \mbra{1 - \frac{1}{2} \sbra{\alpha_+^2 + \alpha_-^2}} B_z \frac{d B_{x}}{dz} = 0
    \end{equation}
\end{subequations}
Here we note that all the background fields are functions of coordinate $z$ only. The first equation is the pressure balance condition which constrains the thermal pressure $P(z)$ and the second equation is the tension force balance condition which leads to the requirement
\begin{equation}
    \alpha_+^2 + \alpha_-^2 = 2
\end{equation}
Without loss of generality, we assume $1 \leq \alpha_+ \leq \sqrt{2}$ and thus $\alpha_- = \pm \sqrt{2 - \alpha_+^2}$. The positive ($+$) branch of $\alpha_-$ corresponds to the case that the two ion flows are counter-streaming and the negative ($-$) branch of $\alpha_-$ corresponds to the case that the two ion flows are of the same direction. We define $\mu = (\alpha_+ - \alpha_-)/2$ such that the average flow speed $\mb{V} = (\mb{V_+} + \mb{V_-})/2 = \mu \mb{V_A}$, i.e. $\mu = \mb{V}/\mb{V_A}$ is the amplitude of the average ion velocity normalized by the Alfv\'en velocity. In Figure \ref{fig:alpha-_mu}, we plot the $\alpha_-(\alpha_+)$ and $\mu(\alpha_+)$ curves for $\alpha_+ \in [1,\sqrt{2}]$. By varying $\alpha_+$ and selecting either the positive or negative branch of $\alpha_-$, we are able to regulate the average flow speed such that $0 \leq \mu \leq 1$. Specifically, $\alpha_+ = \alpha_- = 1$ corresponds to $\mu = 0$ as in this case the two ion populations are counter-streaming with the same speed $V_A$. On the other hand, $\alpha_\pm = \pm 1$ leads to $\mu = 1$ as in this case the two ion populations are streaming with the same velocity $\mb{V} = \mb{V_A}$ and this model converges to the one-fluid MHD model.

\subsection{Perturbation equations}
To linearize the equation set, we write the perturbed, or first-order, fields as $\mb{u_{\pm,e}}$, $p_\pm$, $\mb{b}$, $\mb{j}$, and $\mb{E_1}$ respectively and assume all perturbations are in the form:
\begin{equation}
    f(x,z,t) = f(z) \exp(ikx + \gamma t).
\end{equation}
The linearized momentum equations are:
\begin{equation}\label{eq:linearized_momentum}
    \frac{1}{2} \rho_0 \sbra{\gamma \mb{u_\pm} + \mb{u_\pm}\cdot \nabla\mb{V_\pm} + \mb{V_\pm}\cdot \nabla\mb{u_\pm}} = - \nabla p_\pm + \frac{1}{2} \frac{q}{m_p} \rho_0 \sbra{\mb{E_1} + \frac{1}{c} \mb{u_\pm} \times \mb{B} + \frac{1}{c}\mb{V_\pm} \times \mb{b}}
\end{equation}
where $\mb{E_1} \approx - \mb{u_e}\times \mb{B}/c - \mb{V_e} \times \mb{b}/c$. For convenience, We further define 
\begin{equation}
     \mb{u} = \frac{1}{2} \sbra{\mb{u_+} + \mb{u_-}}, \mb{w} = \frac{1}{2} \sbra{\mb{u_+} - \mb{u_-}}
\end{equation}
where $\mb{u}$ is the perturbation of the ion net flow. By adding up and taking difference between the two equations of Eq (\ref{eq:linearized_momentum}), we get
\begin{subequations}
    \begin{equation}\label{eq:1st_order_u}
        \gamma \mb{u} + \mb{f_-} \cdot \nabla \mb{V_A} + \mb{V_A} \cdot \nabla \mb{f_-} = - \frac{1}{\rho_0} \nabla \sbra{p_+ + p_-}  + \frac{1}{c\rho_0} \sbra{\mb{j} \times \mb{B} + \mb{J} \times \mb{b}}
    \end{equation}
    \begin{equation}\label{eq:1st_order_w}
        \gamma \mb{w} + \mb{f_+} \cdot \nabla \mb{V_A} + \mb{V_A} \cdot \nabla \mb{f_+} = - \frac{1}{\rho_0} \nabla \sbra{p_+ - p_-}  + \frac{q}{m_p c} \mbra{\mb{w} \times \mb{B} + \frac{1}{2} \sbra{\alpha_+ + \alpha_-} \mb{V_A} \times \mb{b}}
    \end{equation}
\end{subequations}
where $\mb{f_\pm} =  \sbra{\alpha_+ \mb{u_+} \pm \alpha_- \mb{u_-}}/2$, and we have used the assumption $\mb{V_\pm} = \pm \alpha_\pm \mb{V_A}$. If we normalize length to the thickness of the current sheet $a$, normalize magnetic field to $B_0$, normalize speed to the Alfv\'en speed $V_{A0} = B_0/ \sqrt{4 \pi \rho_0}$, Eq (\ref{eq:1st_order_w}) becomes:
\begin{equation}
    \gamma \mb{w} + \mb{f_+} \cdot \nabla \mb{B} + \mb{B} \cdot \nabla \mb{f_+} = -\nabla \sbra{p_+ - p_-}  + \frac{a}{d_i} \mbra{\mb{w} \times \mb{B} + \frac{1}{2} \sbra{\alpha_+ + \alpha_-} \mb{V_A} \times \mb{b}}
\end{equation}
where $d_i = c/\sqrt{4\pi n e^2/m_p}$ is the ion skin depth and $n$ is the plasma number density. We can see that, in the MHD limit where the typical length scale $a$ is much larger than the ion skin depth $d_i$, this equation reduces to 
\begin{equation}
    \mb{w} \times \mb{B} + \frac{1}{2} \sbra{\alpha_+ + \alpha_-} \mb{V_A} \times \mb{b} = 0
\end{equation}
or alternatively
\begin{equation}\label{eq:1st_order_w_reduced}
    \mb{w} = \frac{1}{2} \sbra{\alpha_+ + \alpha_-} \mb{b}
\end{equation}
which correlates the magnetic field perturbation and the difference between the perturbations of the two ion flow velocities. In the special case $\alpha_\pm = \pm 1$, we have $\mb{w} \equiv 0$, which is consistent with the fact that the two ion populations merge into one fluid.
The induction equation is acquired by inserting $\mb{E_1} = - \mb{u_e}\times \mb{B}/c - \mb{V_e} \times \mb{b}/c + \eta \mb{j}$, where $\eta$ is the resistivity, into the linearized Maxwell-Faraday equation $\partial \mb{b}/\partial t = -c \nabla \times \mb{E_1}$. As we are analyzing the MHD case and hence the Hall effect is neglected, we can write $\mb{E_1} \approx - \mb{u}\times \mb{B}/c - \mb{V} \times \mb{b}/c + \eta \mb{j}$. After normalization, the linearized induction equation is written as
\begin{equation}\label{eq:1st_order_b}
    \gamma \mb{b} = \mb{b} \cdot \nabla  \mb{V} - \mb{V} \cdot \nabla \mb{b} + \mb{B} \cdot \nabla \mb{u} - \mb{u} \cdot \nabla \mb{B}  + \frac{1}{S} \nabla^2 \mb{b}
\end{equation}
where $S= a V_{A0}/ \eta^\prime$ is the Lundquist number and $\eta^\prime = c^2 \eta/4\pi$ is the magnetic diffusivity. Here we have used the divergence-free conditions for both the magnetic field and the flow velocity. 

After some algebra starting from Eqs (\ref{eq:1st_order_u}, \ref{eq:1st_order_w_reduced}, \& \ref{eq:1st_order_b}), a two-equation set for $u_z$ and $b_z$ is written as
\begin{subequations}\label{eq:uz_bz}
    \begin{equation}\label{eq:uz}
    \begin{aligned}
        \gamma \sbra{u_z^{\p\p} - k^2 u_z} + & \mu \lbra{B_z \sbra{u_z^{\p\p\p} - k^2 u_z^\p} + ik \mbra{B_x \sbra{u_z^{\p\p} - k^2 u_z} - B_x^{\p\p} u_z}} = \\
        & \sigma \lbra{ B_z \sbra{b_z^{\p\p\p} - k^2 b_z^\p} + ik \mbra{B_x \sbra{b_z^{\p\p} - k^2 b_z} - B_x^{\p\p} b_z} }
    \end{aligned}
\end{equation}
\begin{equation}\label{eq:bz}
\begin{aligned}
    \gamma b_z = \sbra{ikB_x u_z + B_z u_z^\p} - \mu \sbra{ikB_x b_z + B_z b_z^\p} + \frac{1}{S} \sbra{b_z^{\p\p} - k^2 b_z}
\end{aligned}
\end{equation}
\end{subequations}
where $\mu = (\alpha_+ - \alpha_-)/2$, $\sigma = 1 - \sbra{\alpha_+ + \alpha_-}^2/4$ and prime indicates $d/dz$. From Eq (\ref{eq:uz_bz}), we see that $B_z$ is a singular parameter as it increases the order of the equation for $u_z$ from two to three. In addition, compared with the tearing mode with an anti-parallel magnetic field, the growth rate is in general complex in this model rather than purely real, implying propagating perturbations. It is also worth noting that, if $B_z=0$ and $\alpha_+ = \alpha_- = 0$ such that $\mu=0$, $\sigma = 1$, Eq (\ref{eq:uz_bz}) reduces to the classical tearing mode equation:
\begin{subequations}
    \begin{equation}
    \begin{aligned}
        \gamma \sbra{u_z^{\p\p} - k^2 u_z}  =   ik \mbra{B_x \sbra{b_z^{\p\p} - k^2 b_z} - B_x^{\p\p} b_z} 
    \end{aligned}
\end{equation}
\begin{equation}
\begin{aligned}
    \gamma b_z = ikB_x u_z + \frac{1}{S} \sbra{b_z^{\p\p} - k^2 b_z}
\end{aligned}
\end{equation}
\end{subequations}

\section{Stability analysis}\label{sec:results}
The system of Eq (\ref{eq:uz_bz}) is a boundary-value eigen-problem. The boundary conditions are that $u_z$ and $b_z$ vanish far from the current sheet: $u_z, b_z (z \rightarrow \pm \infty) \rightarrow 0$. In practice, we cannot set the boundaries to infinity. However, far from the current sheet we have $B_x\approx \pm B_0$, i.e. $B_x$ is approximately constant, so it is observed that $u_z,b_z \propto \exp(-k|z|)$ are solutions to Eq (\ref{eq:uz}). Plugging the condition $u_z,b_z \propto \exp(-k|z|)$ into Eq (\ref{eq:bz}), we get the ratio between $u_z$ and $b_z$ at the boundaries. We use the boundary-value-problem solver implemented in the Python library SciPy \cite{Virtanenetal2020Scipy} to solve Eq (\ref{eq:uz_bz}). The solver adopts a 4th order collocation algorithm with the control of residuals \cite<ref.>{Kierzenkaetal2001BVP,Ascheretal1994numerical} and is able to solve the eigenvalue and eigen-functions simultaneously. The solver has been successfully utilized to analyze the linear stability of the oblique tearing mode with a guide field \cite{shi2020oblique}.

\subsection{Effect of $B_z\ne0$}
\begin{figure}[htb!]
    \centering
    \includegraphics[width=\hsize]{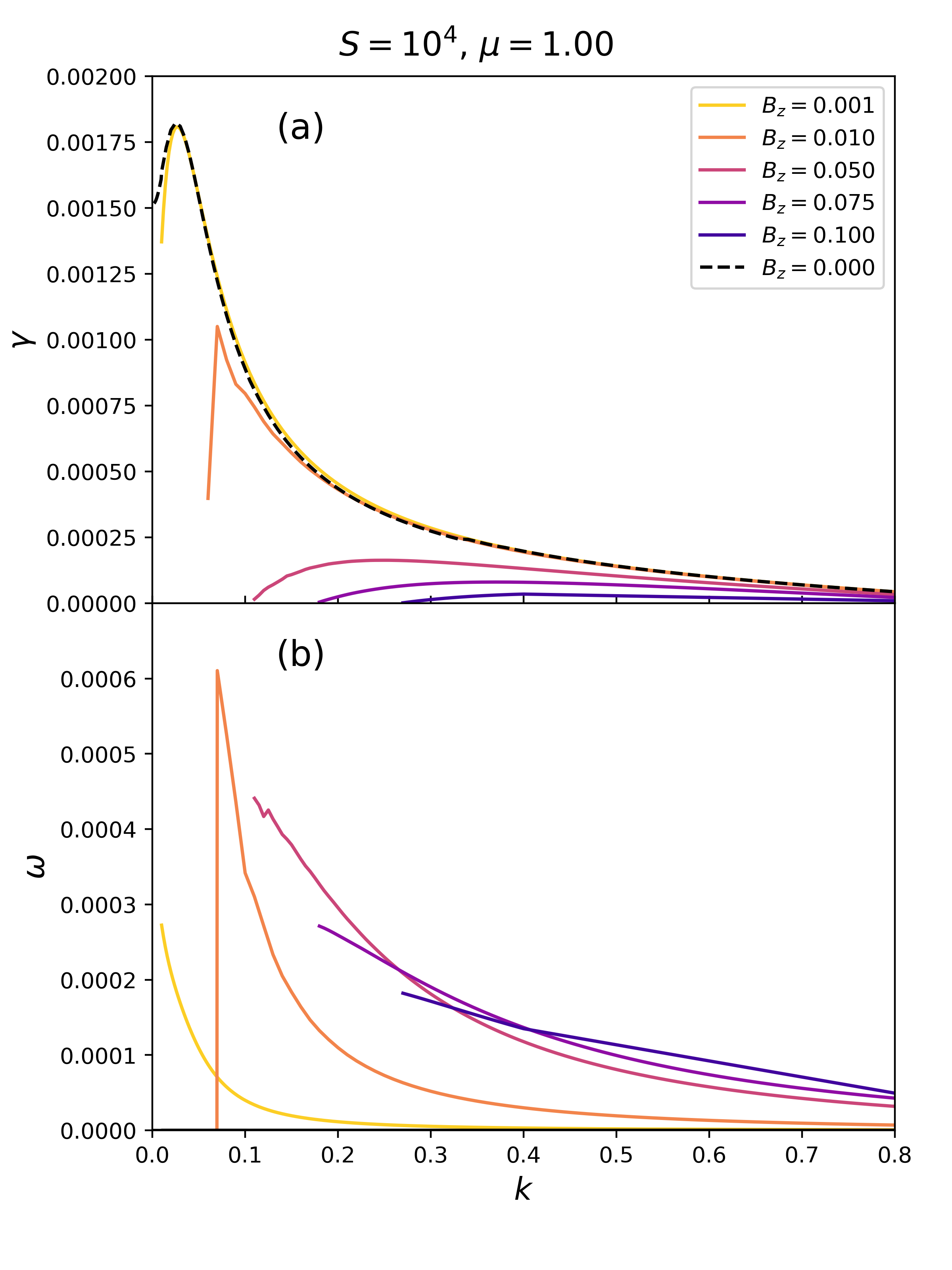}
    \caption{(a) $\gamma-k$ and (b) $\omega-k$ curves for $S=10^4$, $\mu=1$, and varying $B_z$. The light (yellow) to dark (purple) colors correspond to $B_z$ increasing from 0.001 to 0.1. The black dashed curve is $B_z = 0$.}
    \label{fig:disp_rel_different_Bz}
\end{figure}

In Figure \ref{fig:disp_rel_different_Bz}, we plot the dispersion relations $\gamma-k$ and $\omega-k$ in the top (panel (a)) and bottom (panel (b)) panels respectively. We note that, here $\gamma$ and $\omega$ are the real and opposite-imaginary parts of the complex $\gamma$ appearing in Eq (\ref{eq:uz_bz}), i.e. $\gamma_{complex} = \gamma - i \omega$. For this figure, the Lundquist number $S$ is fixed at $10^4$ and $\mu$, i.e. the average ion speed normalized to $V_{A0}$, is fixed at $1.0$, corresponding to an Alfv\'enic one-fluid model as discussed in Section \ref{sec:backgroundfields}. In each panel, curves of different colors correspond to different values of $B_z$, ranging from 0.001 to 0.1. To validate the solver and to testify whether the solution to Eq (\ref{eq:uz_bz}) converges as the singular parameter $B_z \rightarrow 0$, we also solve the problem with $B_z = 0$ at which the equation set degenerates to a lower order. The solved $\gamma$ and $\omega$ for $B_z=0$ are plotted as dashed curves and we can see that as $B_z$ approaches to 0 from a finite value, the $\gamma-k$ curve converges to the dashed one. We note that if $B_z$ is exactly zero, $\omega$ is also zero, i.e. the mode is not propagating due to the fact that the background shear flow is symmetric in $z$. From panel (a) of Figure \ref{fig:disp_rel_different_Bz}, we see that the growth rate is monotinically decreasing with $B_z$. That is to say, the existence of a finite $B_z$ quenches the growth of the instability.

\begin{figure}[htb!]
    \centering
    \includegraphics[width=\hsize]{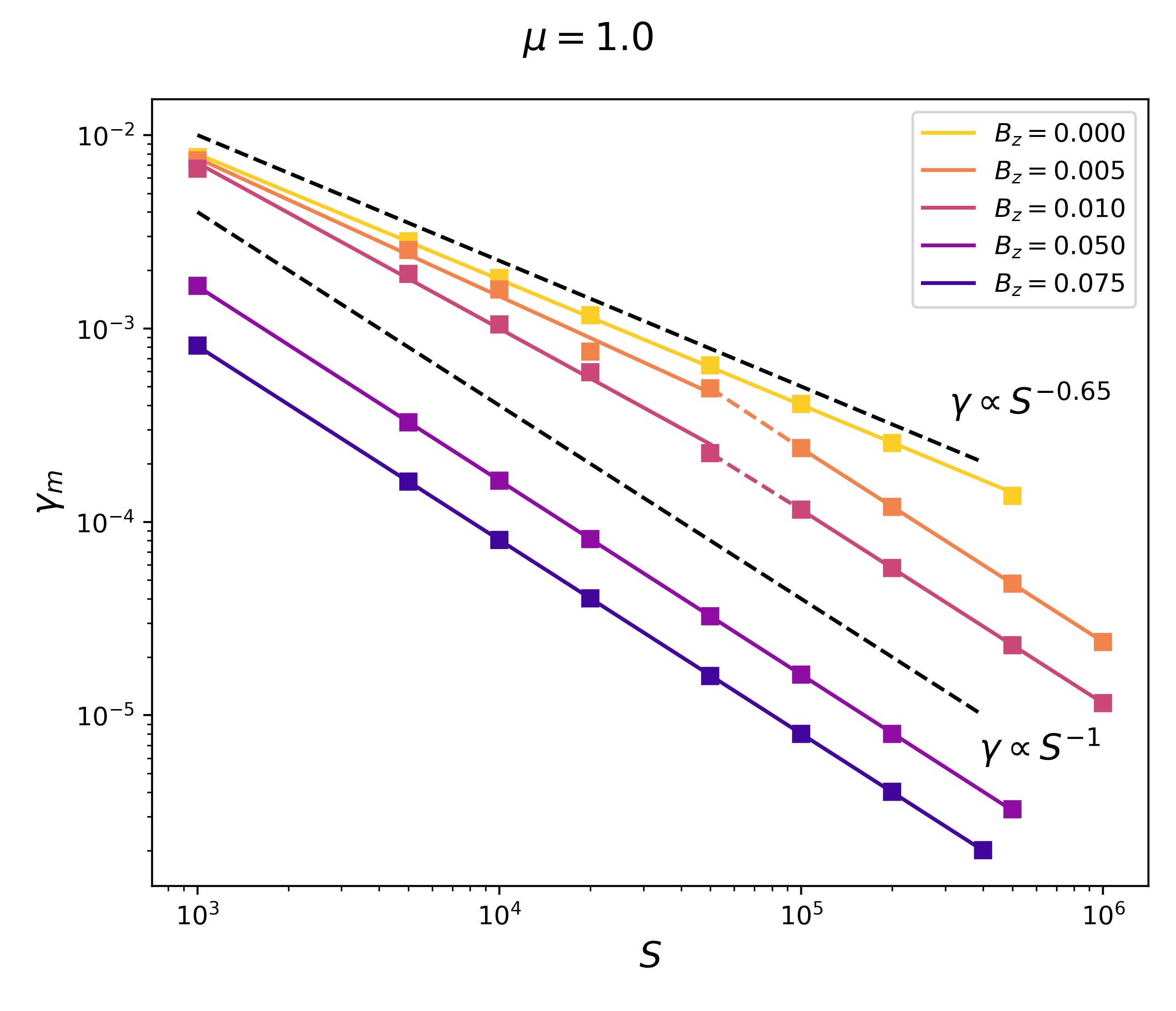}
    \caption{Maximum growth rate $\gamma_m$ as a function of $S$ for $\mu=1$ and varying $B_z$ in log-log scale. The squares are the numerical results and the solid lines are linear-fittings of the squares. For $B_z=0.005$ and $B_z=0.01$ we apply a two-segment linear-fitting as the slope of the curve is clearly non-constant. The two black dashed lines are $\gamma \propto S^{-0.65}$ and $\gamma \propto S^{-1}$ for reference.}
    \label{fig:max_gamma_different_Bz}
\end{figure}

In analysis of the tearing instability, how the maximum growth rate scales with the Lundquist number is important. In Figure \ref{fig:max_gamma_different_Bz}, we show the maximum growth rate $\gamma_m$, which is the peak value of each $\gamma-k$ curve such as those shown in Figure \ref{fig:disp_rel_different_Bz}, as a function of the Lundquist number $S$ for $\mu = 1$ and different values of $B_z$ in log-log scale. In the figure, squares are the numerical results and solid lines are linear-fittings of the squares. Obviously, $\gamma_m$ decreases with increasing $B_z$, consistent with the results shown in Figure \ref{fig:disp_rel_different_Bz}. For $B_z = 0$, a clear power-law relation $\gamma_m \propto S^{-0.65}$ is observed. We note that in a current sheet without any flow, the fastest growing tearing mode has a growth rate $\gamma_m \propto S^{-0.5}$. Thus, the scaling $\gamma_m \propto S^{-0.65}$ indicates that the current sheet is more stable with the Alfv\'enic background flow than without the flow. As we increase $B_z$, the $\gamma_m - S$ line is no longer straight, as can be seen from the curves $B_z=0.005$ and $B_z = 0.01$, whose slopes gradually transit to $-1$ at large Lundquist numbers. When $B_z$ is large enough, i.e. $B_z=0.05$ and $B_z = 0.075$, the whole $\gamma_m -S$ curve in the regime $S \geq 10^3$ is straight with a slope $-1$. We note that, a growth rate $\gamma \propto S^{-1}$ implies that $\gamma \propto \eta/a^2$, that is to say, there is actually no ``growth'' of instability in the system because the growth rate is supported purely by the diffusion of the background magnetic field. Thus, Figure \ref{fig:max_gamma_different_Bz} shows that, a finite $B_z$ significantly stabilizes the current sheet: when either $B_z$ or $S$ is large enough, the tearing instability vanishes and only the diffusion is taking effect in transferring the energy from the background magnetic field to the perturbed fields.

\subsection{Effect of $\mu<1$}
\begin{figure}[htb!]
    \centering
    \includegraphics[width=\hsize]{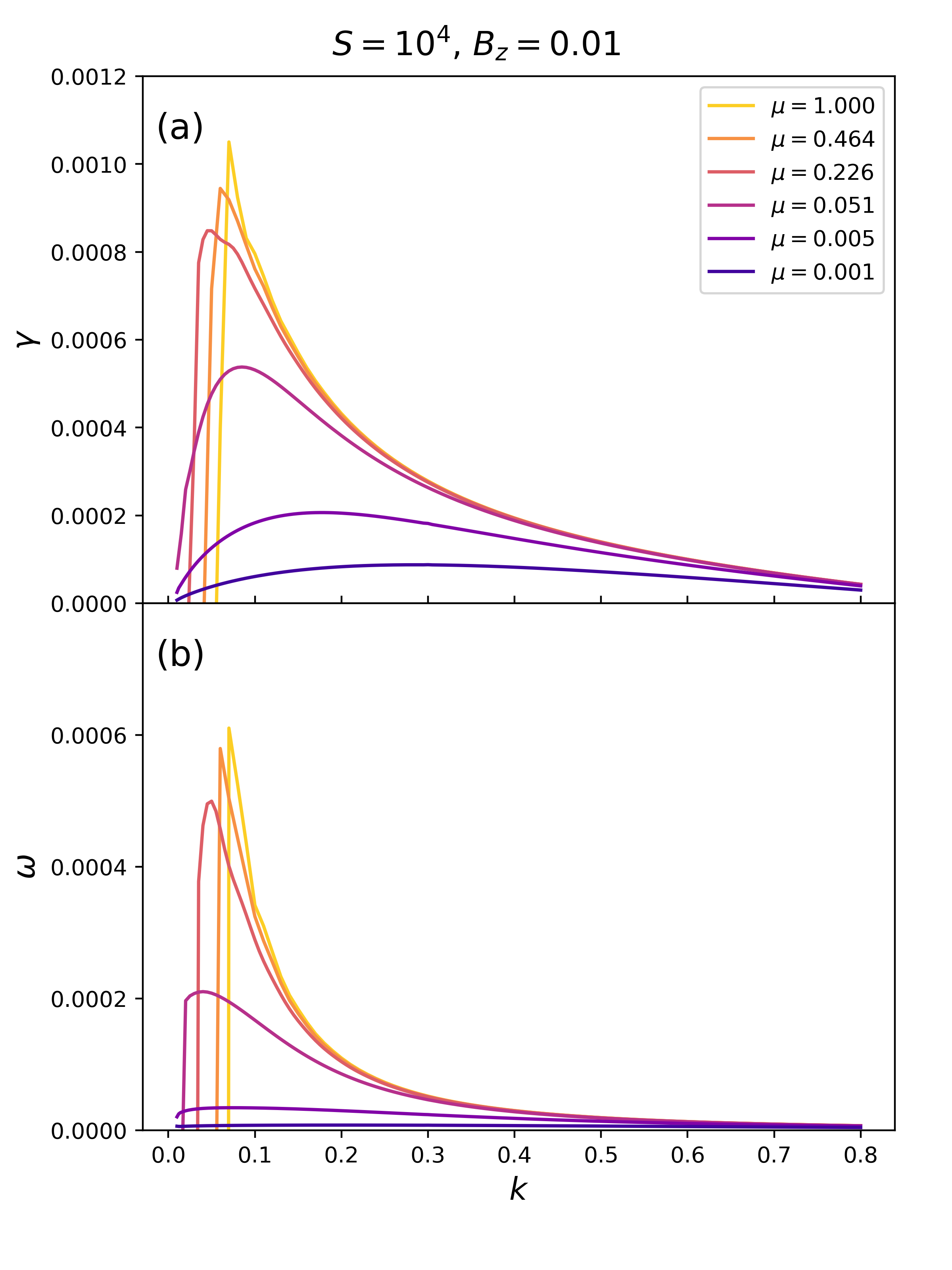}
    \caption{(a) $\gamma-k$ and (b) $\omega-k$ curves for $S=10^4$, $B_z=0.01$ and varying $\mu$. The light (yellow) to dark (purple) colors correspond to $\mu$ decreasing from 1.0 to 0.001.}
    \label{fig:disp_rel_different_mu}
\end{figure}

\begin{figure}[htb!]
    \centering
    \includegraphics[width=\hsize]{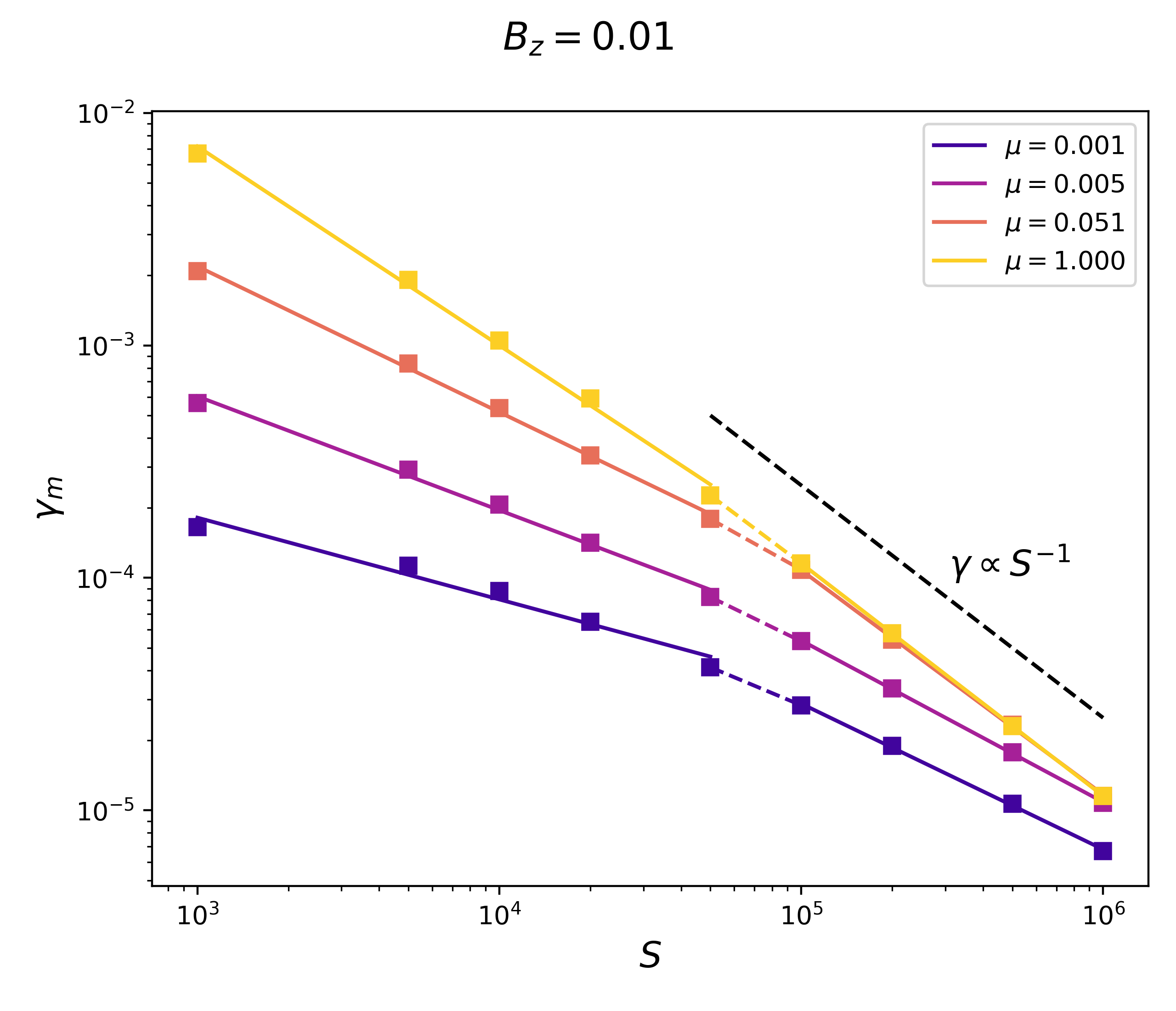}
    \caption{Maximum growth rate $\gamma_m$ as a function of $S$ for $B_z=0.01$ and varying $\mu$. The squares are the numerical results and the solid lines are linear-fittings of the squares. The black dashed line is $\gamma \propto S^{-1}$ for reference.}
    \label{fig:max_gamma_different_mu}
\end{figure}

In Figure \ref{fig:disp_rel_different_mu}, we plot the $\gamma-k$ (panel (a)) and $\omega-k$ (panel (b)) curves for $S=10^4$, $B_z=0.01$, and varying $\mu$, i.e. the average ion flow speed. From panel (a), we see that the growth rate decreases with $\mu$ in general. In Figure \ref{fig:max_gamma_different_mu}, we plot the maximum growth rate $\gamma_m$ as a function of the Lundquist number $S$ for $B_z=0.01$ and different values of $\mu$ in log-log scale. Apparently, the $\gamma_m-S$ relation is not a single power law and the $\gamma_m-S$ curve steepens with $S$. Similar to what is shown in Figure \ref{fig:max_gamma_different_Bz}, for very large $S$ the relation converges to $\gamma_m \propto S^{-1}$, i.e. a status where the instability is merely a result of diffusion.

\begin{figure}[htb!]
    \centering
    \includegraphics[width=\hsize]{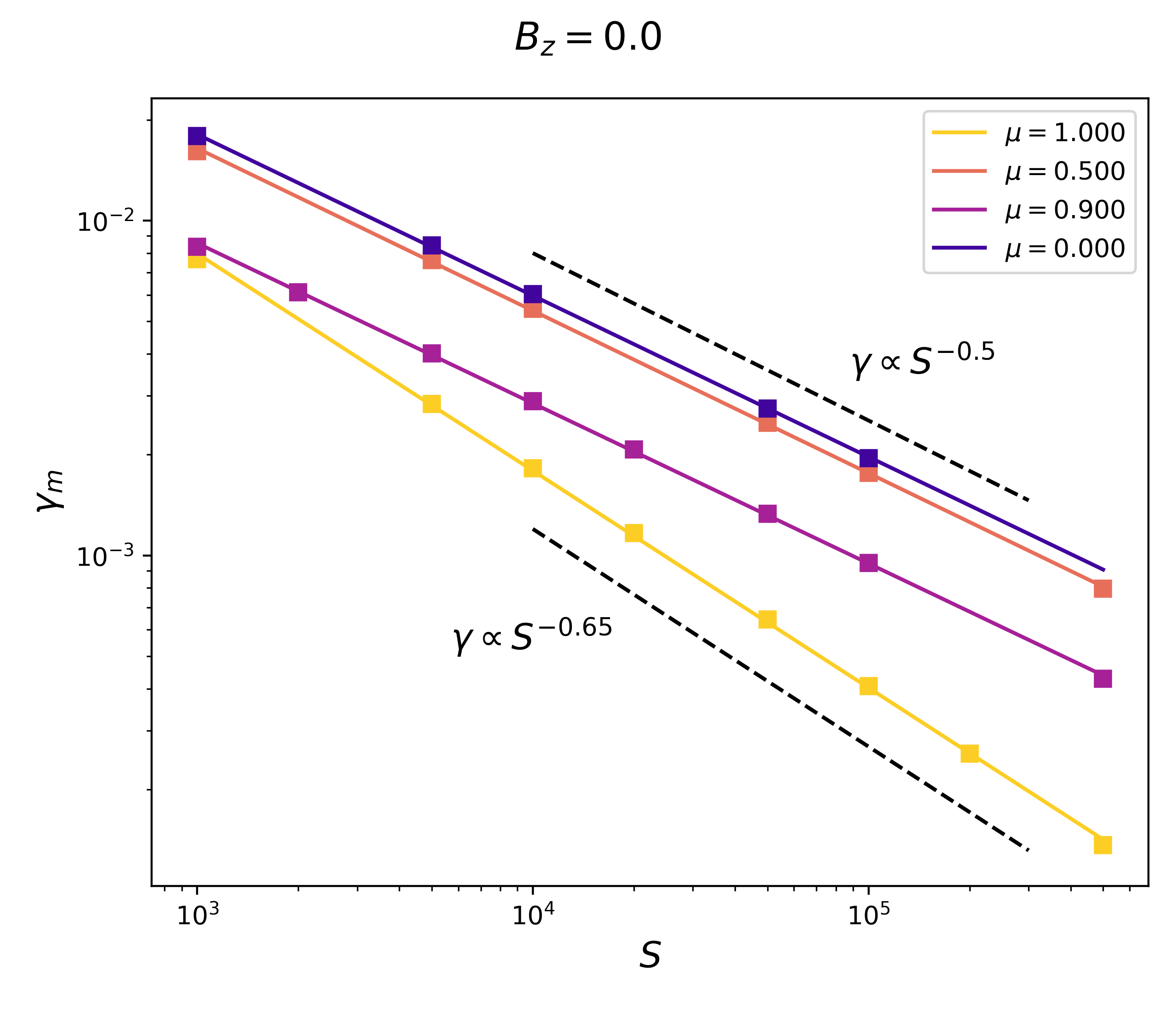}
    \caption{Maximum growth rate $\gamma_m$ as a function of $S$ for $B_z=0$ and varying $\mu$ based on one-fluid MHD model. The squares are the numerical results and the solid lines are linear-fittings of the squares. The black dashed lines are $\gamma \propto S^{-0.5}$ and $\gamma \propto S^{-0.65}$ for reference.}
    \label{fig:max_gamma_different_mu_Bz=0}
\end{figure}

From Figures \ref{fig:disp_rel_different_mu} \& \ref{fig:max_gamma_different_mu}, we can see that as $\mu \rightarrow 0$, the growth rate converges to zero. This is the case $\alpha_+ = \alpha_- = 1$ where the two ion populations counter stream at exactly the same speed $V_{A0}$. In this case, $\mu =\sigma = 0$ and Eq (\ref{eq:uz}) reduces to $\gamma(u_z^{\prime\prime} - k^2 u_z) = 0$, which naturally gives the solution $\gamma  =0$ since otherwise we are left with $u_z^{\prime\prime} - k^2 u_z = 0$ which does not have a solution of $u_z$ that decays to zero at $|z| \rightarrow \pm \infty$. However, we should point out that, the average speed $\mu$ in this two-proton model is not equivalent to the flow speed in the one-fluid MHD model. Consider the case $B_z = 0$ so that we are freely to add a one-fluid shear flow of any speed as long as its velocity is parallel to the background magnetic field. In Figure \ref{fig:max_gamma_different_mu_Bz=0}, we show the maximum growth rate $\gamma_m$ as a function of $S$ for $B_z=0$ and different flow speed $\mu$. However, different from Figure \ref{fig:max_gamma_different_mu}, the results are calculated based on the one-fluid MHD model, i.e. there is only one proton population. Figure \ref{fig:max_gamma_different_mu_Bz=0} shows that, in the MHD case, $\gamma_m$ decreases with $\mu$, at least for $\mu \leq 1$ \cite{chen1997tearing}. For $\mu > 1$, the shear flow is super-Alfv\'enic and thus Kelvin-Helmholtz instability will arise \cite{wang1988streaming}. Except for $\mu = 1.0$, $\gamma_m$ scales with $S$ as $\gamma_m \propto S^{-0.5}$. That is to say, although the background flow suppresses the tearing instability, the scaling relation of the maximum growth rate with $S$ does not change. However, as $\mu$ approaches unity, the scaling relation changes to $\gamma_m \propto S^{-0.65}$, implying that the system becomes more stable, consistent with previous studies (\cite{dahlburg1997evolution} and \cite{einaudi1986resistive}) which show that the growth rate scales as $S^{-\beta}$ with $1/2<\beta<2/3$. But We note here that, these studies claim that in resistive-MHD regime the current sheet with Alfv\'enic plasma flow (i.e., our $\mu=1$ case) is stable. However, our numerical results indicate that this case is not ideally stable but has finite positive growth rate as shown in Figure \ref{fig:disp_rel_different_Bz} \& \ref{fig:max_gamma_different_mu_Bz=0}. In general, we can write $\gamma_m \tau_a \sim S^{-\beta}$ where $S = a V_{A0} / \eta$ and $\tau_a = a/V_{A0}$. We note that this expression is for a 1D current sheet which is infinitely long. In practice, consider a current sheet of finite length $L$, we usually need to measure the time scale by $\tau_L = L / V_{A0}$ and we define the Lundquist number by $S_L = LV_{A0}/ \eta$. This transforms the scaling relation to $\gamma_m \tau_L \sim S_L^{-\beta} (a/L)^{-\beta-1}$. Apparently, the inverse aspect ratio $a/L$ is a key parameter determining the growth rate $\gamma_m \tau_L$. For an arbitrary inverse aspect ratio $a/L \sim S_L^{-\delta}$, we get $\gamma_m \tau_L \sim S_L^{-\beta+\delta(\beta+1)}$, implying a threshold $\delta = \beta/(\beta+1)$, at which $\gamma_m \tau_L \sim O(1)$ is achieved and this is the so-called ``ideal tearing'' \cite{PucciandVelli2013,tenerani2016ideally,pucci2020onset}. For the classical tearing $\beta = 1/2$, we have $\delta = 1/3$, that is to say when a macroscopic current thins to $a/L \sim S_L^{-1/3}$, the growth rate of the tearing mode transits from extremely small (assuming $S_L \rightarrow \infty $) to unity and thus the current sheet breaks up rapidly. As $\beta$ increases, such as in the MHD case with Alfv\'enic flow shown in Figure \ref{fig:max_gamma_different_mu_Bz=0} or in the two-ion case shown in Figures \ref{fig:max_gamma_different_Bz} \& \ref{fig:max_gamma_different_mu}, the critical value of $\delta$ also increases, indicating that the current sheet must be thinner in order to achieve fast growth of the tearing instability. For example, for $\beta \approx 2/3$ as in the Alfv\'enic flow case, we have $\delta = 2/5>1/3$ and especially, for $\beta = 1$, we get the threshold $a/L \sim S_L^{-1/2}$, indicating that fast growth of instability happens only when Sweet-Parker type current sheet is formed. This is because, as we discussed before, $\gamma_m \tau_a \sim S^{-1}$ implies that the growth of instability is fully supported by diffusion.

\subsection{Eigen-functions}
\begin{figure}[htb!]
    \centering
    \includegraphics[width=\hsize]{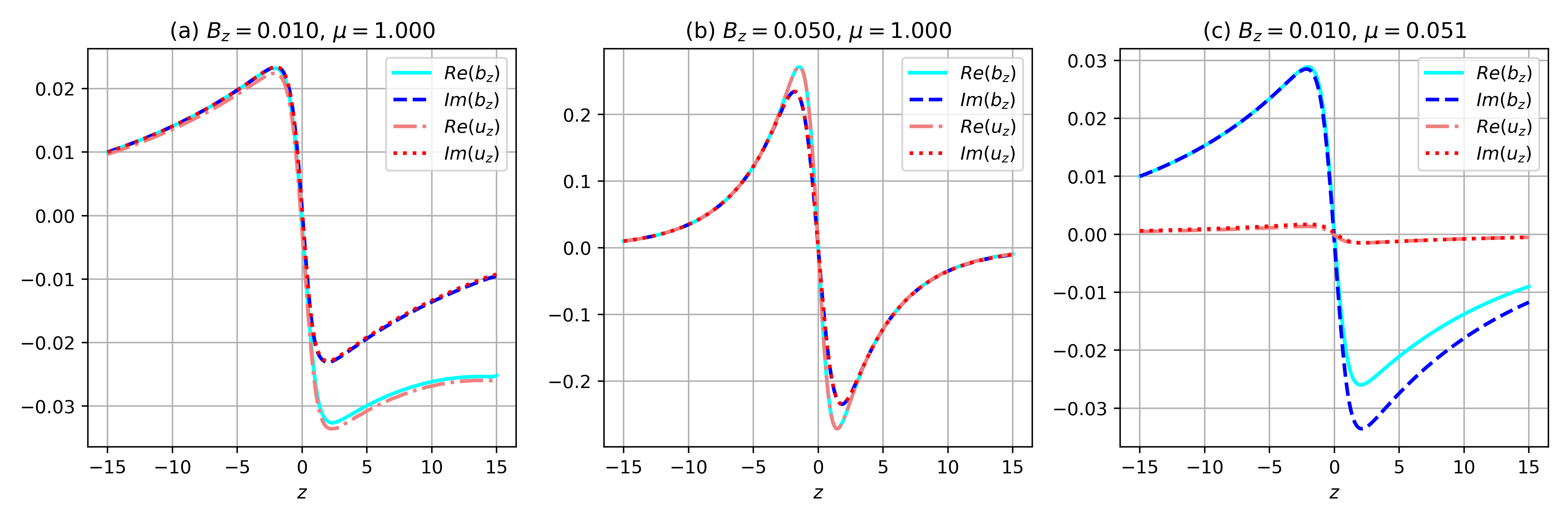}
    \caption{Numerically solved eigen-functions $u_z$ and $b_z$ for three sets of parameters: (a) $B_z=0.01$, $\mu=1.0$. (b) $B_z=0.05$, $\mu=1.0$. (c) $B_z=0.01$, $\mu=0.051$. Cyan solid curves are the real parts of $b_z$, blue dashed curves are the imaginary parts of $b_z$, coral dashed-dotted curves are the real parts of $u_z$, and red dotted curves are the imaginary parts of $u_z$.}
    \label{fig:eigenfunctions}
\end{figure}

In Figure \ref{fig:eigenfunctions}, we show the eigen-functions $u_z$ and $b_z$ for three sets of parameters. Panel (a) is $B_z=0.01$, $\mu=1.0$, panel (b) is $B_z=0.05$, $\mu=1.0$, and panel (c) is $B_z=0.01$, $\mu=0.051$. The shapes of the eigen-functions do not differ significantly among the three cases, though the relative amplitudes of $u_z$ and $b_z$ change with the parameters. The most outstanding feature of these eigen-functions is the quasi-odd function $b_z(z)$. It is known that for the classical tearing mode, $b_z$ is symmetric in $z$ such that $b_z(z=0)>0$, which provides the magnetic flux necessary for the reconnection to happen. Meanwhile, $u_z$ is naturally asymmetric in $z$ as $z=0$ is the stagnation point of the flow. From Figure \ref{fig:eigenfunctions}, we can see that $b_z$ is nearly asymmetric in $z$ so that $b_z(z=0)$ is very small, limiting the reconnection rate. In addition, as the function $b_z(z)$ is very different from the classical tearing case, plasmoids will not be generated at the center of the current sheet but on the two sides of the the current sheet instead.

\section{Discussion}\label{sec:discussion}

In this study we investigate the stability of 1D current sheet with a finite normal component of magnetic field $B_z\ne0$ and the tension force $j_yB_z/c$ balanced by the plasma flows. Two main results are: (1) strong stabilizing effect of $B_z$ in such 1D current sheets and (2) possible destabilization by imbalance of counter-streaming flows, i.e. by field-aligned net flow. Next, we will discuss these results in the context of the Earth’s magnetotail dynamics.

\subsection{Current sheet instability at substorm onset}
As observations suggest the formation of thin (almost 1D) current sheets during substorm growth phase \cite{Petrukovich13,Artemyev15:grl:2dCS,Sitnov21}, there is a natural interest to investigate the instability of such current sheet configuration. The main idea is that 1D current sheets with nongyrotropic pressure tensor (i.e., with $B_zj_y/c\approx\partial p_{xz}/\partial z$) would be more unstable than classical 2D current sheets (with $B_zj_y/c\approx\partial p/\partial x$), and this resolves the issue of magnetic reconnection onset during weak substorms without strong external drivers (see discussion in \citeA{Zelenyi08JASTP}). We use multi-fluid current sheet model with the counter-streaming plasma flows generating such nongyrotropy ($p_{xz}\ne 0$). Our results show that we cannot make current sheet with $B_z\ne0$ more unstable solely by substituting $\partial p/\partial x$ by $\partial p_{xz}/\partial z$. This result suggests that the investigation of spontaneous (undriven) instability in the magnetotail current sheet requires more specific current sheet configurations. One of the ideas has been proposed by \citeA{Sitnov10} for 2D current sheets and confirmed in a set of MHD \cite{Merkin15, Birn18} and kinetic \cite{Sitnov14, Bessho&Bhattacharjee14, Pritchett15:pop} simulations which show that current sheets with non-monotonical $B_z(x)$ profiles can be more unstable. Alternative ideas would be needed to explain spontaneous reconnection in 1D current sheets (with $B_z=const$), and we showed that such ideas should include more than pressure nongyrotropy.

\subsection{Field-aligned plasma flows}
We generalize the 1D current sheet model of \cite{Steinhauer08} by including a net plasma flow. Such flow is shown be able to reduce the stabilization effect of $B_z\ne 0$. This is an interesting effect, because similar field-aligned flows can be included into 2D current sheet equilibria in application to the magnetotail and solar flares \cite{Birn92,Hau96:pop,Nickeler&Wiegelmann10}. Using an approach proposed by \citeA{Birn91:pop, Birn92}, one can introduce arbitrary flow along magnetic field lines in a single MHD, and such flow may influence the stability of 2D current sheets, i.e. altering the magnitude of inverse $\partial B_z/\partial x$ gradient that is considered to destabilize such current sheets \cite{Merkin15,Birn18}. Note, this destabilizing effect of field-aligned plasma flow differs from the stabilizing effect of cross-field diamagnetic flow considered by \citeA{Swisdak03}. Indeed, we focus here on the stability conditions of structures with a finite normal component, whereas the \cite{Swisdak03} model and its observational confirmations \cite{Phan10, Phan14} were obtained for much more unstable tangential discontinuities without a normal component $B_n$. Our results show that current sheets with $B_n\ne0$ and plasma flow equal to the Alfv\'en speed are completely stable to tearing, and can therefore survive for a long time, e.g., during the long lasting substorm growth phase. The same result explains the  absence of magnetic reconnection signatures in multiple rotational discontinuities observed in the near-Sun region \cite{Phan20}. 

Obtained results generalize the previous investigations of the plasma flow effects on the current sheet instability in the absence of $B_z$, i.e. for 1D tangential discontinuities with 1D \cite{chen1997tearing,wang1988streaming,Hoshino&Higashimori15} and 2D plasma flows \cite{Phan&Sonnerup91,Ip&Sonnerup96}. The effect of decrease of the growth rate for the long wavelength tearing mode due to the sub-Alfv\'enic plasma flow, found for such tangential discontinuities \cite{Bulanov78}, has been revealed for the 1D current sheet with $B_z\ne 0$. However, contrast to generally stabilizing effect of the sub-Alfv\'enic plasma flow on the tearing more in tangential discontinuity \cite{dahlburg1997evolution,einaudi1986resistive}, in the current sheet with $B_z\ne 0$ the field-aligned flow increases the peak growth rate, i.e. the most stable current sheet with $B_z\ne 0$ is formed by two counter-streaming flows, whereas any imbalance between these two flows provide an addition free energy for the tearing mode. This effect resembles the effect of super-Alfv\'enic flow driving the tearing mode in the tangential discontinuities \cite{wang1988streaming}.

\subsection{Speculation on possible drivers of magnetic reconnection in the magnetotail}
Our results show that 1D current sheet, such as ones formed at the late stage of the substorm growth phase, is well stable relative to the tearing mode. Together with results of 2D current sheet stability (see discussion in \citeA{Sitnov02}), our results rise the question of initialization of substorm reconnection in absence of strong external driver (i.e. spontaneous reconnection). One of the solutions of this issue has been proposed by \cite{Sitnov10,Sitnov13} for 2D current sheets with nonmonotonical $B_z(x)$ profiles. This solution has been shown to work for current sheets with $\partial p/\partial x=B_zj_y/c$, and one of the prospective directions of further investigation is the merging of models of such 2D current sheets and current sheet with field-aligned plasma flows. Inclusion of such flows (in multi-fluid approach) will relax $\partial p/\partial x=B_zj_y/c$ condition by addition $\partial p_{xz}/\partial z$ term, and thus realistic quasi-1D current sheets with a weak $\partial p/\partial x$ can be described. From other point of view, even weak $\partial p/\partial x$ allows inclusion of a nonmonotonical $B_z(x)$ profiles as a possible source of free energy for the tearing mode. Another prospective approach consists further generalization of multi-fluid models with inclusion of realistic anisotropic electron components (see discussion in \citeA{Artemyev20:jgr:electrons}). Intense electron currents in high-beta plasma of thin current sheets may provide an additional free energy source for the tearing mode excitation.

\section{Conclusions}\label{sec:conclusion}
In this work, we carried out linear stability analysis of the 1D MHD current sheet with a uniform normal magnetic field ($B_z$ in the magnetotail coordinates). In our model, the magnetic tension force is balanced by the field-aligned ion flow, which is generalized by assuming that ions shape two populations of the same density but with different field-aligned velocities. The main results of our calculation are 
\begin{enumerate}
    \item The existence of a finite $B_z$ significantly stabilizes the tearing instability. With a finite $B_z$, the scaling relation between the maximum growth rate of the instability ($\gamma_m$) and the Lundquist number ($S$) converges to $\gamma_m \propto S^{-1}$ in the limit $S \rightarrow \infty$, indicating the instability is purely diffusion-supported.
    
    \item In this two-ion model, the most stable case is when the two ion populations counter-stream at the same speed $V = B/\sqrt{4\pi \rho}$ such that the net flow velocity is zero. In this case the instability does not grow at all. An imbalance between the two ion populations, which lead to a non-zero net flow, destabilizes the current sheet. The most unstable case is when the two ion populations stream in the same direction at the same speed $V = B/\sqrt{4\pi \rho}$. This is the one-fluid MHD case where the average flow velocity is $\mb{B}/\sqrt{4\pi \rho}$.
\end{enumerate}
Our results show that, in the magnetotail, a 1D, i.e. very long and thin, current sheet which has a finite $B_z$ component is very stable to the tearing instability. To answer the question of the onset of fast spontaneous reconnection in the magnetotail, more ingredients are required, such as the ion/electron kinetic effects.

\acknowledgments
This work was supported by the NASA HERMES DRIVE Science Center grant No. 80NSSC20K0604. The boundary-value-problem solver used in this study can be found at {https://docs.scipy.org/doc/scipy} and is described in \cite{Virtanenetal2020Scipy}.



\begin{thebibliography}{}

\bibitem [\protect \citeauthoryear {%
{Angelopoulos}%
, {Artemyev}%
, {Phan}%
\BCBL {}\ \BBA {} {Miyashita}%
}{%
{Angelopoulos}%
\ \protect \BOthers {.}}{%
{\protect \APACyear {2020}}%
}]{%
Angelopoulos20}
\APACinsertmetastar {%
Angelopoulos20}%
\begin{APACrefauthors}%
{Angelopoulos}, V.%
, {Artemyev}, A.%
, {Phan}, T\BPBI D.%
\BCBL {}\ \BBA {} {Miyashita}, Y.%
\end{APACrefauthors}%
\unskip\
\newblock
\APACrefYearMonthDay{2020}{{\APACmonth{01}}}{}.
\newblock
{\BBOQ}\APACrefatitle {{Near-Earth magnetotail reconnection powers space
  storms}} {{Near-Earth magnetotail reconnection powers space storms}}.{\BBCQ}
\newblock
\APACjournalVolNumPages{Nature Physics}{16}{3}{317-321}.
\newblock
\begin{APACrefDOI} \doi{10.1038/s41567-019-0749-4} \end{APACrefDOI}
\PrintBackRefs{\CurrentBib}

\bibitem [\protect \citeauthoryear {%
{Angelopoulos}%
\ \protect \BOthers {.}}{%
{Angelopoulos}%
\ \protect \BOthers {.}}{%
{\protect \APACyear {2008}}%
}]{%
Angelopoulos08}
\APACinsertmetastar {%
Angelopoulos08}%
\begin{APACrefauthors}%
{Angelopoulos}, V.%
, {McFadden}, J\BPBI P.%
, {Larson}, D.%
, {Carlson}, C\BPBI W.%
, {Mende}, S\BPBI B.%
, {Frey}, H.%
\BDBL {}{Kepko}, L.%
\end{APACrefauthors}%
\unskip\
\newblock
\APACrefYearMonthDay{2008}{{\APACmonth{08}}}{}.
\newblock
{\BBOQ}\APACrefatitle {{Tail Reconnection Triggering Substorm Onset}} {{Tail
  Reconnection Triggering Substorm Onset}}.{\BBCQ}
\newblock
\APACjournalVolNumPages{Science}{321}{}{931-935}.
\newblock
\begin{APACrefDOI} \doi{10.1126/science.1160495} \end{APACrefDOI}
\PrintBackRefs{\CurrentBib}

\bibitem [\protect \citeauthoryear {%
{Angelopoulos}%
\ \protect \BOthers {.}}{%
{Angelopoulos}%
\ \protect \BOthers {.}}{%
{\protect \APACyear {2013}}%
}]{%
Angelopoulos13}
\APACinsertmetastar {%
Angelopoulos13}%
\begin{APACrefauthors}%
{Angelopoulos}, V.%
, {Runov}, A.%
, {Zhou}, X\BPBI Z.%
, {Turner}, D\BPBI L.%
, {Kiehas}, S\BPBI A.%
, {Li}, S\BPBI S.%
\BCBL {}\ \BBA {} {Shinohara}, I.%
\end{APACrefauthors}%
\unskip\
\newblock
\APACrefYearMonthDay{2013}{}{}.
\newblock
{\BBOQ}\APACrefatitle {{Electromagnetic Energy Conversion at Reconnection
  Fronts}} {{Electromagnetic Energy Conversion at Reconnection Fronts}}.{\BBCQ}
\newblock
\APACjournalVolNumPages{Science}{341}{}{1478-1482}.
\newblock
\begin{APACrefDOI} \doi{10.1126/science.1236992} \end{APACrefDOI}
\PrintBackRefs{\CurrentBib}

\bibitem [\protect \citeauthoryear {%
{Artemyev}%
, {Angelopoulos}%
, {Runov}%
\BCBL {}\ \BBA {} {Zhang}%
}{%
{Artemyev}%
, {Angelopoulos}%
, {Runov}%
\BCBL {}\ \BBA {} {Zhang}%
}{%
{\protect \APACyear {2020}}%
}]{%
Artemyev20:jgr:outflow}
\APACinsertmetastar {%
Artemyev20:jgr:outflow}%
\begin{APACrefauthors}%
{Artemyev}, A\BPBI V.%
, {Angelopoulos}, V.%
, {Runov}, A.%
\BCBL {}\ \BBA {} {Zhang}, X\BPBI J.%
\end{APACrefauthors}%
\unskip\
\newblock
\APACrefYearMonthDay{2020}{{\APACmonth{07}}}{}.
\newblock
{\BBOQ}\APACrefatitle {{Ionospheric Outflow During the Substorm Growth Phase:
  THEMIS Observations of Oxygen Ions at the Plasma Sheet Boundary}}
  {{Ionospheric Outflow During the Substorm Growth Phase: THEMIS Observations
  of Oxygen Ions at the Plasma Sheet Boundary}}.{\BBCQ}
\newblock
\APACjournalVolNumPages{Journal of Geophysical Research (Space
  Physics)}{125}{7}{e27612}.
\newblock
\begin{APACrefDOI} \doi{10.1029/2019JA027612} \end{APACrefDOI}
\PrintBackRefs{\CurrentBib}

\bibitem [\protect \citeauthoryear {%
{Artemyev}%
, {Angelopoulos}%
, {Vasko}%
\BCBL {}\ \protect \BOthers {.}}{%
{Artemyev}%
, {Angelopoulos}%
, {Vasko}%
\BCBL {}\ \protect \BOthers {.}}{%
{\protect \APACyear {2020}}%
}]{%
Artemyev20:jgr:electrons}
\APACinsertmetastar {%
Artemyev20:jgr:electrons}%
\begin{APACrefauthors}%
{Artemyev}, A\BPBI V.%
, {Angelopoulos}, V.%
, {Vasko}, I\BPBI Y.%
, {Petrukovich}, A\BPBI A.%
, {Runov}, A.%
, {Saito}, Y.%
\BDBL {}{Strangeway}, R\BPBI J.%
\end{APACrefauthors}%
\unskip\
\newblock
\APACrefYearMonthDay{2020}{}{}.
\newblock
{\BBOQ}\APACrefatitle {Contribution of Anisotropic Electron Current to the
  Magnetotail Current Sheet as a Function of Location and Plasma Conditions}
  {Contribution of anisotropic electron current to the magnetotail current
  sheet as a function of location and plasma conditions}.{\BBCQ}
\newblock
\APACjournalVolNumPages{Journal of Geophysical Research: Space
  Physics}{125}{1}{e2019JA027251}.
\newblock
\begin{APACrefURL}
  \url{https://agupubs.onlinelibrary.wiley.com/doi/abs/10.1029/2019JA027251}
  \end{APACrefURL}
\newblock
\APACrefnote{e2019JA027251 10.1029/2019JA027251}
\newblock
\begin{APACrefDOI} \doi{10.1029/2019JA027251} \end{APACrefDOI}
\PrintBackRefs{\CurrentBib}

\bibitem [\protect \citeauthoryear {%
{Artemyev}%
\ \protect \BOthers {.}}{%
{Artemyev}%
\ \protect \BOthers {.}}{%
{\protect \APACyear {2019}}%
}]{%
Artemyev19:jgr:ions}
\APACinsertmetastar {%
Artemyev19:jgr:ions}%
\begin{APACrefauthors}%
{Artemyev}, A\BPBI V.%
, {Angelopoulos}, V.%
, {Vasko}, I\BPBI Y.%
, {Zhang}, X\BPBI J.%
, {Runov}, A.%
\BCBL {}\ \BBA {} {Zelenyi}, L\BPBI M.%
\end{APACrefauthors}%
\unskip\
\newblock
\APACrefYearMonthDay{2019}{May}{}.
\newblock
{\BBOQ}\APACrefatitle {{Ion Anisotropy in Earth's Magnetotail Current Sheet:
  Multicomponent Ion Population}} {{Ion Anisotropy in Earth's Magnetotail
  Current Sheet: Multicomponent Ion Population}}.{\BBCQ}
\newblock
\APACjournalVolNumPages{Journal of Geophysical Research (Space
  Physics)}{124}{5}{3454-3467}.
\newblock
\begin{APACrefDOI} \doi{10.1029/2019JA026604} \end{APACrefDOI}
\PrintBackRefs{\CurrentBib}

\bibitem [\protect \citeauthoryear {%
{Artemyev}%
, {Petrukovich}%
, {Nakamura}%
\BCBL {}\ \BBA {} {Zelenyi}%
}{%
{Artemyev}%
\ \protect \BOthers {.}}{%
{\protect \APACyear {2010}}%
}]{%
Artemyev10:jgr}
\APACinsertmetastar {%
Artemyev10:jgr}%
\begin{APACrefauthors}%
{Artemyev}, A\BPBI V.%
, {Petrukovich}, A\BPBI A.%
, {Nakamura}, R.%
\BCBL {}\ \BBA {} {Zelenyi}, L\BPBI M.%
\end{APACrefauthors}%
\unskip\
\newblock
\APACrefYearMonthDay{2010}{{\APACmonth{12}}}{}.
\newblock
{\BBOQ}\APACrefatitle {{Proton velocity distribution in thin current sheets:
  Cluster observations and theory of transient trajectories}} {{Proton velocity
  distribution in thin current sheets: Cluster observations and theory of
  transient trajectories}}.{\BBCQ}
\newblock
\APACjournalVolNumPages{J. Geophys. Res.}{115}{}{A12255}.
\newblock
\begin{APACrefDOI} \doi{10.1029/2010JA015702} \end{APACrefDOI}
\PrintBackRefs{\CurrentBib}

\bibitem [\protect \citeauthoryear {%
{Artemyev}%
, {Petrukovich}%
, {Nakamura}%
\BCBL {}\ \BBA {} {Zelenyi}%
}{%
{Artemyev}%
\ \protect \BOthers {.}}{%
{\protect \APACyear {2015}}%
}]{%
Artemyev15:grl:2dCS}
\APACinsertmetastar {%
Artemyev15:grl:2dCS}%
\begin{APACrefauthors}%
{Artemyev}, A\BPBI V.%
, {Petrukovich}, A\BPBI A.%
, {Nakamura}, R.%
\BCBL {}\ \BBA {} {Zelenyi}, L\BPBI M.%
\end{APACrefauthors}%
\unskip\
\newblock
\APACrefYearMonthDay{2015}{{\APACmonth{05}}}{}.
\newblock
{\BBOQ}\APACrefatitle {{Two-dimensional configuration of the magnetotail
  current sheet: THEMIS observations}} {{Two-dimensional configuration of the
  magnetotail current sheet: THEMIS observations}}.{\BBCQ}
\newblock
\APACjournalVolNumPages{\grl}{42}{}{3662-3667}.
\newblock
\begin{APACrefDOI} \doi{10.1002/2015GL063994} \end{APACrefDOI}
\PrintBackRefs{\CurrentBib}

\bibitem [\protect \citeauthoryear {%
{Artemyev}%
\ \protect \BOthers {.}}{%
{Artemyev}%
\ \protect \BOthers {.}}{%
{\protect \APACyear {2008}}%
}]{%
Artemyev08:angeo}
\APACinsertmetastar {%
Artemyev08:angeo}%
\begin{APACrefauthors}%
{Artemyev}, A\BPBI V.%
, {Petrukovich}, A\BPBI A.%
, {Zelenyi}, L\BPBI M.%
, {Malova}, H\BPBI V.%
, {Popov}, V\BPBI Y.%
, {Nakamura}, R.%
\BDBL {}{Apatenkov}, S.%
\end{APACrefauthors}%
\unskip\
\newblock
\APACrefYearMonthDay{2008}{{\APACmonth{09}}}{}.
\newblock
{\BBOQ}\APACrefatitle {{Comparison of multi-point measurements of current sheet
  structure and analytical models}} {{Comparison of multi-point measurements of
  current sheet structure and analytical models}}.{\BBCQ}
\newblock
\APACjournalVolNumPages{Annales Geophysicae}{26}{}{2749-2758}.
\PrintBackRefs{\CurrentBib}

\bibitem [\protect \citeauthoryear {%
{Artemyev}%
, {Vasko}%
, {Angelopoulos}%
\BCBL {}\ \BBA {} {Runov}%
}{%
{Artemyev}%
\ \protect \BOthers {.}}{%
{\protect \APACyear {2016}}%
}]{%
Artemyev16:pop:cs}
\APACinsertmetastar {%
Artemyev16:pop:cs}%
\begin{APACrefauthors}%
{Artemyev}, A\BPBI V.%
, {Vasko}, I\BPBI Y.%
, {Angelopoulos}, V.%
\BCBL {}\ \BBA {} {Runov}, A.%
\end{APACrefauthors}%
\unskip\
\newblock
\APACrefYearMonthDay{2016}{{\APACmonth{09}}}{}.
\newblock
{\BBOQ}\APACrefatitle {{Effects of electron pressure anisotropy on current
  sheet configuration}} {{Effects of electron pressure anisotropy on current
  sheet configuration}}.{\BBCQ}
\newblock
\APACjournalVolNumPages{Physics of Plasmas}{23}{9}{092901}.
\newblock
\begin{APACrefDOI} \doi{10.1063/1.4961926} \end{APACrefDOI}
\PrintBackRefs{\CurrentBib}

\bibitem [\protect \citeauthoryear {%
{Artemyev}%
\ \protect \BOthers {.}}{%
{Artemyev}%
\ \protect \BOthers {.}}{%
{\protect \APACyear {2014}}%
}]{%
Artemyev14:jgr}
\APACinsertmetastar {%
Artemyev14:jgr}%
\begin{APACrefauthors}%
{Artemyev}, A\BPBI V.%
, {Walsh}, A\BPBI P.%
, {Petrukovich}, A\BPBI A.%
, {Baumjohann}, W.%
, {Nakamura}, R.%
\BCBL {}\ \BBA {} {Fazakerley}, A\BPBI N.%
\end{APACrefauthors}%
\unskip\
\newblock
\APACrefYearMonthDay{2014}{{\APACmonth{09}}}{}.
\newblock
{\BBOQ}\APACrefatitle {{Electron pitch angle/energy distribution in the
  magnetotail}} {{Electron pitch angle/energy distribution in the
  magnetotail}}.{\BBCQ}
\newblock
\APACjournalVolNumPages{\jgr}{119}{}{7214-7227}.
\newblock
\begin{APACrefDOI} \doi{10.1002/2014JA020350} \end{APACrefDOI}
\PrintBackRefs{\CurrentBib}

\bibitem [\protect \citeauthoryear {%
{Artemyev}%
\ \BBA {} {Zelenyi}%
}{%
{Artemyev}%
\ \BBA {} {Zelenyi}%
}{%
{\protect \APACyear {2013}}%
}]{%
Artemyev&Zelenyi13}
\APACinsertmetastar {%
Artemyev&Zelenyi13}%
\begin{APACrefauthors}%
{Artemyev}, A\BPBI V.%
\BCBT {}\ \BBA {} {Zelenyi}, L\BPBI M.%
\end{APACrefauthors}%
\unskip\
\newblock
\APACrefYearMonthDay{2013}{}{}.
\newblock
{\BBOQ}\APACrefatitle {{Kinetic Structure of Current Sheets in the Earth
  Magnetotail}} {{Kinetic Structure of Current Sheets in the Earth
  Magnetotail}}.{\BBCQ}
\newblock
\APACjournalVolNumPages{Space Sci. Rev.}{178}{}{419-440}.
\newblock
\begin{APACrefDOI} \doi{10.1007/s11214-012-9954-5} \end{APACrefDOI}
\PrintBackRefs{\CurrentBib}

\bibitem [\protect \citeauthoryear {%
{Artemyev}%
, {Zelenyi}%
, {Petrukovich}%
\BCBL {}\ \BBA {} {Nakamura}%
}{%
{Artemyev}%
\ \protect \BOthers {.}}{%
{\protect \APACyear {2011}}%
}]{%
Artemyev11:grl}
\APACinsertmetastar {%
Artemyev11:grl}%
\begin{APACrefauthors}%
{Artemyev}, A\BPBI V.%
, {Zelenyi}, L\BPBI M.%
, {Petrukovich}, A\BPBI A.%
\BCBL {}\ \BBA {} {Nakamura}, R.%
\end{APACrefauthors}%
\unskip\
\newblock
\APACrefYearMonthDay{2011}{{\APACmonth{07}}}{}.
\newblock
{\BBOQ}\APACrefatitle {{Hot electrons as tracers of large-scale structure of
  magnetotail current sheets}} {{Hot electrons as tracers of large-scale
  structure of magnetotail current sheets}}.{\BBCQ}
\newblock
\APACjournalVolNumPages{\grl}{38}{}{L14102}.
\newblock
\begin{APACrefDOI} \doi{10.1029/2011GL047979} \end{APACrefDOI}
\PrintBackRefs{\CurrentBib}

\bibitem [\protect \citeauthoryear {%
Ascher%
, Mattheij%
\BCBL {}\ \BBA {} Russell%
}{%
Ascher%
\ \protect \BOthers {.}}{%
{\protect \APACyear {1994}}%
}]{%
Ascheretal1994numerical}
\APACinsertmetastar {%
Ascheretal1994numerical}%
\begin{APACrefauthors}%
Ascher, U\BPBI M.%
, Mattheij, R\BPBI M.%
\BCBL {}\ \BBA {} Russell, R\BPBI D.%
\end{APACrefauthors}%
\unskip\
\newblock
\APACrefYear{1994}.
\newblock
\APACrefbtitle {Numerical solution of boundary value problems for ordinary
  differential equations} {Numerical solution of boundary value problems for
  ordinary differential equations}\ (\BVOL~13).
\newblock
\APACaddressPublisher{}{Siam}.
\PrintBackRefs{\CurrentBib}

\bibitem [\protect \citeauthoryear {%
{Ashour-Abdalla}%
, {Berchem}%
, {Buechner}%
\BCBL {}\ \BBA {} {Zelenyi}%
}{%
{Ashour-Abdalla}%
\ \protect \BOthers {.}}{%
{\protect \APACyear {1993}}%
}]{%
Maha93}
\APACinsertmetastar {%
Maha93}%
\begin{APACrefauthors}%
{Ashour-Abdalla}, M.%
, {Berchem}, J\BPBI P.%
, {Buechner}, J.%
\BCBL {}\ \BBA {} {Zelenyi}, L\BPBI M.%
\end{APACrefauthors}%
\unskip\
\newblock
\APACrefYearMonthDay{1993}{{\APACmonth{04}}}{}.
\newblock
{\BBOQ}\APACrefatitle {{Shaping of the magnetotail from the mantle - Global and
  local structuring}} {{Shaping of the magnetotail from the mantle - Global and
  local structuring}}.{\BBCQ}
\newblock
\APACjournalVolNumPages{\jgr}{98}{}{5651-5676}.
\newblock
\begin{APACrefDOI} \doi{10.1029/92JA01662} \end{APACrefDOI}
\PrintBackRefs{\CurrentBib}

\bibitem [\protect \citeauthoryear {%
{Ashour-Abdalla}%
, {Frank}%
, {Paterson}%
, {Peroomian}%
\BCBL {}\ \BBA {} {Zelenyi}%
}{%
{Ashour-Abdalla}%
\ \protect \BOthers {.}}{%
{\protect \APACyear {1996}}%
}]{%
Maha96}
\APACinsertmetastar {%
Maha96}%
\begin{APACrefauthors}%
{Ashour-Abdalla}, M.%
, {Frank}, L\BPBI A.%
, {Paterson}, W\BPBI R.%
, {Peroomian}, V.%
\BCBL {}\ \BBA {} {Zelenyi}, L\BPBI M.%
\end{APACrefauthors}%
\unskip\
\newblock
\APACrefYearMonthDay{1996}{{\APACmonth{02}}}{}.
\newblock
{\BBOQ}\APACrefatitle {{Proton velocity distributions in the magnetotail:
  Theory and observations}} {{Proton velocity distributions in the magnetotail:
  Theory and observations}}.{\BBCQ}
\newblock
\APACjournalVolNumPages{\jgr}{101}{}{2587-2598}.
\newblock
\begin{APACrefDOI} \doi{10.1029/95JA02539} \end{APACrefDOI}
\PrintBackRefs{\CurrentBib}

\bibitem [\protect \citeauthoryear {%
{Ashour-Abdalla}%
\ \protect \BOthers {.}}{%
{Ashour-Abdalla}%
\ \protect \BOthers {.}}{%
{\protect \APACyear {2006}}%
}]{%
Maha06}
\APACinsertmetastar {%
Maha06}%
\begin{APACrefauthors}%
{Ashour-Abdalla}, M.%
, {Leboeuf}, J\BPBI N.%
, {Schriver}, D.%
, {Bosqued}, J.%
, {Cornilleau-Wehrlin}, N.%
, {Sotnikov}, V.%
\BDBL {}{Fazakerley}, A\BPBI N.%
\end{APACrefauthors}%
\unskip\
\newblock
\APACrefYearMonthDay{2006}{{\APACmonth{10}}}{}.
\newblock
{\BBOQ}\APACrefatitle {{Instabilities driven by ion shell distributions
  observed by Cluster in the midaltitude plasma sheet boundary layer}}
  {{Instabilities driven by ion shell distributions observed by Cluster in the
  midaltitude plasma sheet boundary layer}}.{\BBCQ}
\newblock
\APACjournalVolNumPages{J. Geophys. Res.}{111}{}{10223}.
\newblock
\begin{APACrefDOI} \doi{10.1029/2005JA011490} \end{APACrefDOI}
\PrintBackRefs{\CurrentBib}

\bibitem [\protect \citeauthoryear {%
{Ashour-Abdalla}%
, {Zelenyi}%
, {Bosqued}%
\BCBL {}\ \BBA {} {Kovrazhkin}%
}{%
{Ashour-Abdalla}%
\ \protect \BOthers {.}}{%
{\protect \APACyear {1992}}%
}]{%
Maha92}
\APACinsertmetastar {%
Maha92}%
\begin{APACrefauthors}%
{Ashour-Abdalla}, M.%
, {Zelenyi}, L\BPBI M.%
, {Bosqued}, J\BPBI M.%
\BCBL {}\ \BBA {} {Kovrazhkin}, R\BPBI A.%
\end{APACrefauthors}%
\unskip\
\newblock
\APACrefYearMonthDay{1992}{{\APACmonth{03}}}{}.
\newblock
{\BBOQ}\APACrefatitle {{Precipitation of fast ion beams from the plasma sheet
  boundary layer}} {{Precipitation of fast ion beams from the plasma sheet
  boundary layer}}.{\BBCQ}
\newblock
\APACjournalVolNumPages{\grl}{19}{}{617-620}.
\newblock
\begin{APACrefDOI} \doi{10.1029/92GL00048} \end{APACrefDOI}
\PrintBackRefs{\CurrentBib}

\bibitem [\protect \citeauthoryear {%
{Baker}%
, {Pulkkinen}%
, {Angelopoulos}%
, {Baumjohann}%
\BCBL {}\ \BBA {} {McPherron}%
}{%
{Baker}%
\ \protect \BOthers {.}}{%
{\protect \APACyear {1996}}%
}]{%
Baker96}
\APACinsertmetastar {%
Baker96}%
\begin{APACrefauthors}%
{Baker}, D\BPBI N.%
, {Pulkkinen}, T\BPBI I.%
, {Angelopoulos}, V.%
, {Baumjohann}, W.%
\BCBL {}\ \BBA {} {McPherron}, R\BPBI L.%
\end{APACrefauthors}%
\unskip\
\newblock
\APACrefYearMonthDay{1996}{{\APACmonth{06}}}{}.
\newblock
{\BBOQ}\APACrefatitle {{Neutral line model of substorms: Past results and
  present view}} {{Neutral line model of substorms: Past results and present
  view}}.{\BBCQ}
\newblock
\APACjournalVolNumPages{\jgr}{101}{}{12975-13010}.
\newblock
\begin{APACrefDOI} \doi{10.1029/95JA03753} \end{APACrefDOI}
\PrintBackRefs{\CurrentBib}

\bibitem [\protect \citeauthoryear {%
{Baumjohann}%
, {Paschmann}%
\BCBL {}\ \BBA {} {Luehr}%
}{%
{Baumjohann}%
\ \protect \BOthers {.}}{%
{\protect \APACyear {1990}}%
}]{%
Baumjohann90:pressure}
\APACinsertmetastar {%
Baumjohann90:pressure}%
\begin{APACrefauthors}%
{Baumjohann}, W.%
, {Paschmann}, G.%
\BCBL {}\ \BBA {} {Luehr}, H.%
\end{APACrefauthors}%
\unskip\
\newblock
\APACrefYearMonthDay{1990}{{\APACmonth{01}}}{}.
\newblock
{\BBOQ}\APACrefatitle {{Pressure balance between lobe and plasma sheet}}
  {{Pressure balance between lobe and plasma sheet}}.{\BBCQ}
\newblock
\APACjournalVolNumPages{\grl}{17}{}{45-48}.
\newblock
\begin{APACrefDOI} \doi{10.1029/GL017i001p00045} \end{APACrefDOI}
\PrintBackRefs{\CurrentBib}

\bibitem [\protect \citeauthoryear {%
{Baumjohann}%
\ \protect \BOthers {.}}{%
{Baumjohann}%
\ \protect \BOthers {.}}{%
{\protect \APACyear {2007}}%
}]{%
Baumjohann07}
\APACinsertmetastar {%
Baumjohann07}%
\begin{APACrefauthors}%
{Baumjohann}, W.%
, {Roux}, A.%
, {Le Contel}, O.%
, {Nakamura}, R.%
, {Birn}, J.%
, {Hoshino}, M.%
\BDBL {}{Runov}, A.%
\end{APACrefauthors}%
\unskip\
\newblock
\APACrefYearMonthDay{2007}{{\APACmonth{06}}}{}.
\newblock
{\BBOQ}\APACrefatitle {{Dynamics of thin current sheets: Cluster observations}}
  {{Dynamics of thin current sheets: Cluster observations}}.{\BBCQ}
\newblock
\APACjournalVolNumPages{Annales Geophysicae}{25}{}{1365-1389}.
\PrintBackRefs{\CurrentBib}

\bibitem [\protect \citeauthoryear {%
{Bessho}%
\ \BBA {} {Bhattacharjee}%
}{%
{Bessho}%
\ \BBA {} {Bhattacharjee}%
}{%
{\protect \APACyear {2014}}%
}]{%
Bessho&Bhattacharjee14}
\APACinsertmetastar {%
Bessho&Bhattacharjee14}%
\begin{APACrefauthors}%
{Bessho}, N.%
\BCBT {}\ \BBA {} {Bhattacharjee}, A.%
\end{APACrefauthors}%
\unskip\
\newblock
\APACrefYearMonthDay{2014}{{\APACmonth{10}}}{}.
\newblock
{\BBOQ}\APACrefatitle {{Instability of the current sheet in the Earth's
  magnetotail with normal magnetic field}} {{Instability of the current sheet
  in the Earth's magnetotail with normal magnetic field}}.{\BBCQ}
\newblock
\APACjournalVolNumPages{Physics of Plasmas}{21}{10}{102905}.
\newblock
\begin{APACrefDOI} \doi{10.1063/1.4899043} \end{APACrefDOI}
\PrintBackRefs{\CurrentBib}

\bibitem [\protect \citeauthoryear {%
{Birn}%
}{%
{Birn}%
}{%
{\protect \APACyear {1991}}%
}]{%
Birn91:pop}
\APACinsertmetastar {%
Birn91:pop}%
\begin{APACrefauthors}%
{Birn}, J.%
\end{APACrefauthors}%
\unskip\
\newblock
\APACrefYearMonthDay{1991}{{\APACmonth{02}}}{}.
\newblock
{\BBOQ}\APACrefatitle {{Stretched three-dimensional plasma equilibria with
  field-aligned flow}} {{Stretched three-dimensional plasma equilibria with
  field-aligned flow}}.{\BBCQ}
\newblock
\APACjournalVolNumPages{Physics of Fluids B}{3}{2}{479-484}.
\newblock
\begin{APACrefDOI} \doi{10.1063/1.859891} \end{APACrefDOI}
\PrintBackRefs{\CurrentBib}

\bibitem [\protect \citeauthoryear {%
{Birn}%
}{%
{Birn}%
}{%
{\protect \APACyear {1992}}%
}]{%
Birn92}
\APACinsertmetastar {%
Birn92}%
\begin{APACrefauthors}%
{Birn}, J.%
\end{APACrefauthors}%
\unskip\
\newblock
\APACrefYearMonthDay{1992}{{\APACmonth{11}}}{}.
\newblock
{\BBOQ}\APACrefatitle {{Quasi-steady current sheet structures with
  field-aligned flow}} {{Quasi-steady current sheet structures with
  field-aligned flow}}.{\BBCQ}
\newblock
\APACjournalVolNumPages{\jgr}{97}{A11}{16817-16826}.
\newblock
\begin{APACrefDOI} \doi{10.1029/92JA01527} \end{APACrefDOI}
\PrintBackRefs{\CurrentBib}

\bibitem [\protect \citeauthoryear {%
{Birn}%
, {Dorelli}%
, {Hesse}%
\BCBL {}\ \BBA {} {Schindler}%
}{%
{Birn}%
, {Dorelli}%
\BCBL {}\ \protect \BOthers {.}}{%
{\protect \APACyear {2004}}%
}]{%
Birn04MHD}
\APACinsertmetastar {%
Birn04MHD}%
\begin{APACrefauthors}%
{Birn}, J.%
, {Dorelli}, J\BPBI C.%
, {Hesse}, M.%
\BCBL {}\ \BBA {} {Schindler}, K.%
\end{APACrefauthors}%
\unskip\
\newblock
\APACrefYearMonthDay{2004}{{\APACmonth{02}}}{}.
\newblock
{\BBOQ}\APACrefatitle {{Thin current sheets and loss of equilibrium:
  Three-dimensional theory and simulations}} {{Thin current sheets and loss of
  equilibrium: Three-dimensional theory and simulations}}.{\BBCQ}
\newblock
\APACjournalVolNumPages{J. Geophys. Res.}{109}{}{2215}.
\newblock
\begin{APACrefDOI} \doi{10.1029/2003JA010275} \end{APACrefDOI}
\PrintBackRefs{\CurrentBib}

\bibitem [\protect \citeauthoryear {%
{Birn}%
, {Hesse}%
\BCBL {}\ \BBA {} {Schindler}%
}{%
{Birn}%
\ \protect \BOthers {.}}{%
{\protect \APACyear {1998}}%
}]{%
Birn98:cs}
\APACinsertmetastar {%
Birn98:cs}%
\begin{APACrefauthors}%
{Birn}, J.%
, {Hesse}, M.%
\BCBL {}\ \BBA {} {Schindler}, K.%
\end{APACrefauthors}%
\unskip\
\newblock
\APACrefYearMonthDay{1998}{{\APACmonth{04}}}{}.
\newblock
{\BBOQ}\APACrefatitle {{Formation of thin current sheets in space plasmas}}
  {{Formation of thin current sheets in space plasmas}}.{\BBCQ}
\newblock
\APACjournalVolNumPages{\jgr}{103}{}{6843-6852}.
\newblock
\begin{APACrefDOI} \doi{10.1029/97JA03602} \end{APACrefDOI}
\PrintBackRefs{\CurrentBib}

\bibitem [\protect \citeauthoryear {%
{Birn}%
, {Merkin}%
, {Sitnov}%
\BCBL {}\ \BBA {} {Otto}%
}{%
{Birn}%
\ \protect \BOthers {.}}{%
{\protect \APACyear {2018}}%
}]{%
Birn18}
\APACinsertmetastar {%
Birn18}%
\begin{APACrefauthors}%
{Birn}, J.%
, {Merkin}, V\BPBI G.%
, {Sitnov}, M\BPBI I.%
\BCBL {}\ \BBA {} {Otto}, A.%
\end{APACrefauthors}%
\unskip\
\newblock
\APACrefYearMonthDay{2018}{{\APACmonth{05}}}{}.
\newblock
{\BBOQ}\APACrefatitle {{MHD Stability of Magnetotail Configurations With a
  $B_{z}$ Hump}} {{MHD Stability of Magnetotail Configurations With a $B_{z}$
  Hump}}.{\BBCQ}
\newblock
\APACjournalVolNumPages{\jgr}{123}{}{3477-3492}.
\newblock
\begin{APACrefDOI} \doi{10.1029/2018JA025290} \end{APACrefDOI}
\PrintBackRefs{\CurrentBib}

\bibitem [\protect \citeauthoryear {%
{Birn}%
\ \BBA {} {Priest}%
}{%
{Birn}%
\ \BBA {} {Priest}%
}{%
{\protect \APACyear {2007}}%
}]{%
bookBirn&Priest07}
\APACinsertmetastar {%
bookBirn&Priest07}%
\begin{APACrefauthors}%
{Birn}, J.%
\BCBT {}\ \BBA {} {Priest}, E\BPBI R.%
\end{APACrefauthors}%
\unskip\
\newblock
\APACrefYear{2007}.
\newblock
\APACrefbtitle {{Reconnection of magnetic fields : magnetohydrodynamics and
  collisionless theory and observations}} {{Reconnection of magnetic fields :
  magnetohydrodynamics and collisionless theory and observations}}\ ({Birn,
  J.~\& Priest, E.~R.}, \BED{}).
\PrintBackRefs{\CurrentBib}

\bibitem [\protect \citeauthoryear {%
{Birn}%
, {Schindler}%
\BCBL {}\ \BBA {} {Hesse}%
}{%
{Birn}%
, {Schindler}%
\BCBL {}\ \BBA {} {Hesse}%
}{%
{\protect \APACyear {2004}}%
}]{%
Birn04}
\APACinsertmetastar {%
Birn04}%
\begin{APACrefauthors}%
{Birn}, J.%
, {Schindler}, K.%
\BCBL {}\ \BBA {} {Hesse}, M.%
\end{APACrefauthors}%
\unskip\
\newblock
\APACrefYearMonthDay{2004}{{\APACmonth{02}}}{}.
\newblock
{\BBOQ}\APACrefatitle {{Thin electron current sheets and their relation to
  auroral potentials}} {{Thin electron current sheets and their relation to
  auroral potentials}}.{\BBCQ}
\newblock
\APACjournalVolNumPages{J. Geophys. Res.}{109}{}{2217}.
\newblock
\begin{APACrefDOI} \doi{10.1029/2003JA010303} \end{APACrefDOI}
\PrintBackRefs{\CurrentBib}

\bibitem [\protect \citeauthoryear {%
{Biskamp}%
}{%
{Biskamp}%
}{%
{\protect \APACyear {2000}}%
}]{%
bookBiskamp00}
\APACinsertmetastar {%
bookBiskamp00}%
\begin{APACrefauthors}%
{Biskamp}, D.%
\end{APACrefauthors}%
\unskip\
\newblock
\APACrefYear{2000}.
\newblock
\APACrefbtitle {{Magnetic Reconnection in Plasmas}} {{Magnetic Reconnection in
  Plasmas}}.
\PrintBackRefs{\CurrentBib}

\bibitem [\protect \citeauthoryear {%
{Bulanov}%
, {Syrovatski{\v{i}}}%
\BCBL {}\ \BBA {} {Sakai}%
}{%
{Bulanov}%
\ \protect \BOthers {.}}{%
{\protect \APACyear {1978}}%
}]{%
Bulanov78}
\APACinsertmetastar {%
Bulanov78}%
\begin{APACrefauthors}%
{Bulanov}, S\BPBI V.%
, {Syrovatski{\v{i}}}, S\BPBI I.%
\BCBL {}\ \BBA {} {Sakai}, J.%
\end{APACrefauthors}%
\unskip\
\newblock
\APACrefYearMonthDay{1978}{{\APACmonth{08}}}{}.
\newblock
{\BBOQ}\APACrefatitle {{Stabilizing influence of plasma flow on dissipative
  tearing instability}} {{Stabilizing influence of plasma flow on dissipative
  tearing instability}}.{\BBCQ}
\newblock
\APACjournalVolNumPages{Soviet Journal of Experimental and Theoretical Physics
  Letters}{28}{}{177-179}.
\PrintBackRefs{\CurrentBib}

\bibitem [\protect \citeauthoryear {%
{Burkhart}%
, {Drake}%
, {Dusenbery}%
\BCBL {}\ \BBA {} {Speiser}%
}{%
{Burkhart}%
\ \protect \BOthers {.}}{%
{\protect \APACyear {1992}}%
}]{%
Burkhart92TCS}
\APACinsertmetastar {%
Burkhart92TCS}%
\begin{APACrefauthors}%
{Burkhart}, G\BPBI R.%
, {Drake}, J\BPBI F.%
, {Dusenbery}, P\BPBI B.%
\BCBL {}\ \BBA {} {Speiser}, T\BPBI W.%
\end{APACrefauthors}%
\unskip\
\newblock
\APACrefYearMonthDay{1992}{{\APACmonth{09}}}{}.
\newblock
{\BBOQ}\APACrefatitle {{A particle model for magnetotail neutral sheet
  equilibria}} {{A particle model for magnetotail neutral sheet
  equilibria}}.{\BBCQ}
\newblock
\APACjournalVolNumPages{J. Geophys. Res.}{97}{}{13799-13815}.
\newblock
\begin{APACrefDOI} \doi{10.1029/92JA00495} \end{APACrefDOI}
\PrintBackRefs{\CurrentBib}

\bibitem [\protect \citeauthoryear {%
{Carmichael}%
}{%
{Carmichael}%
}{%
{\protect \APACyear {1964}}%
}]{%
Carmichael64}
\APACinsertmetastar {%
Carmichael64}%
\begin{APACrefauthors}%
{Carmichael}, H.%
\end{APACrefauthors}%
\unskip\
\newblock
\APACrefYearMonthDay{1964}{}{}.
\newblock
{\BBOQ}\APACrefatitle {{A Process for Flares}} {{A Process for Flares}}.{\BBCQ}
\newblock
\APACjournalVolNumPages{NASA Special Publication}{50}{}{451-456}.
\PrintBackRefs{\CurrentBib}

\bibitem [\protect \citeauthoryear {%
Chen%
, Otto%
\BCBL {}\ \BBA {} Lee%
}{%
Chen%
\ \protect \BOthers {.}}{%
{\protect \APACyear {1997}}%
}]{%
chen1997tearing}
\APACinsertmetastar {%
chen1997tearing}%
\begin{APACrefauthors}%
Chen, Q.%
, Otto, A.%
\BCBL {}\ \BBA {} Lee, L.%
\end{APACrefauthors}%
\unskip\
\newblock
\APACrefYearMonthDay{1997}{}{}.
\newblock
{\BBOQ}\APACrefatitle {Tearing instability, Kelvin-Helmholtz instability, and
  magnetic reconnection} {Tearing instability, kelvin-helmholtz instability,
  and magnetic reconnection}.{\BBCQ}
\newblock
\APACjournalVolNumPages{Journal of Geophysical Research: Space
  Physics}{102}{A1}{151--161}.
\PrintBackRefs{\CurrentBib}

\bibitem [\protect \citeauthoryear {%
{Coppi}%
, {Laval}%
\BCBL {}\ \BBA {} {Pellat}%
}{%
{Coppi}%
\ \protect \BOthers {.}}{%
{\protect \APACyear {1966}}%
}]{%
Coppi66}
\APACinsertmetastar {%
Coppi66}%
\begin{APACrefauthors}%
{Coppi}, B.%
, {Laval}, G.%
\BCBL {}\ \BBA {} {Pellat}, R.%
\end{APACrefauthors}%
\unskip\
\newblock
\APACrefYearMonthDay{1966}{{\APACmonth{06}}}{}.
\newblock
{\BBOQ}\APACrefatitle {{Dynamics of the Geomagnetic Tail}} {{Dynamics of the
  Geomagnetic Tail}}.{\BBCQ}
\newblock
\APACjournalVolNumPages{Physical Review Letters}{16}{}{1207-1210}.
\newblock
\begin{APACrefDOI} \doi{10.1103/PhysRevLett.16.1207} \end{APACrefDOI}
\PrintBackRefs{\CurrentBib}

\bibitem [\protect \citeauthoryear {%
{Cowley}%
}{%
{Cowley}%
}{%
{\protect \APACyear {1978}}%
}]{%
Cowley78}
\APACinsertmetastar {%
Cowley78}%
\begin{APACrefauthors}%
{Cowley}, S\BPBI W\BPBI H.%
\end{APACrefauthors}%
\unskip\
\newblock
\APACrefYearMonthDay{1978}{{\APACmonth{11}}}{}.
\newblock
{\BBOQ}\APACrefatitle {{The effect of pressure anisotropy on the equilibrium
  structure of magnetic current sheets}} {{The effect of pressure anisotropy on
  the equilibrium structure of magnetic current sheets}}.{\BBCQ}
\newblock
\APACjournalVolNumPages{Planetary and Space Science}{26}{}{1037-1061}.
\newblock
\begin{APACrefDOI} \doi{10.1016/0032-0633(78)90028-4} \end{APACrefDOI}
\PrintBackRefs{\CurrentBib}

\bibitem [\protect \citeauthoryear {%
{Cowley}%
\ \BBA {} {Pellat}%
}{%
{Cowley}%
\ \BBA {} {Pellat}%
}{%
{\protect \APACyear {1979}}%
}]{%
Cowley&Pellat79}
\APACinsertmetastar {%
Cowley&Pellat79}%
\begin{APACrefauthors}%
{Cowley}, S\BPBI W\BPBI H.%
\BCBT {}\ \BBA {} {Pellat}, R.%
\end{APACrefauthors}%
\unskip\
\newblock
\APACrefYearMonthDay{1979}{{\APACmonth{03}}}{}.
\newblock
{\BBOQ}\APACrefatitle {{A note on adiabatic solutions of the one-dimensional
  current sheet problem}} {{A note on adiabatic solutions of the
  one-dimensional current sheet problem}}.{\BBCQ}
\newblock
\APACjournalVolNumPages{\planss}{27}{}{265-271}.
\newblock
\begin{APACrefDOI} \doi{10.1016/0032-0633(79)90069-2} \end{APACrefDOI}
\PrintBackRefs{\CurrentBib}

\bibitem [\protect \citeauthoryear {%
{Cowley}%
\ \BBA {} {Shull}%
}{%
{Cowley}%
\ \BBA {} {Shull}%
}{%
{\protect \APACyear {1983}}%
}]{%
Cowley83}
\APACinsertmetastar {%
Cowley83}%
\begin{APACrefauthors}%
{Cowley}, S\BPBI W\BPBI H.%
\BCBT {}\ \BBA {} {Shull}, P., Jr.%
\end{APACrefauthors}%
\unskip\
\newblock
\APACrefYearMonthDay{1983}{{\APACmonth{02}}}{}.
\newblock
{\BBOQ}\APACrefatitle {{Current sheet acceleration of ions in the geomagnetic
  tail and the properties of ion bursts observed at the lunar distance}}
  {{Current sheet acceleration of ions in the geomagnetic tail and the
  properties of ion bursts observed at the lunar distance}}.{\BBCQ}
\newblock
\APACjournalVolNumPages{Planetary and Space Science}{31}{}{235-245}.
\newblock
\begin{APACrefDOI} \doi{10.1016/0032-0633(83)90058-2} \end{APACrefDOI}
\PrintBackRefs{\CurrentBib}

\bibitem [\protect \citeauthoryear {%
Dahlburg%
, Boncinelli%
\BCBL {}\ \BBA {} Einaudi%
}{%
Dahlburg%
\ \protect \BOthers {.}}{%
{\protect \APACyear {1997}}%
}]{%
dahlburg1997evolution}
\APACinsertmetastar {%
dahlburg1997evolution}%
\begin{APACrefauthors}%
Dahlburg, R.%
, Boncinelli, P.%
\BCBL {}\ \BBA {} Einaudi, G.%
\end{APACrefauthors}%
\unskip\
\newblock
\APACrefYearMonthDay{1997}{}{}.
\newblock
{\BBOQ}\APACrefatitle {The evolution of plane current--vortex sheets} {The
  evolution of plane current--vortex sheets}.{\BBCQ}
\newblock
\APACjournalVolNumPages{Physics of Plasmas}{4}{5}{1213--1226}.
\PrintBackRefs{\CurrentBib}

\bibitem [\protect \citeauthoryear {%
{Del Sarto}%
, {Pucci}%
, {Tenerani}%
\BCBL {}\ \BBA {} {Velli}%
}{%
{Del Sarto}%
\ \protect \BOthers {.}}{%
{\protect \APACyear {2016}}%
}]{%
DelSarto16}
\APACinsertmetastar {%
DelSarto16}%
\begin{APACrefauthors}%
{Del Sarto}, D.%
, {Pucci}, F.%
, {Tenerani}, A.%
\BCBL {}\ \BBA {} {Velli}, M.%
\end{APACrefauthors}%
\unskip\
\newblock
\APACrefYearMonthDay{2016}{{\APACmonth{03}}}{}.
\newblock
{\BBOQ}\APACrefatitle {{``Ideal'' tearing and the transition to fast
  reconnection in the weakly collisional MHD and EMHD regimes}} {{``Ideal''
  tearing and the transition to fast reconnection in the weakly collisional MHD
  and EMHD regimes}}.{\BBCQ}
\newblock
\APACjournalVolNumPages{Journal of Geophysical Research (Space
  Physics)}{121}{3}{1857-1873}.
\newblock
\begin{APACrefDOI} \doi{10.1002/2015JA021975} \end{APACrefDOI}
\PrintBackRefs{\CurrentBib}

\bibitem [\protect \citeauthoryear {%
{Dobrowolny}%
, {Veltri}%
\BCBL {}\ \BBA {} {Mangeney}%
}{%
{Dobrowolny}%
\ \protect \BOthers {.}}{%
{\protect \APACyear {1983}}%
}]{%
Dobrowolny83}
\APACinsertmetastar {%
Dobrowolny83}%
\begin{APACrefauthors}%
{Dobrowolny}, M.%
, {Veltri}, P.%
\BCBL {}\ \BBA {} {Mangeney}, A.%
\end{APACrefauthors}%
\unskip\
\newblock
\APACrefYearMonthDay{1983}{{\APACmonth{06}}}{}.
\newblock
{\BBOQ}\APACrefatitle {{Dissipative instabilities of magnetic neutral layers
  with velocity shear}} {{Dissipative instabilities of magnetic neutral layers
  with velocity shear}}.{\BBCQ}
\newblock
\APACjournalVolNumPages{Journal of Plasma Physics}{29}{3}{393-407}.
\newblock
\begin{APACrefDOI} \doi{10.1017/S0022377800000854} \end{APACrefDOI}
\PrintBackRefs{\CurrentBib}

\bibitem [\protect \citeauthoryear {%
{Eastwood}%
}{%
{Eastwood}%
}{%
{\protect \APACyear {1972}}%
}]{%
Eastwood72}
\APACinsertmetastar {%
Eastwood72}%
\begin{APACrefauthors}%
{Eastwood}, J\BPBI W.%
\end{APACrefauthors}%
\unskip\
\newblock
\APACrefYearMonthDay{1972}{{\APACmonth{10}}}{}.
\newblock
{\BBOQ}\APACrefatitle {{Consistency of fields and particle motion in the
  `Speiser' model of the current sheet}} {{Consistency of fields and particle
  motion in the `Speiser' model of the current sheet}}.{\BBCQ}
\newblock
\APACjournalVolNumPages{Planetary and Space Science}{20}{}{1555-1568}.
\newblock
\begin{APACrefDOI} \doi{10.1016/0032-0633(72)90182-1} \end{APACrefDOI}
\PrintBackRefs{\CurrentBib}

\bibitem [\protect \citeauthoryear {%
{Eastwood}%
}{%
{Eastwood}%
}{%
{\protect \APACyear {1974}}%
}]{%
Eastwood74}
\APACinsertmetastar {%
Eastwood74}%
\begin{APACrefauthors}%
{Eastwood}, J\BPBI W.%
\end{APACrefauthors}%
\unskip\
\newblock
\APACrefYearMonthDay{1974}{{\APACmonth{12}}}{}.
\newblock
{\BBOQ}\APACrefatitle {{The warm current sheet model, and its implications on
  the temporal behaviour of the geomagnetic tail}} {{The warm current sheet
  model, and its implications on the temporal behaviour of the geomagnetic
  tail}}.{\BBCQ}
\newblock
\APACjournalVolNumPages{Planetary and Space Science}{22}{}{1641-1668}.
\newblock
\begin{APACrefDOI} \doi{10.1016/0032-0633(74)90108-1} \end{APACrefDOI}
\PrintBackRefs{\CurrentBib}

\bibitem [\protect \citeauthoryear {%
{Edmondson}%
\ \BBA {} {Lynch}%
}{%
{Edmondson}%
\ \BBA {} {Lynch}%
}{%
{\protect \APACyear {2017}}%
}]{%
Edmondson&Lynch17}
\APACinsertmetastar {%
Edmondson&Lynch17}%
\begin{APACrefauthors}%
{Edmondson}, J\BPBI K.%
\BCBT {}\ \BBA {} {Lynch}, B\BPBI J.%
\end{APACrefauthors}%
\unskip\
\newblock
\APACrefYearMonthDay{2017}{{\APACmonth{11}}}{}.
\newblock
{\BBOQ}\APACrefatitle {{Formation and Reconnection of Three-dimensional Current
  Sheets with a Guide Field in the Solar Corona}} {{Formation and Reconnection
  of Three-dimensional Current Sheets with a Guide Field in the Solar
  Corona}}.{\BBCQ}
\newblock
\APACjournalVolNumPages{\apj}{849}{1}{28}.
\newblock
\begin{APACrefDOI} \doi{10.3847/1538-4357/aa83ba} \end{APACrefDOI}
\PrintBackRefs{\CurrentBib}

\bibitem [\protect \citeauthoryear {%
Einaudi%
\ \BBA {} Rubini%
}{%
Einaudi%
\ \BBA {} Rubini%
}{%
{\protect \APACyear {1986}}%
}]{%
einaudi1986resistive}
\APACinsertmetastar {%
einaudi1986resistive}%
\begin{APACrefauthors}%
Einaudi, G.%
\BCBT {}\ \BBA {} Rubini, F.%
\end{APACrefauthors}%
\unskip\
\newblock
\APACrefYearMonthDay{1986}{}{}.
\newblock
{\BBOQ}\APACrefatitle {Resistive instabilities in a flowing plasma: I. Inviscid
  case} {Resistive instabilities in a flowing plasma: I. inviscid case}.{\BBCQ}
\newblock
\APACjournalVolNumPages{The Physics of fluids}{29}{8}{2563--2568}.
\PrintBackRefs{\CurrentBib}

\bibitem [\protect \citeauthoryear {%
{Francfort}%
\ \BBA {} {Pellat}%
}{%
{Francfort}%
\ \BBA {} {Pellat}%
}{%
{\protect \APACyear {1976}}%
}]{%
FP76}
\APACinsertmetastar {%
FP76}%
\begin{APACrefauthors}%
{Francfort}, P.%
\BCBT {}\ \BBA {} {Pellat}, R.%
\end{APACrefauthors}%
\unskip\
\newblock
\APACrefYearMonthDay{1976}{{\APACmonth{08}}}{}.
\newblock
{\BBOQ}\APACrefatitle {{Magnetic merging in collisionless plasmas}} {{Magnetic
  merging in collisionless plasmas}}.{\BBCQ}
\newblock
\APACjournalVolNumPages{Geophys. Res. Lett.}{3}{}{433-436}.
\newblock
\begin{APACrefDOI} \doi{10.1029/GL003i008p00433} \end{APACrefDOI}
\PrintBackRefs{\CurrentBib}

\bibitem [\protect \citeauthoryear {%
{Galeev}%
\ \BBA {} {Sudan}%
}{%
{Galeev}%
\ \BBA {} {Sudan}%
}{%
{\protect \APACyear {1985}}%
}]{%
bookGaleev85:vol2}
\APACinsertmetastar {%
bookGaleev85:vol2}%
\begin{APACrefauthors}%
{Galeev}, A\BPBI A.%
\BCBT {}\ \BBA {} {Sudan}, R\BPBI N.%
\end{APACrefauthors}%
\unskip\
\newblock
\APACrefYear{1985}.
\newblock
\APACrefbtitle {{Handbook of plasma physics. Vol. 2: Basic plasma physics II.}}
  {{Handbook of plasma physics. Vol. 2: Basic plasma physics II.}}
\PrintBackRefs{\CurrentBib}

\bibitem [\protect \citeauthoryear {%
{Galeev}%
\ \BBA {} {Zelenyi}%
}{%
{Galeev}%
\ \BBA {} {Zelenyi}%
}{%
{\protect \APACyear {1976}}%
}]{%
Galeev76}
\APACinsertmetastar {%
Galeev76}%
\begin{APACrefauthors}%
{Galeev}, A\BPBI A.%
\BCBT {}\ \BBA {} {Zelenyi}, L\BPBI M.%
\end{APACrefauthors}%
\unskip\
\newblock
\APACrefYearMonthDay{1976}{{\APACmonth{06}}}{}.
\newblock
{\BBOQ}\APACrefatitle {{Tearing instability in plasma configurations}}
  {{Tearing instability in plasma configurations}}.{\BBCQ}
\newblock
\APACjournalVolNumPages{Soviet Journal of Experimental and Theoretical
  Physics}{43}{}{1113}.
\PrintBackRefs{\CurrentBib}

\bibitem [\protect \citeauthoryear {%
{Goldstein}%
\ \BBA {} {Schindler}%
}{%
{Goldstein}%
\ \BBA {} {Schindler}%
}{%
{\protect \APACyear {1982}}%
}]{%
Goldstein82}
\APACinsertmetastar {%
Goldstein82}%
\begin{APACrefauthors}%
{Goldstein}, H.%
\BCBT {}\ \BBA {} {Schindler}, K.%
\end{APACrefauthors}%
\unskip\
\newblock
\APACrefYearMonthDay{1982}{{\APACmonth{05}}}{}.
\newblock
{\BBOQ}\APACrefatitle {{Large-Scale Collision-Free Instability of
  Two-Dimensional Plasma Sheets}} {{Large-Scale Collision-Free Instability of
  Two-Dimensional Plasma Sheets}}.{\BBCQ}
\newblock
\APACjournalVolNumPages{Physical Review Letters}{48}{}{1468-1471}.
\newblock
\begin{APACrefDOI} \doi{10.1103/PhysRevLett.48.1468} \end{APACrefDOI}
\PrintBackRefs{\CurrentBib}

\bibitem [\protect \citeauthoryear {%
{Gonzalez}%
\ \BBA {} {Parker}%
}{%
{Gonzalez}%
\ \BBA {} {Parker}%
}{%
{\protect \APACyear {2016}}%
}]{%
book:Gonzalez&Parker}
\APACinsertmetastar {%
book:Gonzalez&Parker}%
\begin{APACrefauthors}%
{Gonzalez}, W.%
\BCBT {}\ \BBA {} {Parker}, E.%
\end{APACrefauthors}%
\unskip\
\newblock
\APACrefYear{2016}.
\newblock
\APACrefbtitle {{Magnetic Reconnection}} {{Magnetic Reconnection}}\
  (\BVOL~427).
\newblock
\begin{APACrefDOI} \doi{10.1007/978-3-319-26432-5} \end{APACrefDOI}
\PrintBackRefs{\CurrentBib}

\bibitem [\protect \citeauthoryear {%
{Gosling}%
}{%
{Gosling}%
}{%
{\protect \APACyear {2012}}%
}]{%
Gosling12}
\APACinsertmetastar {%
Gosling12}%
\begin{APACrefauthors}%
{Gosling}, J\BPBI T.%
\end{APACrefauthors}%
\unskip\
\newblock
\APACrefYearMonthDay{2012}{{\APACmonth{11}}}{}.
\newblock
{\BBOQ}\APACrefatitle {{Magnetic Reconnection in the Solar Wind}} {{Magnetic
  Reconnection in the Solar Wind}}.{\BBCQ}
\newblock
\APACjournalVolNumPages{\ssr}{172}{}{187-200}.
\newblock
\begin{APACrefDOI} \doi{10.1007/s11214-011-9747-2} \end{APACrefDOI}
\PrintBackRefs{\CurrentBib}

\bibitem [\protect \citeauthoryear {%
{Grigorenko}%
, {Hoshino}%
, {Hirai}%
, {Mukai}%
\BCBL {}\ \BBA {} {Zelenyi}%
}{%
{Grigorenko}%
\ \protect \BOthers {.}}{%
{\protect \APACyear {2009}}%
}]{%
Grigorenko09}
\APACinsertmetastar {%
Grigorenko09}%
\begin{APACrefauthors}%
{Grigorenko}, E\BPBI E.%
, {Hoshino}, M.%
, {Hirai}, M.%
, {Mukai}, T.%
\BCBL {}\ \BBA {} {Zelenyi}, L\BPBI M.%
\end{APACrefauthors}%
\unskip\
\newblock
\APACrefYearMonthDay{2009}{{\APACmonth{03}}}{}.
\newblock
{\BBOQ}\APACrefatitle {{``Geography'' of ion acceleration in the magnetotail:
  X-line versus current sheet effects}} {{``Geography'' of ion acceleration in
  the magnetotail: X-line versus current sheet effects}}.{\BBCQ}
\newblock
\APACjournalVolNumPages{J. Geophys. Res.}{114}{}{3203}.
\newblock
\begin{APACrefDOI} \doi{10.1029/2008JA013811} \end{APACrefDOI}
\PrintBackRefs{\CurrentBib}

\bibitem [\protect \citeauthoryear {%
{Grigorenko}%
\ \protect \BOthers {.}}{%
{Grigorenko}%
\ \protect \BOthers {.}}{%
{\protect \APACyear {2011}}%
}]{%
Grigorenko11:SSR}
\APACinsertmetastar {%
Grigorenko11:SSR}%
\begin{APACrefauthors}%
{Grigorenko}, E\BPBI E.%
, {Zelenyi}, L\BPBI M.%
, {Dolgonosov}, M\BPBI S.%
, {Artemiev}, A\BPBI V.%
, {Owen}, C\BPBI J.%
, {Sauvaud}, J\BHBI A.%
\BDBL {}{Hirai}, M.%
\end{APACrefauthors}%
\unskip\
\newblock
\APACrefYearMonthDay{2011}{{\APACmonth{12}}}{}.
\newblock
{\BBOQ}\APACrefatitle {{Non-adiabatic Ion Acceleration in the Earth Magnetotail
  and Its Various Manifestations in the Plasma Sheet Boundary Layer}}
  {{Non-adiabatic Ion Acceleration in the Earth Magnetotail and Its Various
  Manifestations in the Plasma Sheet Boundary Layer}}.{\BBCQ}
\newblock
\APACjournalVolNumPages{\ssr}{164}{}{133-181}.
\newblock
\begin{APACrefDOI} \doi{10.1007/s11214-011-9858-9} \end{APACrefDOI}
\PrintBackRefs{\CurrentBib}

\bibitem [\protect \citeauthoryear {%
{Hau}%
}{%
{Hau}%
}{%
{\protect \APACyear {1993}}%
}]{%
Hau93}
\APACinsertmetastar {%
Hau93}%
\begin{APACrefauthors}%
{Hau}, L\BHBI N.%
\end{APACrefauthors}%
\unskip\
\newblock
\APACrefYearMonthDay{1993}{{\APACmonth{04}}}{}.
\newblock
{\BBOQ}\APACrefatitle {{Anisotropic magnetotail equilibrium and convection}}
  {{Anisotropic magnetotail equilibrium and convection}}.{\BBCQ}
\newblock
\APACjournalVolNumPages{\grl}{20}{}{555-558}.
\newblock
\begin{APACrefDOI} \doi{10.1029/93GL00243} \end{APACrefDOI}
\PrintBackRefs{\CurrentBib}

\bibitem [\protect \citeauthoryear {%
{Hau}%
}{%
{Hau}%
}{%
{\protect \APACyear {1996}}%
}]{%
Hau96:pop}
\APACinsertmetastar {%
Hau96:pop}%
\begin{APACrefauthors}%
{Hau}, L\BPBI N.%
\end{APACrefauthors}%
\unskip\
\newblock
\APACrefYearMonthDay{1996}{{\APACmonth{03}}}{}.
\newblock
{\BBOQ}\APACrefatitle {{General formulation and exact solution for
  two-dimensional field-aligned magnetohydrodynamic equilibrium flows}}
  {{General formulation and exact solution for two-dimensional field-aligned
  magnetohydrodynamic equilibrium flows}}.{\BBCQ}
\newblock
\APACjournalVolNumPages{Physics of Plasmas}{3}{3}{1113-1119}.
\newblock
\begin{APACrefDOI} \doi{10.1063/1.871767} \end{APACrefDOI}
\PrintBackRefs{\CurrentBib}

\bibitem [\protect \citeauthoryear {%
{Hau}%
, {Wolf}%
, {Voigt}%
\BCBL {}\ \BBA {} {Wu}%
}{%
{Hau}%
\ \protect \BOthers {.}}{%
{\protect \APACyear {1989}}%
}]{%
Hau89}
\APACinsertmetastar {%
Hau89}%
\begin{APACrefauthors}%
{Hau}, L\BPBI N.%
, {Wolf}, R\BPBI A.%
, {Voigt}, G\BPBI H.%
\BCBL {}\ \BBA {} {Wu}, C\BPBI C.%
\end{APACrefauthors}%
\unskip\
\newblock
\APACrefYearMonthDay{1989}{{\APACmonth{02}}}{}.
\newblock
{\BBOQ}\APACrefatitle {{Steady state magnetic field configurations for the
  Earth's magnetotail}} {{Steady state magnetic field configurations for the
  Earth's magnetotail}}.{\BBCQ}
\newblock
\APACjournalVolNumPages{\jgr}{94}{A2}{1303-1316}.
\newblock
\begin{APACrefDOI} \doi{10.1029/JA094iA02p01303} \end{APACrefDOI}
\PrintBackRefs{\CurrentBib}

\bibitem [\protect \citeauthoryear {%
{Hill}%
}{%
{Hill}%
}{%
{\protect \APACyear {1975}}%
}]{%
Hill75}
\APACinsertmetastar {%
Hill75}%
\begin{APACrefauthors}%
{Hill}, T\BPBI W.%
\end{APACrefauthors}%
\unskip\
\newblock
\APACrefYearMonthDay{1975}{{\APACmonth{12}}}{}.
\newblock
{\BBOQ}\APACrefatitle {{Magnetic merging in a collisionless plasma}} {{Magnetic
  merging in a collisionless plasma}}.{\BBCQ}
\newblock
\APACjournalVolNumPages{J. Geophys. Res.}{80}{}{4689-4699}.
\newblock
\begin{APACrefDOI} \doi{10.1029/JA080i034p04689} \end{APACrefDOI}
\PrintBackRefs{\CurrentBib}

\bibitem [\protect \citeauthoryear {%
{Hirayama}%
}{%
{Hirayama}%
}{%
{\protect \APACyear {1974}}%
}]{%
Hirayama74}
\APACinsertmetastar {%
Hirayama74}%
\begin{APACrefauthors}%
{Hirayama}, T.%
\end{APACrefauthors}%
\unskip\
\newblock
\APACrefYearMonthDay{1974}{{\APACmonth{02}}}{}.
\newblock
{\BBOQ}\APACrefatitle {{Theoretical Model of Flares and Prominences. I:
  Evaporating Flare Model}} {{Theoretical Model of Flares and Prominences. I:
  Evaporating Flare Model}}.{\BBCQ}
\newblock
\APACjournalVolNumPages{\solphys}{34}{}{323-338}.
\newblock
\begin{APACrefDOI} \doi{10.1007/BF00153671} \end{APACrefDOI}
\PrintBackRefs{\CurrentBib}

\bibitem [\protect \citeauthoryear {%
{Hoshino}%
\ \BBA {} {Higashimori}%
}{%
{Hoshino}%
\ \BBA {} {Higashimori}%
}{%
{\protect \APACyear {2015}}%
}]{%
Hoshino&Higashimori15}
\APACinsertmetastar {%
Hoshino&Higashimori15}%
\begin{APACrefauthors}%
{Hoshino}, M.%
\BCBT {}\ \BBA {} {Higashimori}, K.%
\end{APACrefauthors}%
\unskip\
\newblock
\APACrefYearMonthDay{2015}{{\APACmonth{05}}}{}.
\newblock
{\BBOQ}\APACrefatitle {{Generation of Alfv{\'e}nic waves and turbulence in
  reconnection jets}} {{Generation of Alfv{\'e}nic waves and turbulence in
  reconnection jets}}.{\BBCQ}
\newblock
\APACjournalVolNumPages{\jgr}{120}{}{3715-3727}.
\newblock
\begin{APACrefDOI} \doi{10.1002/2014JA020520} \end{APACrefDOI}
\PrintBackRefs{\CurrentBib}

\bibitem [\protect \citeauthoryear {%
{Hudson}%
}{%
{Hudson}%
}{%
{\protect \APACyear {1970}}%
}]{%
Hudson70}
\APACinsertmetastar {%
Hudson70}%
\begin{APACrefauthors}%
{Hudson}, P\BPBI D.%
\end{APACrefauthors}%
\unskip\
\newblock
\APACrefYearMonthDay{1970}{{\APACmonth{11}}}{}.
\newblock
{\BBOQ}\APACrefatitle {{Discontinuities in an anisotropic plasma and their
  identification in the solar wind}} {{Discontinuities in an anisotropic plasma
  and their identification in the solar wind}}.{\BBCQ}
\newblock
\APACjournalVolNumPages{\planss}{18}{}{1611-1622}.
\newblock
\begin{APACrefDOI} \doi{10.1016/0032-0633(70)90036-X} \end{APACrefDOI}
\PrintBackRefs{\CurrentBib}

\bibitem [\protect \citeauthoryear {%
{Ip}%
\ \BBA {} {Sonnerup}%
}{%
{Ip}%
\ \BBA {} {Sonnerup}%
}{%
{\protect \APACyear {1996}}%
}]{%
Ip&Sonnerup96}
\APACinsertmetastar {%
Ip&Sonnerup96}%
\begin{APACrefauthors}%
{Ip}, J\BPBI T\BPBI C.%
\BCBT {}\ \BBA {} {Sonnerup}, B\BPBI U\BPBI {\"O}.%
\end{APACrefauthors}%
\unskip\
\newblock
\APACrefYearMonthDay{1996}{{\APACmonth{10}}}{}.
\newblock
{\BBOQ}\APACrefatitle {{Resistive tearing-mode instability in a
  magnetic-field-reversing current sheet with coplanar viscous stagnation-point
  flow}} {{Resistive tearing-mode instability in a magnetic-field-reversing
  current sheet with coplanar viscous stagnation-point flow}}.{\BBCQ}
\newblock
\APACjournalVolNumPages{Journal of Plasma Physics}{56}{2}{265-284}.
\newblock
\begin{APACrefDOI} \doi{10.1017/S0022377800019267} \end{APACrefDOI}
\PrintBackRefs{\CurrentBib}

\bibitem [\protect \citeauthoryear {%
{Kan}%
}{%
{Kan}%
}{%
{\protect \APACyear {1973}}%
}]{%
Kan73}
\APACinsertmetastar {%
Kan73}%
\begin{APACrefauthors}%
{Kan}, J\BPBI R.%
\end{APACrefauthors}%
\unskip\
\newblock
\APACrefYearMonthDay{1973}{}{}.
\newblock
{\BBOQ}\APACrefatitle {{On the structure of the magnetotail current sheet.}}
  {{On the structure of the magnetotail current sheet.}}{\BBCQ}
\newblock
\APACjournalVolNumPages{J. Geophys. Res.}{78}{}{3773-3781}.
\newblock
\begin{APACrefDOI} \doi{10.1029/JA078i019p03773} \end{APACrefDOI}
\PrintBackRefs{\CurrentBib}

\bibitem [\protect \citeauthoryear {%
{Keika}%
, {Kistler}%
\BCBL {}\ \BBA {} {Brandt}%
}{%
{Keika}%
\ \protect \BOthers {.}}{%
{\protect \APACyear {2013}}%
}]{%
Keika13:outflow}
\APACinsertmetastar {%
Keika13:outflow}%
\begin{APACrefauthors}%
{Keika}, K.%
, {Kistler}, L\BPBI M.%
\BCBL {}\ \BBA {} {Brandt}, P\BPBI C.%
\end{APACrefauthors}%
\unskip\
\newblock
\APACrefYearMonthDay{2013}{{\APACmonth{07}}}{}.
\newblock
{\BBOQ}\APACrefatitle {{Energization of O$^{+}$ ions in the Earth's inner
  magnetosphere and the effects on ring current buildup: A review of previous
  observations and possible mechanisms}} {{Energization of O$^{+}$ ions in the
  Earth's inner magnetosphere and the effects on ring current buildup: A review
  of previous observations and possible mechanisms}}.{\BBCQ}
\newblock
\APACjournalVolNumPages{Journal of Geophysical Research (Space
  Physics)}{118}{}{4441-4464}.
\newblock
\begin{APACrefDOI} \doi{10.1002/jgra.50371} \end{APACrefDOI}
\PrintBackRefs{\CurrentBib}

\bibitem [\protect \citeauthoryear {%
Kierzenka%
\ \BBA {} Shampine%
}{%
Kierzenka%
\ \BBA {} Shampine%
}{%
{\protect \APACyear {2001}}%
}]{%
Kierzenkaetal2001BVP}
\APACinsertmetastar {%
Kierzenkaetal2001BVP}%
\begin{APACrefauthors}%
Kierzenka, J.%
\BCBT {}\ \BBA {} Shampine, L\BPBI F.%
\end{APACrefauthors}%
\unskip\
\newblock
\APACrefYearMonthDay{2001}{}{}.
\newblock
{\BBOQ}\APACrefatitle {A BVP solver based on residual control and the Maltab
  PSE} {A bvp solver based on residual control and the maltab pse}.{\BBCQ}
\newblock
\APACjournalVolNumPages{ACM Transactions on Mathematical Software
  (TOMS)}{27}{3}{299--316}.
\PrintBackRefs{\CurrentBib}

\bibitem [\protect \citeauthoryear {%
{Kistler}%
\ \protect \BOthers {.}}{%
{Kistler}%
\ \protect \BOthers {.}}{%
{\protect \APACyear {2005}}%
}]{%
Kistler05}
\APACinsertmetastar {%
Kistler05}%
\begin{APACrefauthors}%
{Kistler}, L\BPBI M.%
, {Mouikis}, C.%
, {M{\"o}Bius}, E.%
, {Klecker}, B.%
, {Sauvaud}, J\BPBI A.%
, {R{\'e}Me}, H.%
\BDBL {}{Balogh}, A.%
\end{APACrefauthors}%
\unskip\
\newblock
\APACrefYearMonthDay{2005}{{\APACmonth{06}}}{}.
\newblock
{\BBOQ}\APACrefatitle {{Contribution of nonadiabatic ions to the cross-tail
  current in an O$^{+}$ dominated thin current sheet}} {{Contribution of
  nonadiabatic ions to the cross-tail current in an O$^{+}$ dominated thin
  current sheet}}.{\BBCQ}
\newblock
\APACjournalVolNumPages{\jgr}{110}{}{6213}.
\newblock
\begin{APACrefDOI} \doi{10.1029/2004JA010653} \end{APACrefDOI}
\PrintBackRefs{\CurrentBib}

\bibitem [\protect \citeauthoryear {%
{Kopp}%
\ \BBA {} {Pneuman}%
}{%
{Kopp}%
\ \BBA {} {Pneuman}%
}{%
{\protect \APACyear {1976}}%
}]{%
Kopp&Pneuman76}
\APACinsertmetastar {%
Kopp&Pneuman76}%
\begin{APACrefauthors}%
{Kopp}, R\BPBI A.%
\BCBT {}\ \BBA {} {Pneuman}, G\BPBI W.%
\end{APACrefauthors}%
\unskip\
\newblock
\APACrefYearMonthDay{1976}{{\APACmonth{10}}}{}.
\newblock
{\BBOQ}\APACrefatitle {{Magnetic reconnection in the corona and the loop
  prominence phenomenon}} {{Magnetic reconnection in the corona and the loop
  prominence phenomenon}}.{\BBCQ}
\newblock
\APACjournalVolNumPages{\solphys}{50}{}{85-98}.
\newblock
\begin{APACrefDOI} \doi{10.1007/BF00206193} \end{APACrefDOI}
\PrintBackRefs{\CurrentBib}

\bibitem [\protect \citeauthoryear {%
{Kronberg}%
\ \protect \BOthers {.}}{%
{Kronberg}%
\ \protect \BOthers {.}}{%
{\protect \APACyear {2015}}%
}]{%
Kronberg15}
\APACinsertmetastar {%
Kronberg15}%
\begin{APACrefauthors}%
{Kronberg}, E\BPBI A.%
, {Grigorenko}, E\BPBI E.%
, {Haaland}, S\BPBI E.%
, {Daly}, P\BPBI W.%
, {Delcourt}, D\BPBI C.%
, {Luo}, H.%
\BDBL {}{Dandouras}, I.%
\end{APACrefauthors}%
\unskip\
\newblock
\APACrefYearMonthDay{2015}{{\APACmonth{05}}}{}.
\newblock
{\BBOQ}\APACrefatitle {{Distribution of energetic oxygen and hydrogen in the
  near-Earth plasma sheet}} {{Distribution of energetic oxygen and hydrogen in
  the near-Earth plasma sheet}}.{\BBCQ}
\newblock
\APACjournalVolNumPages{Journal of Geophysical Research (Space
  Physics)}{120}{}{3415-3431}.
\newblock
\begin{APACrefDOI} \doi{10.1002/2014JA020882} \end{APACrefDOI}
\PrintBackRefs{\CurrentBib}

\bibitem [\protect \citeauthoryear {%
{Kuznetsova}%
\ \BBA {} {Zelenyi}%
}{%
{Kuznetsova}%
\ \BBA {} {Zelenyi}%
}{%
{\protect \APACyear {1991}}%
}]{%
Kuznetsova91}
\APACinsertmetastar {%
Kuznetsova91}%
\begin{APACrefauthors}%
{Kuznetsova}, M\BPBI M.%
\BCBT {}\ \BBA {} {Zelenyi}, L\BPBI M.%
\end{APACrefauthors}%
\unskip\
\newblock
\APACrefYearMonthDay{1991}{{\APACmonth{10}}}{}.
\newblock
{\BBOQ}\APACrefatitle {{Magnetic reconnection in collisionless field reversals
  - The universality of the ion tearing mode}} {{Magnetic reconnection in
  collisionless field reversals - The universality of the ion tearing
  mode}}.{\BBCQ}
\newblock
\APACjournalVolNumPages{\grl}{18}{}{1825-1828}.
\newblock
\begin{APACrefDOI} \doi{10.1029/91GL02245} \end{APACrefDOI}
\PrintBackRefs{\CurrentBib}

\bibitem [\protect \citeauthoryear {%
{Laval}%
, {Pellat}%
\BCBL {}\ \BBA {} {Vuillemin}%
}{%
{Laval}%
\ \protect \BOthers {.}}{%
{\protect \APACyear {1966}}%
}]{%
LPV66}
\APACinsertmetastar {%
LPV66}%
\begin{APACrefauthors}%
{Laval}, G.%
, {Pellat}, R.%
\BCBL {}\ \BBA {} {Vuillemin}, M.%
\end{APACrefauthors}%
\unskip\
\newblock
\APACrefYearMonthDay{1966}{}{}.
\newblock
{\BBOQ}\APACrefatitle {{Instabilit{\'e}s {\'e}lectromagn{\'e}tiques des plasmas
  sans collisions (CN-21/71)}} {{Instabilit{\'e}s {\'e}lectromagn{\'e}tiques
  des plasmas sans collisions (CN-21/71)}}.{\BBCQ}
\newblock
\BIn{} \APACrefbtitle {Plasma Physics and Controlled Nuclear Fusion Research,
  Volume II} {Plasma physics and controlled nuclear fusion research, volume
  ii}\ (\BPG~259-277).
\PrintBackRefs{\CurrentBib}

\bibitem [\protect \citeauthoryear {%
{Lembege}%
\ \BBA {} {Pellat}%
}{%
{Lembege}%
\ \BBA {} {Pellat}%
}{%
{\protect \APACyear {1982}}%
}]{%
LP82}
\APACinsertmetastar {%
LP82}%
\begin{APACrefauthors}%
{Lembege}, B.%
\BCBT {}\ \BBA {} {Pellat}, R.%
\end{APACrefauthors}%
\unskip\
\newblock
\APACrefYearMonthDay{1982}{{\APACmonth{11}}}{}.
\newblock
{\BBOQ}\APACrefatitle {{Stability of a thick two-dimensional quasineutral
  sheet}} {{Stability of a thick two-dimensional quasineutral sheet}}.{\BBCQ}
\newblock
\APACjournalVolNumPages{Physics of Fluids}{25}{}{1995-2004}.
\newblock
\begin{APACrefDOI} \doi{10.1063/1.863677} \end{APACrefDOI}
\PrintBackRefs{\CurrentBib}

\bibitem [\protect \citeauthoryear {%
{Liu}%
, {Birn}%
, {Daughton}%
, {Hesse}%
\BCBL {}\ \BBA {} {Schindler}%
}{%
{Liu}%
\ \protect \BOthers {.}}{%
{\protect \APACyear {2014}}%
}]{%
Liu14:CS}
\APACinsertmetastar {%
Liu14:CS}%
\begin{APACrefauthors}%
{Liu}, Y\BHBI H.%
, {Birn}, J.%
, {Daughton}, W.%
, {Hesse}, M.%
\BCBL {}\ \BBA {} {Schindler}, K.%
\end{APACrefauthors}%
\unskip\
\newblock
\APACrefYearMonthDay{2014}{{\APACmonth{12}}}{}.
\newblock
{\BBOQ}\APACrefatitle {{Onset of reconnection in the near magnetotail: PIC
  simulations}} {{Onset of reconnection in the near magnetotail: PIC
  simulations}}.{\BBCQ}
\newblock
\APACjournalVolNumPages{\jgr}{119}{}{9773-9789}.
\newblock
\begin{APACrefDOI} \doi{10.1002/2014JA020492} \end{APACrefDOI}
\PrintBackRefs{\CurrentBib}

\bibitem [\protect \citeauthoryear {%
{Loureiro}%
, {Samtaney}%
, {Schekochihin}%
\BCBL {}\ \BBA {} {Uzdensky}%
}{%
{Loureiro}%
\ \protect \BOthers {.}}{%
{\protect \APACyear {2012}}%
}]{%
Loureiro12}
\APACinsertmetastar {%
Loureiro12}%
\begin{APACrefauthors}%
{Loureiro}, N\BPBI F.%
, {Samtaney}, R.%
, {Schekochihin}, A\BPBI A.%
\BCBL {}\ \BBA {} {Uzdensky}, D\BPBI A.%
\end{APACrefauthors}%
\unskip\
\newblock
\APACrefYearMonthDay{2012}{{\APACmonth{04}}}{}.
\newblock
{\BBOQ}\APACrefatitle {{Magnetic reconnection and stochastic plasmoid chains in
  high-Lundquist-number plasmas}} {{Magnetic reconnection and stochastic
  plasmoid chains in high-Lundquist-number plasmas}}.{\BBCQ}
\newblock
\APACjournalVolNumPages{Physics of Plasmas}{19}{4}{042303-042303}.
\newblock
\begin{APACrefDOI} \doi{10.1063/1.3703318} \end{APACrefDOI}
\PrintBackRefs{\CurrentBib}

\bibitem [\protect \citeauthoryear {%
{Lu}%
\ \protect \BOthers {.}}{%
{Lu}%
\ \protect \BOthers {.}}{%
{\protect \APACyear {2019}}%
}]{%
Lu19:jgr:cs}
\APACinsertmetastar {%
Lu19:jgr:cs}%
\begin{APACrefauthors}%
{Lu}, S.%
, {Artemyev}, A\BPBI V.%
, {Angelopoulos}, V.%
, {Lin}, Y.%
, {Zhang}, X\BPBI J.%
, {Liu}, J.%
\BDBL {}{Strangeway}, R\BPBI J.%
\end{APACrefauthors}%
\unskip\
\newblock
\APACrefYearMonthDay{2019}{Feb}{}.
\newblock
{\BBOQ}\APACrefatitle {{The Hall Electric Field in Earth's Magnetotail Thin
  Current Sheet}} {{The Hall Electric Field in Earth's Magnetotail Thin Current
  Sheet}}.{\BBCQ}
\newblock
\APACjournalVolNumPages{Journal of Geophysical Research (Space
  Physics)}{124}{2}{1052-1062}.
\newblock
\begin{APACrefDOI} \doi{10.1029/2018JA026202} \end{APACrefDOI}
\PrintBackRefs{\CurrentBib}

\bibitem [\protect \citeauthoryear {%
{Lu}%
, {Pritchett}%
, {Angelopoulos}%
\BCBL {}\ \BBA {} {Artemyev}%
}{%
{Lu}%
\ \protect \BOthers {.}}{%
{\protect \APACyear {2018}}%
}]{%
Lu18:pop}
\APACinsertmetastar {%
Lu18:pop}%
\begin{APACrefauthors}%
{Lu}, S.%
, {Pritchett}, P\BPBI L.%
, {Angelopoulos}, V.%
\BCBL {}\ \BBA {} {Artemyev}, A\BPBI V.%
\end{APACrefauthors}%
\unskip\
\newblock
\APACrefYearMonthDay{2018}{Jan}{}.
\newblock
{\BBOQ}\APACrefatitle {{Magnetic reconnection in Earth's magnetotail: Energy
  conversion and its earthward-tailward asymmetry}} {{Magnetic reconnection in
  Earth's magnetotail: Energy conversion and its earthward-tailward
  asymmetry}}.{\BBCQ}
\newblock
\APACjournalVolNumPages{Physics of Plasmas}{25}{1}{012905}.
\newblock
\begin{APACrefDOI} \doi{10.1063/1.5016435} \end{APACrefDOI}
\PrintBackRefs{\CurrentBib}

\bibitem [\protect \citeauthoryear {%
{Lyons}%
\ \BBA {} {Speiser}%
}{%
{Lyons}%
\ \BBA {} {Speiser}%
}{%
{\protect \APACyear {1982}}%
}]{%
Lyons82}
\APACinsertmetastar {%
Lyons82}%
\begin{APACrefauthors}%
{Lyons}, L\BPBI R.%
\BCBT {}\ \BBA {} {Speiser}, T\BPBI W.%
\end{APACrefauthors}%
\unskip\
\newblock
\APACrefYearMonthDay{1982}{{\APACmonth{04}}}{}.
\newblock
{\BBOQ}\APACrefatitle {{Evidence for current sheet acceleration in the
  geomagnetic tail}} {{Evidence for current sheet acceleration in the
  geomagnetic tail}}.{\BBCQ}
\newblock
\APACjournalVolNumPages{J. Geophys. Res.}{87}{}{2276-2286}.
\newblock
\begin{APACrefDOI} \doi{10.1029/JA087iA04p02276} \end{APACrefDOI}
\PrintBackRefs{\CurrentBib}

\bibitem [\protect \citeauthoryear {%
{Maggiolo}%
\ \BBA {} {Kistler}%
}{%
{Maggiolo}%
\ \BBA {} {Kistler}%
}{%
{\protect \APACyear {2014}}%
}]{%
Maggiolo&Kistler14}
\APACinsertmetastar {%
Maggiolo&Kistler14}%
\begin{APACrefauthors}%
{Maggiolo}, R.%
\BCBT {}\ \BBA {} {Kistler}, L\BPBI M.%
\end{APACrefauthors}%
\unskip\
\newblock
\APACrefYearMonthDay{2014}{{\APACmonth{04}}}{}.
\newblock
{\BBOQ}\APACrefatitle {{Spatial variation in the plasma sheet composition:
  Dependence on geomagnetic and solar activity}} {{Spatial variation in the
  plasma sheet composition: Dependence on geomagnetic and solar
  activity}}.{\BBCQ}
\newblock
\APACjournalVolNumPages{Journal of Geophysical Research (Space
  Physics)}{119}{}{2836-2857}.
\newblock
\begin{APACrefDOI} \doi{10.1002/2013JA019517} \end{APACrefDOI}
\PrintBackRefs{\CurrentBib}

\bibitem [\protect \citeauthoryear {%
{Merkin}%
, {Sitnov}%
\BCBL {}\ \BBA {} {Lyon}%
}{%
{Merkin}%
\ \protect \BOthers {.}}{%
{\protect \APACyear {2015}}%
}]{%
Merkin15}
\APACinsertmetastar {%
Merkin15}%
\begin{APACrefauthors}%
{Merkin}, V\BPBI G.%
, {Sitnov}, M\BPBI I.%
\BCBL {}\ \BBA {} {Lyon}, J\BPBI G.%
\end{APACrefauthors}%
\unskip\
\newblock
\APACrefYearMonthDay{2015}{{\APACmonth{03}}}{}.
\newblock
{\BBOQ}\APACrefatitle {{Evolution of generalized two-dimensional magnetotail
  equilibria in ideal and resistive MHD}} {{Evolution of generalized
  two-dimensional magnetotail equilibria in ideal and resistive MHD}}.{\BBCQ}
\newblock
\APACjournalVolNumPages{\jgr}{120}{}{1993-2014}.
\newblock
\begin{APACrefDOI} \doi{10.1002/2014JA020651} \end{APACrefDOI}
\PrintBackRefs{\CurrentBib}

\bibitem [\protect \citeauthoryear {%
Mingalev%
\ \protect \BOthers {.}}{%
Mingalev%
\ \protect \BOthers {.}}{%
{\protect \APACyear {2018}}%
}]{%
Mingalev18}
\APACinsertmetastar {%
Mingalev18}%
\begin{APACrefauthors}%
Mingalev, O\BPBI V.%
, Malova, H\BPBI V.%
, Mingalev, I\BPBI V.%
, Mel'nik, M\BPBI N.%
, Setsko, P\BPBI V.%
\BCBL {}\ \BBA {} Zelenyi, L\BPBI M.%
\end{APACrefauthors}%
\unskip\
\newblock
\APACrefYearMonthDay{2018}{Oct}{01}.
\newblock
{\BBOQ}\APACrefatitle {Model of a Thin Current Sheet in the Earth's Magnetotail
  with a Kinetic Description of Magnetized Electrons} {Model of a thin current
  sheet in the earth's magnetotail with a kinetic description of magnetized
  electrons}.{\BBCQ}
\newblock
\APACjournalVolNumPages{Plasma Physics Reports}{44}{10}{899--919}.
\newblock
\begin{APACrefURL} \url{https://doi.org/10.1134/S1063780X18100082}
  \end{APACrefURL}
\newblock
\begin{APACrefDOI} \doi{10.1134/S1063780X18100082} \end{APACrefDOI}
\PrintBackRefs{\CurrentBib}

\bibitem [\protect \citeauthoryear {%
{Mingalev}%
, {Mingalev}%
, {Malova}%
\BCBL {}\ \BBA {} {Zelenyi}%
}{%
{Mingalev}%
\ \protect \BOthers {.}}{%
{\protect \APACyear {2007}}%
}]{%
Mingalev07}
\APACinsertmetastar {%
Mingalev07}%
\begin{APACrefauthors}%
{Mingalev}, O\BPBI V.%
, {Mingalev}, I\BPBI V.%
, {Malova}, K\BPBI V.%
\BCBL {}\ \BBA {} {Zelenyi}, L\BPBI M.%
\end{APACrefauthors}%
\unskip\
\newblock
\APACrefYearMonthDay{2007}{{\APACmonth{11}}}{}.
\newblock
{\BBOQ}\APACrefatitle {{Numerical simulations of plasma equilibrium in a
  one-dimensional current sheet with a nonzero normal magnetic field
  component}} {{Numerical simulations of plasma equilibrium in a
  one-dimensional current sheet with a nonzero normal magnetic field
  component}}.{\BBCQ}
\newblock
\APACjournalVolNumPages{Plasma Physics Reports}{33}{}{942-955}.
\newblock
\begin{APACrefDOI} \doi{10.1134/S1063780X07110062} \end{APACrefDOI}
\PrintBackRefs{\CurrentBib}

\bibitem [\protect \citeauthoryear {%
{Nagai}%
\ \BBA {} {Machida}%
}{%
{Nagai}%
\ \BBA {} {Machida}%
}{%
{\protect \APACyear {1998}}%
}]{%
Nagai&Machida98}
\APACinsertmetastar {%
Nagai&Machida98}%
\begin{APACrefauthors}%
{Nagai}, T.%
\BCBT {}\ \BBA {} {Machida}, S.%
\end{APACrefauthors}%
\unskip\
\newblock
\APACrefYearMonthDay{1998}{}{}.
\newblock
{\BBOQ}\APACrefatitle {{Magnetic Reconnection in the Near-Earth Magnetotail}}
  {{Magnetic Reconnection in the Near-Earth Magnetotail}}.{\BBCQ}
\newblock
\APACjournalVolNumPages{Washington DC American Geophysical Union Geophysical
  Monograph Series}{105}{}{211}.
\newblock
\begin{APACrefDOI} \doi{10.1029/GM105p0211} \end{APACrefDOI}
\PrintBackRefs{\CurrentBib}

\bibitem [\protect \citeauthoryear {%
{Nakai}%
, {Kamide}%
\BCBL {}\ \BBA {} {Russell}%
}{%
{Nakai}%
\ \protect \BOthers {.}}{%
{\protect \APACyear {1991}}%
}]{%
Nakai91}
\APACinsertmetastar {%
Nakai91}%
\begin{APACrefauthors}%
{Nakai}, H.%
, {Kamide}, Y.%
\BCBL {}\ \BBA {} {Russell}, C\BPBI T.%
\end{APACrefauthors}%
\unskip\
\newblock
\APACrefYearMonthDay{1991}{{\APACmonth{04}}}{}.
\newblock
{\BBOQ}\APACrefatitle {{Influences of solar wind parameters and geomagnetic
  activity on the tail lobe magnetic field - A statistical study}} {{Influences
  of solar wind parameters and geomagnetic activity on the tail lobe magnetic
  field - A statistical study}}.{\BBCQ}
\newblock
\APACjournalVolNumPages{\jgr}{96}{}{5511-5523}.
\newblock
\begin{APACrefDOI} \doi{10.1029/90JA02361} \end{APACrefDOI}
\PrintBackRefs{\CurrentBib}

\bibitem [\protect \citeauthoryear {%
{Nickeler}%
\ \BBA {} {Wiegelmann}%
}{%
{Nickeler}%
\ \BBA {} {Wiegelmann}%
}{%
{\protect \APACyear {2010}}%
}]{%
Nickeler&Wiegelmann10}
\APACinsertmetastar {%
Nickeler&Wiegelmann10}%
\begin{APACrefauthors}%
{Nickeler}, D\BPBI H.%
\BCBT {}\ \BBA {} {Wiegelmann}, T.%
\end{APACrefauthors}%
\unskip\
\newblock
\APACrefYearMonthDay{2010}{{\APACmonth{08}}}{}.
\newblock
{\BBOQ}\APACrefatitle {{Thin current sheets caused by plasma flow gradients in
  space and astrophysical plasma}} {{Thin current sheets caused by plasma flow
  gradients in space and astrophysical plasma}}.{\BBCQ}
\newblock
\APACjournalVolNumPages{Annales Geophysicae}{28}{}{1523-1532}.
\newblock
\begin{APACrefDOI} \doi{10.5194/angeo-28-1523-2010} \end{APACrefDOI}
\PrintBackRefs{\CurrentBib}

\bibitem [\protect \citeauthoryear {%
{Ofman}%
, {Chen}%
, {Morrison}%
\BCBL {}\ \BBA {} {Steinolfson}%
}{%
{Ofman}%
\ \protect \BOthers {.}}{%
{\protect \APACyear {1991}}%
}]{%
Ofman91}
\APACinsertmetastar {%
Ofman91}%
\begin{APACrefauthors}%
{Ofman}, L.%
, {Chen}, X\BPBI L.%
, {Morrison}, P\BPBI J.%
\BCBL {}\ \BBA {} {Steinolfson}, R\BPBI S.%
\end{APACrefauthors}%
\unskip\
\newblock
\APACrefYearMonthDay{1991}{{\APACmonth{06}}}{}.
\newblock
{\BBOQ}\APACrefatitle {{Resistive tearing mode instability with shear flow and
  viscosity}} {{Resistive tearing mode instability with shear flow and
  viscosity}}.{\BBCQ}
\newblock
\APACjournalVolNumPages{Physics of Fluids B}{3}{6}{1364-1373}.
\newblock
\begin{APACrefDOI} \doi{10.1063/1.859701} \end{APACrefDOI}
\PrintBackRefs{\CurrentBib}

\bibitem [\protect \citeauthoryear {%
{Pellat}%
, {Coroniti}%
\BCBL {}\ \BBA {} {Pritchett}%
}{%
{Pellat}%
\ \protect \BOthers {.}}{%
{\protect \APACyear {1991}}%
}]{%
Pellat91}
\APACinsertmetastar {%
Pellat91}%
\begin{APACrefauthors}%
{Pellat}, R.%
, {Coroniti}, F\BPBI V.%
\BCBL {}\ \BBA {} {Pritchett}, P\BPBI L.%
\end{APACrefauthors}%
\unskip\
\newblock
\APACrefYearMonthDay{1991}{{\APACmonth{02}}}{}.
\newblock
{\BBOQ}\APACrefatitle {{Does ion tearing exist?}} {{Does ion tearing
  exist?}}{\BBCQ}
\newblock
\APACjournalVolNumPages{\grl}{18}{}{143-146}.
\newblock
\begin{APACrefDOI} \doi{10.1029/91GL00123} \end{APACrefDOI}
\PrintBackRefs{\CurrentBib}

\bibitem [\protect \citeauthoryear {%
{Petrukovich}%
, {Artemyev}%
, {Nakamura}%
, {Panov}%
\BCBL {}\ \BBA {} {Baumjohann}%
}{%
{Petrukovich}%
\ \protect \BOthers {.}}{%
{\protect \APACyear {2013}}%
}]{%
Petrukovich13}
\APACinsertmetastar {%
Petrukovich13}%
\begin{APACrefauthors}%
{Petrukovich}, A\BPBI A.%
, {Artemyev}, A\BPBI V.%
, {Nakamura}, R.%
, {Panov}, E\BPBI V.%
\BCBL {}\ \BBA {} {Baumjohann}, W.%
\end{APACrefauthors}%
\unskip\
\newblock
\APACrefYearMonthDay{2013}{{\APACmonth{09}}}{}.
\newblock
{\BBOQ}\APACrefatitle {{Cluster observations of dBz/dx during growth phase
  magnetotail stretching intervals}} {{Cluster observations of dBz/dx during
  growth phase magnetotail stretching intervals}}.{\BBCQ}
\newblock
\APACjournalVolNumPages{\jgr}{118}{}{5720-5730}.
\newblock
\begin{APACrefDOI} \doi{10.1002/jgra.50550} \end{APACrefDOI}
\PrintBackRefs{\CurrentBib}

\bibitem [\protect \citeauthoryear {%
{Petrukovich}%
, {Artemyev}%
, {Vasko}%
, {Nakamura}%
\BCBL {}\ \BBA {} {Zelenyi}%
}{%
{Petrukovich}%
\ \protect \BOthers {.}}{%
{\protect \APACyear {2015}}%
}]{%
Petrukovich15:ssr}
\APACinsertmetastar {%
Petrukovich15:ssr}%
\begin{APACrefauthors}%
{Petrukovich}, A\BPBI A.%
, {Artemyev}, A\BPBI V.%
, {Vasko}, I\BPBI Y.%
, {Nakamura}, R.%
\BCBL {}\ \BBA {} {Zelenyi}, L\BPBI M.%
\end{APACrefauthors}%
\unskip\
\newblock
\APACrefYearMonthDay{2015}{}{}.
\newblock
{\BBOQ}\APACrefatitle {{Current sheets in the Earth magnetotail: plasma and
  magnetic field structure with Cluster project observations}} {{Current sheets
  in the Earth magnetotail: plasma and magnetic field structure with Cluster
  project observations}}.{\BBCQ}
\newblock
\APACjournalVolNumPages{\ssr}{188}{}{311-337}.
\newblock
\begin{APACrefDOI} \doi{10.1007/s11214-014-0126-7} \end{APACrefDOI}
\PrintBackRefs{\CurrentBib}

\bibitem [\protect \citeauthoryear {%
{Petrukovich}%
\ \protect \BOthers {.}}{%
{Petrukovich}%
\ \protect \BOthers {.}}{%
{\protect \APACyear {1999}}%
}]{%
Petrukovich99}
\APACinsertmetastar {%
Petrukovich99}%
\begin{APACrefauthors}%
{Petrukovich}, A\BPBI A.%
, {Mukai}, T.%
, {Kokubun}, S.%
, {Romanov}, S\BPBI A.%
, {Saito}, Y.%
, {Yamamoto}, T.%
\BCBL {}\ \BBA {} {Zelenyi}, L\BPBI M.%
\end{APACrefauthors}%
\unskip\
\newblock
\APACrefYearMonthDay{1999}{{\APACmonth{03}}}{}.
\newblock
{\BBOQ}\APACrefatitle {{Substorm-associated pressure variations in the
  magnetotail plasma sheet and lobe}} {{Substorm-associated pressure variations
  in the magnetotail plasma sheet and lobe}}.{\BBCQ}
\newblock
\APACjournalVolNumPages{\jgr}{104}{}{4501-4514}.
\newblock
\begin{APACrefDOI} \doi{10.1029/98JA02418} \end{APACrefDOI}
\PrintBackRefs{\CurrentBib}

\bibitem [\protect \citeauthoryear {%
{Petrukovich}%
\ \protect \BOthers {.}}{%
{Petrukovich}%
\ \protect \BOthers {.}}{%
{\protect \APACyear {1998}}%
}]{%
Petrukovich98}
\APACinsertmetastar {%
Petrukovich98}%
\begin{APACrefauthors}%
{Petrukovich}, A\BPBI A.%
, {Sergeev}, V\BPBI A.%
, {Zelenyi}, L\BPBI M.%
, {Mukai}, T.%
, {Yamamoto}, T.%
, {Kokubun}, S.%
\BDBL {}{Sandahl}, I.%
\end{APACrefauthors}%
\unskip\
\newblock
\APACrefYearMonthDay{1998}{{\APACmonth{01}}}{}.
\newblock
{\BBOQ}\APACrefatitle {{Two spacecraft observations of a reconnection pulse
  during an auroral breakup}} {{Two spacecraft observations of a reconnection
  pulse during an auroral breakup}}.{\BBCQ}
\newblock
\APACjournalVolNumPages{\jgr}{103}{}{47-60}.
\newblock
\begin{APACrefDOI} \doi{10.1029/97JA02296} \end{APACrefDOI}
\PrintBackRefs{\CurrentBib}

\bibitem [\protect \citeauthoryear {%
Phan%
\ \protect \BOthers {.}}{%
Phan%
\ \protect \BOthers {.}}{%
{\protect \APACyear {2020}}%
}]{%
Phan20}
\APACinsertmetastar {%
Phan20}%
\begin{APACrefauthors}%
Phan, T.%
, Bale, S.%
, Eastwood, J.%
, Lavraud, B.%
, Drake, J.%
, Oieroset, M.%
\BDBL {}others%
\end{APACrefauthors}%
\unskip\
\newblock
\APACrefYearMonthDay{2020}{}{}.
\newblock
{\BBOQ}\APACrefatitle {Parker solar probe in situ observations of magnetic
  reconnection exhausts during encounter 1} {Parker solar probe in situ
  observations of magnetic reconnection exhausts during encounter 1}.{\BBCQ}
\newblock
\APACjournalVolNumPages{The Astrophysical Journal Supplement
  Series}{246}{2}{34}.
\PrintBackRefs{\CurrentBib}

\bibitem [\protect \citeauthoryear {%
{Phan}%
\ \protect \BOthers {.}}{%
{Phan}%
\ \protect \BOthers {.}}{%
{\protect \APACyear {2014}}%
}]{%
Phan14}
\APACinsertmetastar {%
Phan14}%
\begin{APACrefauthors}%
{Phan}, T\BPBI D.%
, {Drake}, J\BPBI F.%
, {Shay}, M\BPBI A.%
, {Gosling}, J\BPBI T.%
, {Paschmann}, G.%
, {Eastwood}, J\BPBI P.%
\BDBL {}{Angelopoulos}, V.%
\end{APACrefauthors}%
\unskip\
\newblock
\APACrefYearMonthDay{2014}{{\APACmonth{10}}}{}.
\newblock
{\BBOQ}\APACrefatitle {{Ion bulk heating in magnetic reconnection exhausts at
  Earth's magnetopause: Dependence on the inflow Alfv{\'e}n speed and magnetic
  shear angle}} {{Ion bulk heating in magnetic reconnection exhausts at Earth's
  magnetopause: Dependence on the inflow Alfv{\'e}n speed and magnetic shear
  angle}}.{\BBCQ}
\newblock
\APACjournalVolNumPages{\grl}{41}{}{7002-7010}.
\newblock
\begin{APACrefDOI} \doi{10.1002/2014GL061547} \end{APACrefDOI}
\PrintBackRefs{\CurrentBib}

\bibitem [\protect \citeauthoryear {%
{Phan}%
\ \protect \BOthers {.}}{%
{Phan}%
\ \protect \BOthers {.}}{%
{\protect \APACyear {2006}}%
}]{%
Phan06:reconnection}
\APACinsertmetastar {%
Phan06:reconnection}%
\begin{APACrefauthors}%
{Phan}, T\BPBI D.%
, {Gosling}, J\BPBI T.%
, {Davis}, M\BPBI S.%
, {Skoug}, R\BPBI M.%
, {{\O}ieroset}, M.%
, {Lin}, R\BPBI P.%
\BDBL {}{Balogh}, A.%
\end{APACrefauthors}%
\unskip\
\newblock
\APACrefYearMonthDay{2006}{{\APACmonth{01}}}{}.
\newblock
{\BBOQ}\APACrefatitle {{A magnetic reconnection X-line extending more than 390
  Earth radii in the solar wind}} {{A magnetic reconnection X-line extending
  more than 390 Earth radii in the solar wind}}.{\BBCQ}
\newblock
\APACjournalVolNumPages{\nat}{439}{}{175-178}.
\newblock
\begin{APACrefDOI} \doi{10.1038/nature04393} \end{APACrefDOI}
\PrintBackRefs{\CurrentBib}

\bibitem [\protect \citeauthoryear {%
{Phan}%
\ \protect \BOthers {.}}{%
{Phan}%
\ \protect \BOthers {.}}{%
{\protect \APACyear {2010}}%
}]{%
Phan10}
\APACinsertmetastar {%
Phan10}%
\begin{APACrefauthors}%
{Phan}, T\BPBI D.%
, {Gosling}, J\BPBI T.%
, {Paschmann}, G.%
, {Pasma}, C.%
, {Drake}, J\BPBI F.%
, {{\O}ieroset}, M.%
\BDBL {}{Davis}, M\BPBI S.%
\end{APACrefauthors}%
\unskip\
\newblock
\APACrefYearMonthDay{2010}{{\APACmonth{08}}}{}.
\newblock
{\BBOQ}\APACrefatitle {{The Dependence of Magnetic Reconnection on Plasma
  {\ensuremath{\beta}} and Magnetic Shear: Evidence from Solar Wind
  Observations}} {{The Dependence of Magnetic Reconnection on Plasma
  {\ensuremath{\beta}} and Magnetic Shear: Evidence from Solar Wind
  Observations}}.{\BBCQ}
\newblock
\APACjournalVolNumPages{\apjl}{719}{2}{L199-L203}.
\newblock
\begin{APACrefDOI} \doi{10.1088/2041-8205/719/2/L199} \end{APACrefDOI}
\PrintBackRefs{\CurrentBib}

\bibitem [\protect \citeauthoryear {%
{Phan}%
\ \BBA {} {Sonnerup}%
}{%
{Phan}%
\ \BBA {} {Sonnerup}%
}{%
{\protect \APACyear {1991}}%
}]{%
Phan&Sonnerup91}
\APACinsertmetastar {%
Phan&Sonnerup91}%
\begin{APACrefauthors}%
{Phan}, T\BPBI D.%
\BCBT {}\ \BBA {} {Sonnerup}, B\BPBI U\BPBI {\"O}.%
\end{APACrefauthors}%
\unskip\
\newblock
\APACrefYearMonthDay{1991}{{\APACmonth{12}}}{}.
\newblock
{\BBOQ}\APACrefatitle {{Resistive tearing-mode instability in a current sheet
  with equilibrium viscous stagnation-point flow}} {{Resistive tearing-mode
  instability in a current sheet with equilibrium viscous stagnation-point
  flow}}.{\BBCQ}
\newblock
\APACjournalVolNumPages{Journal of Plasma Physics}{46}{3}{407-421}.
\newblock
\begin{APACrefDOI} \doi{10.1017/S0022377800016214} \end{APACrefDOI}
\PrintBackRefs{\CurrentBib}

\bibitem [\protect \citeauthoryear {%
{Pritchett}%
}{%
{Pritchett}%
}{%
{\protect \APACyear {2015}}%
}]{%
Pritchett15:pop}
\APACinsertmetastar {%
Pritchett15:pop}%
\begin{APACrefauthors}%
{Pritchett}, P\BPBI L.%
\end{APACrefauthors}%
\unskip\
\newblock
\APACrefYearMonthDay{2015}{{\APACmonth{06}}}{}.
\newblock
{\BBOQ}\APACrefatitle {{Instability of current sheets with a localized
  accumulation of magnetic flux}} {{Instability of current sheets with a
  localized accumulation of magnetic flux}}.{\BBCQ}
\newblock
\APACjournalVolNumPages{Physics of Plasmas}{22}{6}{062102}.
\newblock
\begin{APACrefDOI} \doi{10.1063/1.4921666} \end{APACrefDOI}
\PrintBackRefs{\CurrentBib}

\bibitem [\protect \citeauthoryear {%
{Pritchett}%
\ \BBA {} {Buchner}%
}{%
{Pritchett}%
\ \BBA {} {Buchner}%
}{%
{\protect \APACyear {1995}}%
}]{%
Pritchett95}
\APACinsertmetastar {%
Pritchett95}%
\begin{APACrefauthors}%
{Pritchett}, P\BPBI L.%
\BCBT {}\ \BBA {} {Buchner}, J.%
\end{APACrefauthors}%
\unskip\
\newblock
\APACrefYearMonthDay{1995}{{\APACmonth{03}}}{}.
\newblock
{\BBOQ}\APACrefatitle {{Collisionless reconnection in configurations with a
  minimum in the equatorial magnetic field and with magnetic shear}}
  {{Collisionless reconnection in configurations with a minimum in the
  equatorial magnetic field and with magnetic shear}}.{\BBCQ}
\newblock
\APACjournalVolNumPages{\jgr}{100}{}{3601-3611}.
\newblock
\begin{APACrefDOI} \doi{10.1029/94JA03028} \end{APACrefDOI}
\PrintBackRefs{\CurrentBib}

\bibitem [\protect \citeauthoryear {%
{Pritchett}%
\ \BBA {} {Coroniti}%
}{%
{Pritchett}%
\ \BBA {} {Coroniti}%
}{%
{\protect \APACyear {1992}}%
}]{%
Pritchett92}
\APACinsertmetastar {%
Pritchett92}%
\begin{APACrefauthors}%
{Pritchett}, P\BPBI L.%
\BCBT {}\ \BBA {} {Coroniti}, F\BPBI V.%
\end{APACrefauthors}%
\unskip\
\newblock
\APACrefYearMonthDay{1992}{{\APACmonth{11}}}{}.
\newblock
{\BBOQ}\APACrefatitle {{Formation and stability of the self-consistent
  one-dimensional tail current sheet}} {{Formation and stability of the
  self-consistent one-dimensional tail current sheet}}.{\BBCQ}
\newblock
\APACjournalVolNumPages{J. Geophys. Res.}{97}{}{16773-16787}.
\newblock
\begin{APACrefDOI} \doi{10.1029/92JA01550} \end{APACrefDOI}
\PrintBackRefs{\CurrentBib}

\bibitem [\protect \citeauthoryear {%
{Pritchett}%
\ \BBA {} {Coroniti}%
}{%
{Pritchett}%
\ \BBA {} {Coroniti}%
}{%
{\protect \APACyear {1994}}%
}]{%
Pritchett&Coroniti94}
\APACinsertmetastar {%
Pritchett&Coroniti94}%
\begin{APACrefauthors}%
{Pritchett}, P\BPBI L.%
\BCBT {}\ \BBA {} {Coroniti}, F\BPBI V.%
\end{APACrefauthors}%
\unskip\
\newblock
\APACrefYearMonthDay{1994}{{\APACmonth{07}}}{}.
\newblock
{\BBOQ}\APACrefatitle {{Convection and the formation of thin current sheets in
  the near-Earth plasma sheet}} {{Convection and the formation of thin current
  sheets in the near-Earth plasma sheet}}.{\BBCQ}
\newblock
\APACjournalVolNumPages{\grl}{21}{}{1587-1590}.
\newblock
\begin{APACrefDOI} \doi{10.1029/94GL01364} \end{APACrefDOI}
\PrintBackRefs{\CurrentBib}

\bibitem [\protect \citeauthoryear {%
{Pritchett}%
\ \BBA {} {Coroniti}%
}{%
{Pritchett}%
\ \BBA {} {Coroniti}%
}{%
{\protect \APACyear {1995}}%
}]{%
Pritchett&Coroniti95}
\APACinsertmetastar {%
Pritchett&Coroniti95}%
\begin{APACrefauthors}%
{Pritchett}, P\BPBI L.%
\BCBT {}\ \BBA {} {Coroniti}, F\BPBI V.%
\end{APACrefauthors}%
\unskip\
\newblock
\APACrefYearMonthDay{1995}{{\APACmonth{12}}}{}.
\newblock
{\BBOQ}\APACrefatitle {{Formation of thin current sheets during plasma sheet
  convection}} {{Formation of thin current sheets during plasma sheet
  convection}}.{\BBCQ}
\newblock
\APACjournalVolNumPages{\jgr}{100}{}{23551-23566}.
\newblock
\begin{APACrefDOI} \doi{10.1029/95JA02540} \end{APACrefDOI}
\PrintBackRefs{\CurrentBib}

\bibitem [\protect \citeauthoryear {%
{Pritchett}%
, {Coroniti}%
\BCBL {}\ \BBA {} {Pellat}%
}{%
{Pritchett}%
\ \protect \BOthers {.}}{%
{\protect \APACyear {1997}}%
}]{%
Pritchett97}
\APACinsertmetastar {%
Pritchett97}%
\begin{APACrefauthors}%
{Pritchett}, P\BPBI L.%
, {Coroniti}, F\BPBI V.%
\BCBL {}\ \BBA {} {Pellat}, R.%
\end{APACrefauthors}%
\unskip\
\newblock
\APACrefYearMonthDay{1997}{}{}.
\newblock
{\BBOQ}\APACrefatitle {{Convection-driven reconnection and the stability of the
  near-Earth plasma sheet}} {{Convection-driven reconnection and the stability
  of the near-Earth plasma sheet}}.{\BBCQ}
\newblock
\APACjournalVolNumPages{\grl}{24}{}{873-876}.
\newblock
\begin{APACrefDOI} \doi{10.1029/97GL00672} \end{APACrefDOI}
\PrintBackRefs{\CurrentBib}

\bibitem [\protect \citeauthoryear {%
{Pritchett}%
, {Coroniti}%
, {Pellat}%
\BCBL {}\ \BBA {} {Karimabadi}%
}{%
{Pritchett}%
\ \protect \BOthers {.}}{%
{\protect \APACyear {1991}}%
}]{%
Pritchett91}
\APACinsertmetastar {%
Pritchett91}%
\begin{APACrefauthors}%
{Pritchett}, P\BPBI L.%
, {Coroniti}, F\BPBI V.%
, {Pellat}, R.%
\BCBL {}\ \BBA {} {Karimabadi}, H.%
\end{APACrefauthors}%
\unskip\
\newblock
\APACrefYearMonthDay{1991}{{\APACmonth{07}}}{}.
\newblock
{\BBOQ}\APACrefatitle {{Collisionless reconnection in two-dimensional
  magnetotail equilibria}} {{Collisionless reconnection in two-dimensional
  magnetotail equilibria}}.{\BBCQ}
\newblock
\APACjournalVolNumPages{\jgr}{96}{}{11}.
\newblock
\begin{APACrefDOI} \doi{10.1029/91JA01094} \end{APACrefDOI}
\PrintBackRefs{\CurrentBib}

\bibitem [\protect \citeauthoryear {%
Pucci%
\ \BBA {} Velli%
}{%
Pucci%
\ \BBA {} Velli%
}{%
{\protect \APACyear {2013}}%
}]{%
PucciandVelli2013}
\APACinsertmetastar {%
PucciandVelli2013}%
\begin{APACrefauthors}%
Pucci, F.%
\BCBT {}\ \BBA {} Velli, M.%
\end{APACrefauthors}%
\unskip\
\newblock
\APACrefYearMonthDay{2013}{}{}.
\newblock
{\BBOQ}\APACrefatitle {Reconnection of quasi-singular current sheets: the
  “ideal” tearing mode} {Reconnection of quasi-singular current sheets: the
  “ideal” tearing mode}.{\BBCQ}
\newblock
\APACjournalVolNumPages{The Astrophysical Journal Letters}{780}{2}{L19}.
\PrintBackRefs{\CurrentBib}

\bibitem [\protect \citeauthoryear {%
Pucci%
\ \protect \BOthers {.}}{%
Pucci%
\ \protect \BOthers {.}}{%
{\protect \APACyear {2020}}%
}]{%
pucci2020onset}
\APACinsertmetastar {%
pucci2020onset}%
\begin{APACrefauthors}%
Pucci, F.%
, Velli, M.%
, Shi, C.%
, Singh, K.%
, Tenerani, A.%
, Alladio, F.%
\BDBL {}others%
\end{APACrefauthors}%
\unskip\
\newblock
\APACrefYearMonthDay{2020}{}{}.
\newblock
{\BBOQ}\APACrefatitle {Onset of fast magnetic reconnection and particle
  energization in laboratory and space plasmas} {Onset of fast magnetic
  reconnection and particle energization in laboratory and space
  plasmas}.{\BBCQ}
\newblock
\APACjournalVolNumPages{Journal of Plasma Physics}{86}{6}{}.
\PrintBackRefs{\CurrentBib}

\bibitem [\protect \citeauthoryear {%
{Quest}%
, {Karimabadi}%
\BCBL {}\ \BBA {} {Brittnacher}%
}{%
{Quest}%
\ \protect \BOthers {.}}{%
{\protect \APACyear {1996}}%
}]{%
Quest96}
\APACinsertmetastar {%
Quest96}%
\begin{APACrefauthors}%
{Quest}, K\BPBI B.%
, {Karimabadi}, H.%
\BCBL {}\ \BBA {} {Brittnacher}, M.%
\end{APACrefauthors}%
\unskip\
\newblock
\APACrefYearMonthDay{1996}{{\APACmonth{01}}}{}.
\newblock
{\BBOQ}\APACrefatitle {{Consequences of particle conservation along a flux
  surface for magnetotail tearing}} {{Consequences of particle conservation
  along a flux surface for magnetotail tearing}}.{\BBCQ}
\newblock
\APACjournalVolNumPages{\jgr}{101}{}{179-184}.
\newblock
\begin{APACrefDOI} \doi{10.1029/95JA02986} \end{APACrefDOI}
\PrintBackRefs{\CurrentBib}

\bibitem [\protect \citeauthoryear {%
{Reeves}%
\ \protect \BOthers {.}}{%
{Reeves}%
\ \protect \BOthers {.}}{%
{\protect \APACyear {2008}}%
}]{%
Reeves08:solar}
\APACinsertmetastar {%
Reeves08:solar}%
\begin{APACrefauthors}%
{Reeves}, K\BPBI K.%
, {Guild}, T\BPBI B.%
, {Hughes}, W\BPBI J.%
, {Korreck}, K\BPBI E.%
, {Lin}, J.%
, {Raymond}, J.%
\BDBL {}{Wiltberger}, M.%
\end{APACrefauthors}%
\unskip\
\newblock
\APACrefYearMonthDay{2008}{{\APACmonth{09}}}{}.
\newblock
{\BBOQ}\APACrefatitle {{Posteruptive phenomena in coronal mass ejections and
  substorms: Indicators of a universal process?}} {{Posteruptive phenomena in
  coronal mass ejections and substorms: Indicators of a universal
  process?}}{\BBCQ}
\newblock
\APACjournalVolNumPages{\jgr}{113}{}{0}.
\newblock
\begin{APACrefDOI} \doi{10.1029/2008JA013049} \end{APACrefDOI}
\PrintBackRefs{\CurrentBib}

\bibitem [\protect \citeauthoryear {%
{R{\'e}ville}%
\ \protect \BOthers {.}}{%
{R{\'e}ville}%
\ \protect \BOthers {.}}{%
{\protect \APACyear {2020}}%
}]{%
Reville20}
\APACinsertmetastar {%
Reville20}%
\begin{APACrefauthors}%
{R{\'e}ville}, V.%
, {Velli}, M.%
, {Rouillard}, A\BPBI P.%
, {Lavraud}, B.%
, {Tenerani}, A.%
, {Shi}, C.%
\BCBL {}\ \BBA {} {Strugarek}, A.%
\end{APACrefauthors}%
\unskip\
\newblock
\APACrefYearMonthDay{2020}{{\APACmonth{05}}}{}.
\newblock
{\BBOQ}\APACrefatitle {{Tearing Instability and Periodic Density Perturbations
  in the Slow Solar Wind}} {{Tearing Instability and Periodic Density
  Perturbations in the Slow Solar Wind}}.{\BBCQ}
\newblock
\APACjournalVolNumPages{\apjl}{895}{1}{L20}.
\newblock
\begin{APACrefDOI} \doi{10.3847/2041-8213/ab911d} \end{APACrefDOI}
\PrintBackRefs{\CurrentBib}

\bibitem [\protect \citeauthoryear {%
{Rich}%
, {Vasyliunas}%
\BCBL {}\ \BBA {} {Wolf}%
}{%
{Rich}%
\ \protect \BOthers {.}}{%
{\protect \APACyear {1972}}%
}]{%
Rich72}
\APACinsertmetastar {%
Rich72}%
\begin{APACrefauthors}%
{Rich}, F\BPBI J.%
, {Vasyliunas}, V\BPBI M.%
\BCBL {}\ \BBA {} {Wolf}, R\BPBI A.%
\end{APACrefauthors}%
\unskip\
\newblock
\APACrefYearMonthDay{1972}{}{}.
\newblock
{\BBOQ}\APACrefatitle {{On the Balance of Stresses in the Plasma Sheet}} {{On
  the Balance of Stresses in the Plasma Sheet}}.{\BBCQ}
\newblock
\APACjournalVolNumPages{J. Geophys. Res.}{77}{}{4670-4676}.
\newblock
\begin{APACrefDOI} \doi{10.1029/JA077i025p04670} \end{APACrefDOI}
\PrintBackRefs{\CurrentBib}

\bibitem [\protect \citeauthoryear {%
{Riley}%
\ \BBA {} {Luhmann}%
}{%
{Riley}%
\ \BBA {} {Luhmann}%
}{%
{\protect \APACyear {2012}}%
}]{%
Riley&Luhmann12}
\APACinsertmetastar {%
Riley&Luhmann12}%
\begin{APACrefauthors}%
{Riley}, P.%
\BCBT {}\ \BBA {} {Luhmann}, J\BPBI G.%
\end{APACrefauthors}%
\unskip\
\newblock
\APACrefYearMonthDay{2012}{{\APACmonth{04}}}{}.
\newblock
{\BBOQ}\APACrefatitle {{Interplanetary Signatures of Unipolar Streamers and the
  Origin of the Slow Solar Wind}} {{Interplanetary Signatures of Unipolar
  Streamers and the Origin of the Slow Solar Wind}}.{\BBCQ}
\newblock
\APACjournalVolNumPages{\solphys}{277}{2}{355-373}.
\newblock
\begin{APACrefDOI} \doi{10.1007/s11207-011-9909-0} \end{APACrefDOI}
\PrintBackRefs{\CurrentBib}

\bibitem [\protect \citeauthoryear {%
{Runov}%
\ \protect \BOthers {.}}{%
{Runov}%
\ \protect \BOthers {.}}{%
{\protect \APACyear {2006}}%
}]{%
Runov06}
\APACinsertmetastar {%
Runov06}%
\begin{APACrefauthors}%
{Runov}, A.%
, {Sergeev}, V\BPBI A.%
, {Nakamura}, R.%
, {Baumjohann}, W.%
, {Apatenkov}, S.%
, {Asano}, Y.%
\BDBL {}{Balogh}, A.%
\end{APACrefauthors}%
\unskip\
\newblock
\APACrefYearMonthDay{2006}{{\APACmonth{03}}}{}.
\newblock
{\BBOQ}\APACrefatitle {{Local structure of the magnetotail current sheet: 2001
  Cluster observations}} {{Local structure of the magnetotail current sheet:
  2001 Cluster observations}}.{\BBCQ}
\newblock
\APACjournalVolNumPages{Annales Geophysicae}{24}{}{247-262}.
\PrintBackRefs{\CurrentBib}

\bibitem [\protect \citeauthoryear {%
{Sakai}%
\ \BBA {} {de Jager}%
}{%
{Sakai}%
\ \BBA {} {de Jager}%
}{%
{\protect \APACyear {1996}}%
}]{%
Sakai&deJager96}
\APACinsertmetastar {%
Sakai&deJager96}%
\begin{APACrefauthors}%
{Sakai}, J\BHBI I.%
\BCBT {}\ \BBA {} {de Jager}, C.%
\end{APACrefauthors}%
\unskip\
\newblock
\APACrefYearMonthDay{1996}{{\APACmonth{07}}}{}.
\newblock
{\BBOQ}\APACrefatitle {{Solar Flares and Collisions Between Current-Carrying
  Loops Types and Mechanisms of Solar Flares and Coronal Loop Heating}} {{Solar
  Flares and Collisions Between Current-Carrying Loops Types and Mechanisms of
  Solar Flares and Coronal Loop Heating}}.{\BBCQ}
\newblock
\APACjournalVolNumPages{\ssr}{77}{}{1-192}.
\newblock
\begin{APACrefDOI} \doi{10.1007/BF00227866} \end{APACrefDOI}
\PrintBackRefs{\CurrentBib}

\bibitem [\protect \citeauthoryear {%
{Sauvaud}%
\ \protect \BOthers {.}}{%
{Sauvaud}%
\ \protect \BOthers {.}}{%
{\protect \APACyear {2004}}%
}]{%
Sauvaud04:oxygen}
\APACinsertmetastar {%
Sauvaud04:oxygen}%
\begin{APACrefauthors}%
{Sauvaud}, J\BPBI A.%
, {Louarn}, P.%
, {Fruit}, G.%
, {Stenuit}, H.%
, {Vallat}, C.%
, {Dandouras}, J.%
\BDBL {}{McCarthy}, M.%
\end{APACrefauthors}%
\unskip\
\newblock
\APACrefYearMonthDay{2004}{Jan}{}.
\newblock
{\BBOQ}\APACrefatitle {{Case studies of the dynamics of ionospheric ions in the
  Earth's magnetotail}} {{Case studies of the dynamics of ionospheric ions in
  the Earth's magnetotail}}.{\BBCQ}
\newblock
\APACjournalVolNumPages{Journal of Geophysical Research (Space
  Physics)}{109}{A1}{A01212}.
\newblock
\begin{APACrefDOI} \doi{10.1029/2003JA009996} \end{APACrefDOI}
\PrintBackRefs{\CurrentBib}

\bibitem [\protect \citeauthoryear {%
{Schindler}%
}{%
{Schindler}%
}{%
{\protect \APACyear {1972}}%
}]{%
Schindler72}
\APACinsertmetastar {%
Schindler72}%
\begin{APACrefauthors}%
{Schindler}, K.%
\end{APACrefauthors}%
\unskip\
\newblock
\APACrefYearMonthDay{1972}{}{}.
\newblock
{\BBOQ}\APACrefatitle {{A Self-Consistent Theory of the Tail of the
  Magnetosphere}} {{A Self-Consistent Theory of the Tail of the
  Magnetosphere}}.{\BBCQ}
\newblock
\BIn{} {B.~M.~McCormac}\ (\BED), \APACrefbtitle {Earth's Magnetospheric
  Processes} {Earth's magnetospheric processes}\ (\BVOL~32, \BPG~200).
\PrintBackRefs{\CurrentBib}

\bibitem [\protect \citeauthoryear {%
{Schindler}%
}{%
{Schindler}%
}{%
{\protect \APACyear {1974}}%
}]{%
Schindler74}
\APACinsertmetastar {%
Schindler74}%
\begin{APACrefauthors}%
{Schindler}, K.%
\end{APACrefauthors}%
\unskip\
\newblock
\APACrefYearMonthDay{1974}{}{}.
\newblock
{\BBOQ}\APACrefatitle {{A theory of the substorm mechanism}} {{A theory of the
  substorm mechanism}}.{\BBCQ}
\newblock
\APACjournalVolNumPages{\jgr}{79}{}{2803-2810}.
\newblock
\begin{APACrefDOI} \doi{10.1029/JA079i019p02803} \end{APACrefDOI}
\PrintBackRefs{\CurrentBib}

\bibitem [\protect \citeauthoryear {%
{Schindler}%
}{%
{Schindler}%
}{%
{\protect \APACyear {2006}}%
}]{%
bookSchindler06}
\APACinsertmetastar {%
bookSchindler06}%
\begin{APACrefauthors}%
{Schindler}, K.%
\end{APACrefauthors}%
\unskip\
\newblock
\APACrefYear{2006}.
\newblock
\APACrefbtitle {{Physics of Space Plasma Activity}} {{Physics of Space Plasma
  Activity}}\ ({Schindler, K.}, \BED{}).
\newblock
\APACaddressPublisher{}{Cambridge University Press}.
\newblock
\begin{APACrefDOI} \doi{10.2277/0521858976} \end{APACrefDOI}
\PrintBackRefs{\CurrentBib}

\bibitem [\protect \citeauthoryear {%
{Schindler}%
\ \BBA {} {Birn}%
}{%
{Schindler}%
\ \BBA {} {Birn}%
}{%
{\protect \APACyear {2002}}%
}]{%
SB02}
\APACinsertmetastar {%
SB02}%
\begin{APACrefauthors}%
{Schindler}, K.%
\BCBT {}\ \BBA {} {Birn}, J.%
\end{APACrefauthors}%
\unskip\
\newblock
\APACrefYearMonthDay{2002}{{\APACmonth{08}}}{}.
\newblock
{\BBOQ}\APACrefatitle {{Models of two-dimensional embedded thin current sheets
  from Vlasov theory}} {{Models of two-dimensional embedded thin current sheets
  from Vlasov theory}}.{\BBCQ}
\newblock
\APACjournalVolNumPages{J. Geophys. Res.}{107}{}{1193}.
\newblock
\begin{APACrefDOI} \doi{10.1029/2001JA000304} \end{APACrefDOI}
\PrintBackRefs{\CurrentBib}

\bibitem [\protect \citeauthoryear {%
{Schindler}%
, {Pfirsch}%
\BCBL {}\ \BBA {} {Wobig}%
}{%
{Schindler}%
\ \protect \BOthers {.}}{%
{\protect \APACyear {1973}}%
}]{%
Schindler73}
\APACinsertmetastar {%
Schindler73}%
\begin{APACrefauthors}%
{Schindler}, K.%
, {Pfirsch}, D.%
\BCBL {}\ \BBA {} {Wobig}, H.%
\end{APACrefauthors}%
\unskip\
\newblock
\APACrefYearMonthDay{1973}{{\APACmonth{12}}}{}.
\newblock
{\BBOQ}\APACrefatitle {{Stability of two-dimensional collision-free plasmas}}
  {{Stability of two-dimensional collision-free plasmas}}.{\BBCQ}
\newblock
\APACjournalVolNumPages{Plasma Physics}{15}{}{1165-1184}.
\newblock
\begin{APACrefDOI} \doi{10.1088/0032-1028/15/12/001} \end{APACrefDOI}
\PrintBackRefs{\CurrentBib}

\bibitem [\protect \citeauthoryear {%
{Sergeev}%
, {Angelopoulos}%
, {Mitchell}%
\BCBL {}\ \BBA {} {Russell}%
}{%
{Sergeev}%
\ \protect \BOthers {.}}{%
{\protect \APACyear {1995}}%
}]{%
Sergeev95}
\APACinsertmetastar {%
Sergeev95}%
\begin{APACrefauthors}%
{Sergeev}, V\BPBI A.%
, {Angelopoulos}, V.%
, {Mitchell}, D\BPBI G.%
\BCBL {}\ \BBA {} {Russell}, C\BPBI T.%
\end{APACrefauthors}%
\unskip\
\newblock
\APACrefYearMonthDay{1995}{{\APACmonth{10}}}{}.
\newblock
{\BBOQ}\APACrefatitle {{In situ observations of magnetotail reconnection prior
  to the onset of a small substorm}} {{In situ observations of magnetotail
  reconnection prior to the onset of a small substorm}}.{\BBCQ}
\newblock
\APACjournalVolNumPages{\jgr}{100}{A10}{19121-19134}.
\newblock
\begin{APACrefDOI} \doi{10.1029/95JA01471} \end{APACrefDOI}
\PrintBackRefs{\CurrentBib}

\bibitem [\protect \citeauthoryear {%
{Sergeev}%
, {Gordeev}%
, {Merkin}%
\BCBL {}\ \BBA {} {Sitnov}%
}{%
{Sergeev}%
\ \protect \BOthers {.}}{%
{\protect \APACyear {2018}}%
}]{%
Sergeev18:grl}
\APACinsertmetastar {%
Sergeev18:grl}%
\begin{APACrefauthors}%
{Sergeev}, V\BPBI A.%
, {Gordeev}, E\BPBI I.%
, {Merkin}, V\BPBI G.%
\BCBL {}\ \BBA {} {Sitnov}, M\BPBI I.%
\end{APACrefauthors}%
\unskip\
\newblock
\APACrefYearMonthDay{2018}{{\APACmonth{03}}}{}.
\newblock
{\BBOQ}\APACrefatitle {{Does a Local B-Minimum Appear in the Tail Current Sheet
  During a Substorm Growth Phase?}} {{Does a Local B-Minimum Appear in the Tail
  Current Sheet During a Substorm Growth Phase?}}{\BBCQ}
\newblock
\APACjournalVolNumPages{\grl}{45}{}{2566-2573}.
\newblock
\begin{APACrefDOI} \doi{10.1002/2018GL077183} \end{APACrefDOI}
\PrintBackRefs{\CurrentBib}

\bibitem [\protect \citeauthoryear {%
Shi%
, Velli%
, Pucci%
, Tenerani%
\BCBL {}\ \BBA {} Innocenti%
}{%
Shi%
\ \protect \BOthers {.}}{%
{\protect \APACyear {2020}}%
}]{%
shi2020oblique}
\APACinsertmetastar {%
shi2020oblique}%
\begin{APACrefauthors}%
Shi, C.%
, Velli, M.%
, Pucci, F.%
, Tenerani, A.%
\BCBL {}\ \BBA {} Innocenti, M\BPBI E.%
\end{APACrefauthors}%
\unskip\
\newblock
\APACrefYearMonthDay{2020}{}{}.
\newblock
{\BBOQ}\APACrefatitle {Oblique tearing mode instability: guide field and Hall
  effect} {Oblique tearing mode instability: guide field and hall
  effect}.{\BBCQ}
\newblock
\APACjournalVolNumPages{The Astrophysical Journal}{902}{2}{142}.
\PrintBackRefs{\CurrentBib}

\bibitem [\protect \citeauthoryear {%
{Shukhtina}%
, {Dmitrieva}%
\BCBL {}\ \BBA {} {Sergeev}%
}{%
{Shukhtina}%
\ \protect \BOthers {.}}{%
{\protect \APACyear {2004}}%
}]{%
Shukhtina04}
\APACinsertmetastar {%
Shukhtina04}%
\begin{APACrefauthors}%
{Shukhtina}, M.%
, {Dmitrieva}, N.%
\BCBL {}\ \BBA {} {Sergeev}, V.%
\end{APACrefauthors}%
\unskip\
\newblock
\APACrefYearMonthDay{2004}{{\APACmonth{03}}}{}.
\newblock
{\BBOQ}\APACrefatitle {{Quantitative magnetotail characteristics for different
  magnetospheric states}} {{Quantitative magnetotail characteristics for
  different magnetospheric states}}.{\BBCQ}
\newblock
\APACjournalVolNumPages{Annales Geophysicae}{22}{}{1019-1032}.
\newblock
\begin{APACrefDOI} \doi{10.5194/angeo-22-1019-2004} \end{APACrefDOI}
\PrintBackRefs{\CurrentBib}

\bibitem [\protect \citeauthoryear {%
{Sitnov}%
, {Birn}%
\BCBL {}\ \protect \BOthers {.}}{%
{Sitnov}%
, {Birn}%
\BCBL {}\ \protect \BOthers {.}}{%
{\protect \APACyear {2019}}%
}]{%
Sitnov19}
\APACinsertmetastar {%
Sitnov19}%
\begin{APACrefauthors}%
{Sitnov}, M\BPBI I.%
, {Birn}, J.%
, {Ferdousi}, B.%
, {Gordeev}, E.%
, {Khotyaintsev}, Y.%
, {Merkin}, V.%
\BDBL {}{Zhou}, X.%
\end{APACrefauthors}%
\unskip\
\newblock
\APACrefYearMonthDay{2019}{Jun}{}.
\newblock
{\BBOQ}\APACrefatitle {{Explosive Magnetotail Activity}} {{Explosive
  Magnetotail Activity}}.{\BBCQ}
\newblock
\APACjournalVolNumPages{\ssr}{215}{4}{31}.
\newblock
\begin{APACrefDOI} \doi{10.1007/s11214-019-0599-5} \end{APACrefDOI}
\PrintBackRefs{\CurrentBib}

\bibitem [\protect \citeauthoryear {%
{Sitnov}%
, {Buzulukova}%
, {Swisdak}%
, {Merkin}%
\BCBL {}\ \BBA {} {Moore}%
}{%
{Sitnov}%
\ \protect \BOthers {.}}{%
{\protect \APACyear {2013}}%
}]{%
Sitnov13}
\APACinsertmetastar {%
Sitnov13}%
\begin{APACrefauthors}%
{Sitnov}, M\BPBI I.%
, {Buzulukova}, N.%
, {Swisdak}, M.%
, {Merkin}, V\BPBI G.%
\BCBL {}\ \BBA {} {Moore}, T\BPBI E.%
\end{APACrefauthors}%
\unskip\
\newblock
\APACrefYearMonthDay{2013}{{\APACmonth{01}}}{}.
\newblock
{\BBOQ}\APACrefatitle {{Spontaneous formation of dipolarization fronts and
  reconnection onset in the magnetotail}} {{Spontaneous formation of
  dipolarization fronts and reconnection onset in the magnetotail}}.{\BBCQ}
\newblock
\APACjournalVolNumPages{\grl}{40}{}{22-27}.
\newblock
\begin{APACrefDOI} \doi{10.1029/2012GL054701} \end{APACrefDOI}
\PrintBackRefs{\CurrentBib}

\bibitem [\protect \citeauthoryear {%
{Sitnov}%
, {Guzdar}%
\BCBL {}\ \BBA {} {Swisdak}%
}{%
{Sitnov}%
\ \protect \BOthers {.}}{%
{\protect \APACyear {2003}}%
}]{%
Sitnov03}
\APACinsertmetastar {%
Sitnov03}%
\begin{APACrefauthors}%
{Sitnov}, M\BPBI I.%
, {Guzdar}, P\BPBI N.%
\BCBL {}\ \BBA {} {Swisdak}, M.%
\end{APACrefauthors}%
\unskip\
\newblock
\APACrefYearMonthDay{2003}{{\APACmonth{07}}}{}.
\newblock
{\BBOQ}\APACrefatitle {{A model of the bifurcated current sheet}} {{A model of
  the bifurcated current sheet}}.{\BBCQ}
\newblock
\APACjournalVolNumPages{Geophys. Res. Lett.}{30}{}{45}.
\newblock
\begin{APACrefDOI} \doi{10.1029/2003GL017218} \end{APACrefDOI}
\PrintBackRefs{\CurrentBib}

\bibitem [\protect \citeauthoryear {%
{Sitnov}%
\ \BBA {} {Merkin}%
}{%
{Sitnov}%
\ \BBA {} {Merkin}%
}{%
{\protect \APACyear {2016}}%
}]{%
Sitnov&Merkin16}
\APACinsertmetastar {%
Sitnov&Merkin16}%
\begin{APACrefauthors}%
{Sitnov}, M\BPBI I.%
\BCBT {}\ \BBA {} {Merkin}, V\BPBI G.%
\end{APACrefauthors}%
\unskip\
\newblock
\APACrefYearMonthDay{2016}{{\APACmonth{08}}}{}.
\newblock
{\BBOQ}\APACrefatitle {{Generalized magnetotail equilibria: Effects of the
  dipole field, thin current sheets, and magnetic flux accumulation}}
  {{Generalized magnetotail equilibria: Effects of the dipole field, thin
  current sheets, and magnetic flux accumulation}}.{\BBCQ}
\newblock
\APACjournalVolNumPages{\jgr}{121}{}{7664-7683}.
\newblock
\begin{APACrefDOI} \doi{10.1002/2016JA023001} \end{APACrefDOI}
\PrintBackRefs{\CurrentBib}

\bibitem [\protect \citeauthoryear {%
{Sitnov}%
\ \protect \BOthers {.}}{%
{Sitnov}%
\ \protect \BOthers {.}}{%
{\protect \APACyear {2014}}%
}]{%
Sitnov14}
\APACinsertmetastar {%
Sitnov14}%
\begin{APACrefauthors}%
{Sitnov}, M\BPBI I.%
, {Merkin}, V\BPBI G.%
, {Swisdak}, M.%
, {Motoba}, T.%
, {Buzulukova}, N.%
, {Moore}, T\BPBI E.%
\BDBL {}{Ohtani}, S.%
\end{APACrefauthors}%
\unskip\
\newblock
\APACrefYearMonthDay{2014}{{\APACmonth{09}}}{}.
\newblock
{\BBOQ}\APACrefatitle {{Magnetic reconnection, buoyancy, and flapping motions
  in magnetotail explosions}} {{Magnetic reconnection, buoyancy, and flapping
  motions in magnetotail explosions}}.{\BBCQ}
\newblock
\APACjournalVolNumPages{\jgr}{119}{}{7151-7168}.
\newblock
\begin{APACrefDOI} \doi{10.1002/2014JA020205} \end{APACrefDOI}
\PrintBackRefs{\CurrentBib}

\bibitem [\protect \citeauthoryear {%
{Sitnov}%
\ \BBA {} {Schindler}%
}{%
{Sitnov}%
\ \BBA {} {Schindler}%
}{%
{\protect \APACyear {2010}}%
}]{%
Sitnov10}
\APACinsertmetastar {%
Sitnov10}%
\begin{APACrefauthors}%
{Sitnov}, M\BPBI I.%
\BCBT {}\ \BBA {} {Schindler}, K.%
\end{APACrefauthors}%
\unskip\
\newblock
\APACrefYearMonthDay{2010}{{\APACmonth{04}}}{}.
\newblock
{\BBOQ}\APACrefatitle {{Tearing stability of a multiscale magnetotail current
  sheet}} {{Tearing stability of a multiscale magnetotail current
  sheet}}.{\BBCQ}
\newblock
\APACjournalVolNumPages{\grl}{37}{}{8102}.
\newblock
\begin{APACrefDOI} \doi{10.1029/2010GL042961} \end{APACrefDOI}
\PrintBackRefs{\CurrentBib}

\bibitem [\protect \citeauthoryear {%
{Sitnov}%
, {Sharma}%
, {Guzdar}%
\BCBL {}\ \BBA {} {Yoon}%
}{%
{Sitnov}%
\ \protect \BOthers {.}}{%
{\protect \APACyear {2002}}%
}]{%
Sitnov02}
\APACinsertmetastar {%
Sitnov02}%
\begin{APACrefauthors}%
{Sitnov}, M\BPBI I.%
, {Sharma}, A\BPBI S.%
, {Guzdar}, P\BPBI N.%
\BCBL {}\ \BBA {} {Yoon}, P\BPBI H.%
\end{APACrefauthors}%
\unskip\
\newblock
\APACrefYearMonthDay{2002}{{\APACmonth{09}}}{}.
\newblock
{\BBOQ}\APACrefatitle {{Reconnection onset in the tail of Earth's
  magnetosphere}} {{Reconnection onset in the tail of Earth's
  magnetosphere}}.{\BBCQ}
\newblock
\APACjournalVolNumPages{J. Geophys. Res.}{107}{}{1256}.
\newblock
\begin{APACrefDOI} \doi{10.1029/2001JA009148} \end{APACrefDOI}
\PrintBackRefs{\CurrentBib}

\bibitem [\protect \citeauthoryear {%
{Sitnov}%
, {Stephens}%
, {Motoba}%
\BCBL {}\ \BBA {} {Swisdak}%
}{%
{Sitnov}%
\ \protect \BOthers {.}}{%
{\protect \APACyear {2021}}%
}]{%
Sitnov21}
\APACinsertmetastar {%
Sitnov21}%
\begin{APACrefauthors}%
{Sitnov}, M\BPBI I.%
, {Stephens}, G\BPBI K.%
, {Motoba}, T.%
\BCBL {}\ \BBA {} {Swisdak}, M.%
\end{APACrefauthors}%
\unskip\
\newblock
\APACrefYearMonthDay{2021}{}{}.
\newblock
{\BBOQ}\APACrefatitle {{Data Mining Reconstruction of Magnetotail Reconnection
  and Implications for Its First-Principle Modeling}} {{Data Mining
  Reconstruction of Magnetotail Reconnection and Implications for Its
  First-Principle Modeling}}.{\BBCQ}
\newblock
\APACjournalVolNumPages{Front. Phys.}{}{}{}.
\newblock
\begin{APACrefDOI} \doi{10.3389/fphy.2021.644884} \end{APACrefDOI}
\PrintBackRefs{\CurrentBib}

\bibitem [\protect \citeauthoryear {%
{Sitnov}%
, {Stephens}%
\BCBL {}\ \protect \BOthers {.}}{%
{Sitnov}%
, {Stephens}%
\BCBL {}\ \protect \BOthers {.}}{%
{\protect \APACyear {2019}}%
}]{%
Sitnov19:jgr}
\APACinsertmetastar {%
Sitnov19:jgr}%
\begin{APACrefauthors}%
{Sitnov}, M\BPBI I.%
, {Stephens}, G\BPBI K.%
, {Tsyganenko}, N\BPBI A.%
, {Miyashita}, Y.%
, {Merkin}, V\BPBI G.%
, {Motoba}, T.%
\BDBL {}{Genestreti}, K\BPBI J.%
\end{APACrefauthors}%
\unskip\
\newblock
\APACrefYearMonthDay{2019}{Nov}{}.
\newblock
{\BBOQ}\APACrefatitle {{Signatures of Nonideal Plasma Evolution During
  Substorms Obtained by Mining Multimission Magnetometer Data}} {{Signatures of
  Nonideal Plasma Evolution During Substorms Obtained by Mining Multimission
  Magnetometer Data}}.{\BBCQ}
\newblock
\APACjournalVolNumPages{Journal of Geophysical Research (Space
  Physics)}{124}{11}{8427-8456}.
\newblock
\begin{APACrefDOI} \doi{10.1029/2019JA027037} \end{APACrefDOI}
\PrintBackRefs{\CurrentBib}

\bibitem [\protect \citeauthoryear {%
{Sitnov}%
\ \BBA {} {Swisdak}%
}{%
{Sitnov}%
\ \BBA {} {Swisdak}%
}{%
{\protect \APACyear {2011}}%
}]{%
Sitnov11}
\APACinsertmetastar {%
Sitnov11}%
\begin{APACrefauthors}%
{Sitnov}, M\BPBI I.%
\BCBT {}\ \BBA {} {Swisdak}, M.%
\end{APACrefauthors}%
\unskip\
\newblock
\APACrefYearMonthDay{2011}{{\APACmonth{12}}}{}.
\newblock
{\BBOQ}\APACrefatitle {{Onset of collisionless magnetic reconnection in
  two-dimensional current sheets and formation of dipolarization fronts}}
  {{Onset of collisionless magnetic reconnection in two-dimensional current
  sheets and formation of dipolarization fronts}}.{\BBCQ}
\newblock
\APACjournalVolNumPages{\jgr}{116}{}{12216}.
\newblock
\begin{APACrefDOI} \doi{10.1029/2011JA016920} \end{APACrefDOI}
\PrintBackRefs{\CurrentBib}

\bibitem [\protect \citeauthoryear {%
{Sitnov}%
, {Swisdak}%
\BCBL {}\ \BBA {} {Divin}%
}{%
{Sitnov}%
\ \protect \BOthers {.}}{%
{\protect \APACyear {2009}}%
}]{%
Sitnov09}
\APACinsertmetastar {%
Sitnov09}%
\begin{APACrefauthors}%
{Sitnov}, M\BPBI I.%
, {Swisdak}, M.%
\BCBL {}\ \BBA {} {Divin}, A\BPBI V.%
\end{APACrefauthors}%
\unskip\
\newblock
\APACrefYearMonthDay{2009}{{\APACmonth{04}}}{}.
\newblock
{\BBOQ}\APACrefatitle {{Dipolarization fronts as a signature of transient
  reconnection in the magnetotail}} {{Dipolarization fronts as a signature of
  transient reconnection in the magnetotail}}.{\BBCQ}
\newblock
\APACjournalVolNumPages{\jgr}{114}{}{A04202}.
\newblock
\begin{APACrefDOI} \doi{10.1029/2008JA013980} \end{APACrefDOI}
\PrintBackRefs{\CurrentBib}

\bibitem [\protect \citeauthoryear {%
{Sitnov}%
, {Swisdak}%
, {Guzdar}%
\BCBL {}\ \BBA {} {Runov}%
}{%
{Sitnov}%
\ \protect \BOthers {.}}{%
{\protect \APACyear {2006}}%
}]{%
Sitnov06}
\APACinsertmetastar {%
Sitnov06}%
\begin{APACrefauthors}%
{Sitnov}, M\BPBI I.%
, {Swisdak}, M.%
, {Guzdar}, P\BPBI N.%
\BCBL {}\ \BBA {} {Runov}, A.%
\end{APACrefauthors}%
\unskip\
\newblock
\APACrefYearMonthDay{2006}{{\APACmonth{08}}}{}.
\newblock
{\BBOQ}\APACrefatitle {{Structure and dynamics of a new class of thin current
  sheets}} {{Structure and dynamics of a new class of thin current
  sheets}}.{\BBCQ}
\newblock
\APACjournalVolNumPages{J. Geophys. Res.}{111}{}{8204}.
\newblock
\begin{APACrefDOI} \doi{10.1029/2005JA011517} \end{APACrefDOI}
\PrintBackRefs{\CurrentBib}

\bibitem [\protect \citeauthoryear {%
{Sitnov}%
, {Zelenyi}%
, {Malova}%
\BCBL {}\ \BBA {} {Sharma}%
}{%
{Sitnov}%
\ \protect \BOthers {.}}{%
{\protect \APACyear {2000}}%
}]{%
Sitnov00}
\APACinsertmetastar {%
Sitnov00}%
\begin{APACrefauthors}%
{Sitnov}, M\BPBI I.%
, {Zelenyi}, L\BPBI M.%
, {Malova}, H\BPBI V.%
\BCBL {}\ \BBA {} {Sharma}, A\BPBI S.%
\end{APACrefauthors}%
\unskip\
\newblock
\APACrefYearMonthDay{2000}{{\APACmonth{06}}}{}.
\newblock
{\BBOQ}\APACrefatitle {{Thin current sheet embedded within a thicker plasma
  sheet: Self-consistent kinetic theory}} {{Thin current sheet embedded within
  a thicker plasma sheet: Self-consistent kinetic theory}}.{\BBCQ}
\newblock
\APACjournalVolNumPages{J. Geophys. Res.}{105}{}{13029-13044}.
\newblock
\begin{APACrefDOI} \doi{10.1029/1999JA000431} \end{APACrefDOI}
\PrintBackRefs{\CurrentBib}

\bibitem [\protect \citeauthoryear {%
{Somov}%
\ \BBA {} {Verneta}%
}{%
{Somov}%
\ \BBA {} {Verneta}%
}{%
{\protect \APACyear {1993}}%
}]{%
Somov&Verneta93}
\APACinsertmetastar {%
Somov&Verneta93}%
\begin{APACrefauthors}%
{Somov}, B\BPBI V.%
\BCBT {}\ \BBA {} {Verneta}, A\BPBI I.%
\end{APACrefauthors}%
\unskip\
\newblock
\APACrefYearMonthDay{1993}{{\APACmonth{09}}}{}.
\newblock
{\BBOQ}\APACrefatitle {{Tearing Instability of Reconnecting current Sheets in
  Space Plasmas}} {{Tearing Instability of Reconnecting current Sheets in Space
  Plasmas}}.{\BBCQ}
\newblock
\APACjournalVolNumPages{\ssr}{65}{3-4}{253-288}.
\newblock
\begin{APACrefDOI} \doi{10.1007/BF00754510} \end{APACrefDOI}
\PrintBackRefs{\CurrentBib}

\bibitem [\protect \citeauthoryear {%
{Speiser}%
}{%
{Speiser}%
}{%
{\protect \APACyear {1965}}%
}]{%
Speiser65}
\APACinsertmetastar {%
Speiser65}%
\begin{APACrefauthors}%
{Speiser}, T\BPBI W.%
\end{APACrefauthors}%
\unskip\
\newblock
\APACrefYearMonthDay{1965}{{\APACmonth{09}}}{}.
\newblock
{\BBOQ}\APACrefatitle {{Particle Trajectories in Model Current Sheets, 1,
  Analytical Solutions}} {{Particle Trajectories in Model Current Sheets, 1,
  Analytical Solutions}}.{\BBCQ}
\newblock
\APACjournalVolNumPages{\jgr}{70}{}{4219-4226}.
\newblock
\begin{APACrefDOI} \doi{10.1029/JZ070i017p04219} \end{APACrefDOI}
\PrintBackRefs{\CurrentBib}

\bibitem [\protect \citeauthoryear {%
{Steinhauer}%
, {McCarthy}%
\BCBL {}\ \BBA {} {Whipple}%
}{%
{Steinhauer}%
\ \protect \BOthers {.}}{%
{\protect \APACyear {2008}}%
}]{%
Steinhauer08}
\APACinsertmetastar {%
Steinhauer08}%
\begin{APACrefauthors}%
{Steinhauer}, L\BPBI C.%
, {McCarthy}, M\BPBI P.%
\BCBL {}\ \BBA {} {Whipple}, E\BPBI C.%
\end{APACrefauthors}%
\unskip\
\newblock
\APACrefYearMonthDay{2008}{{\APACmonth{04}}}{}.
\newblock
{\BBOQ}\APACrefatitle {{Multifluid model of a one-dimensional steady state
  magnetotail current sheet}} {{Multifluid model of a one-dimensional steady
  state magnetotail current sheet}}.{\BBCQ}
\newblock
\APACjournalVolNumPages{\jgr}{113}{}{4207}.
\newblock
\begin{APACrefDOI} \doi{10.1029/2007JA012578} \end{APACrefDOI}
\PrintBackRefs{\CurrentBib}

\bibitem [\protect \citeauthoryear {%
{Sturrock}%
}{%
{Sturrock}%
}{%
{\protect \APACyear {1966}}%
}]{%
Sturrock66}
\APACinsertmetastar {%
Sturrock66}%
\begin{APACrefauthors}%
{Sturrock}, P\BPBI A.%
\end{APACrefauthors}%
\unskip\
\newblock
\APACrefYearMonthDay{1966}{{\APACmonth{08}}}{}.
\newblock
{\BBOQ}\APACrefatitle {{Model of the High-Energy Phase of Solar Flares}}
  {{Model of the High-Energy Phase of Solar Flares}}.{\BBCQ}
\newblock
\APACjournalVolNumPages{\nat}{211}{}{695-697}.
\newblock
\begin{APACrefDOI} \doi{10.1038/211695a0} \end{APACrefDOI}
\PrintBackRefs{\CurrentBib}

\bibitem [\protect \citeauthoryear {%
{Swisdak}%
, {Rogers}%
, {Drake}%
\BCBL {}\ \BBA {} {Shay}%
}{%
{Swisdak}%
\ \protect \BOthers {.}}{%
{\protect \APACyear {2003}}%
}]{%
Swisdak03}
\APACinsertmetastar {%
Swisdak03}%
\begin{APACrefauthors}%
{Swisdak}, M.%
, {Rogers}, B\BPBI N.%
, {Drake}, J\BPBI F.%
\BCBL {}\ \BBA {} {Shay}, M\BPBI A.%
\end{APACrefauthors}%
\unskip\
\newblock
\APACrefYearMonthDay{2003}{{\APACmonth{05}}}{}.
\newblock
{\BBOQ}\APACrefatitle {{Diamagnetic suppression of component magnetic
  reconnection at the magnetopause}} {{Diamagnetic suppression of component
  magnetic reconnection at the magnetopause}}.{\BBCQ}
\newblock
\APACjournalVolNumPages{Journal of Geophysical Research (Space
  Physics)}{108}{A5}{1218}.
\newblock
\begin{APACrefDOI} \doi{10.1029/2002JA009726} \end{APACrefDOI}
\PrintBackRefs{\CurrentBib}

\bibitem [\protect \citeauthoryear {%
Tenerani%
, Velli%
, Pucci%
, Landi%
\BCBL {}\ \BBA {} Rappazzo%
}{%
Tenerani%
\ \protect \BOthers {.}}{%
{\protect \APACyear {2016}}%
}]{%
tenerani2016ideally}
\APACinsertmetastar {%
tenerani2016ideally}%
\begin{APACrefauthors}%
Tenerani, A.%
, Velli, M.%
, Pucci, F.%
, Landi, S.%
\BCBL {}\ \BBA {} Rappazzo, A\BPBI F.%
\end{APACrefauthors}%
\unskip\
\newblock
\APACrefYearMonthDay{2016}{}{}.
\newblock
{\BBOQ}\APACrefatitle {‘Ideally’unstable current sheets and the triggering
  of fast magnetic reconnection} {‘ideally’unstable current sheets and the
  triggering of fast magnetic reconnection}.{\BBCQ}
\newblock
\APACjournalVolNumPages{Journal of Plasma Physics}{82}{5}{}.
\PrintBackRefs{\CurrentBib}

\bibitem [\protect \citeauthoryear {%
{Tenerani}%
, {Velli}%
, {Rappazzo}%
\BCBL {}\ \BBA {} {Pucci}%
}{%
{Tenerani}%
\ \protect \BOthers {.}}{%
{\protect \APACyear {2015}}%
}]{%
Tenerani15:apj}
\APACinsertmetastar {%
Tenerani15:apj}%
\begin{APACrefauthors}%
{Tenerani}, A.%
, {Velli}, M.%
, {Rappazzo}, A\BPBI F.%
\BCBL {}\ \BBA {} {Pucci}, F.%
\end{APACrefauthors}%
\unskip\
\newblock
\APACrefYearMonthDay{2015}{{\APACmonth{11}}}{}.
\newblock
{\BBOQ}\APACrefatitle {{Magnetic Reconnection: Recursive Current Sheet Collapse
  Triggered by Ideal Tearing}} {{Magnetic Reconnection: Recursive Current Sheet
  Collapse Triggered by Ideal Tearing}}.{\BBCQ}
\newblock
\APACjournalVolNumPages{\apjl}{813}{}{L32}.
\newblock
\begin{APACrefDOI} \doi{10.1088/2041-8205/813/2/L32} \end{APACrefDOI}
\PrintBackRefs{\CurrentBib}

\bibitem [\protect \citeauthoryear {%
{Terasawa}%
, {Shibata}%
\BCBL {}\ \BBA {} {Scholer}%
}{%
{Terasawa}%
\ \protect \BOthers {.}}{%
{\protect \APACyear {2000}}%
}]{%
Terasawa00:AdSR}
\APACinsertmetastar {%
Terasawa00:AdSR}%
\begin{APACrefauthors}%
{Terasawa}, T.%
, {Shibata}, K.%
\BCBL {}\ \BBA {} {Scholer}, M.%
\end{APACrefauthors}%
\unskip\
\newblock
\APACrefYearMonthDay{2000}{}{}.
\newblock
{\BBOQ}\APACrefatitle {{Comparative Study of Flares and Substorms}}
  {{Comparative Study of Flares and Substorms}}.{\BBCQ}
\newblock
\APACjournalVolNumPages{Advances in Space Research}{26}{}{573-583}.
\newblock
\begin{APACrefDOI} \doi{10.1016/S0273-1177(99)01087-X} \end{APACrefDOI}
\PrintBackRefs{\CurrentBib}

\bibitem [\protect \citeauthoryear {%
{Tsurutani}%
\ \BBA {} {Ho}%
}{%
{Tsurutani}%
\ \BBA {} {Ho}%
}{%
{\protect \APACyear {1999}}%
}]{%
Tsurutani&Ho99}
\APACinsertmetastar {%
Tsurutani&Ho99}%
\begin{APACrefauthors}%
{Tsurutani}, B\BPBI T.%
\BCBT {}\ \BBA {} {Ho}, C\BPBI M.%
\end{APACrefauthors}%
\unskip\
\newblock
\APACrefYearMonthDay{1999}{}{}.
\newblock
{\BBOQ}\APACrefatitle {{A review of discontinuities and Alfv{\'e}n waves in
  interplanetary space: Ulysses results}} {{A review of discontinuities and
  Alfv{\'e}n waves in interplanetary space: Ulysses results}}.{\BBCQ}
\newblock
\APACjournalVolNumPages{Reviews of Geophysics}{37}{}{517-524}.
\newblock
\begin{APACrefDOI} \doi{10.1029/1999RG900010} \end{APACrefDOI}
\PrintBackRefs{\CurrentBib}

\bibitem [\protect \citeauthoryear {%
{Vasko}%
, {Artemyev}%
, {Popov}%
\BCBL {}\ \BBA {} {Malova}%
}{%
{Vasko}%
\ \protect \BOthers {.}}{%
{\protect \APACyear {2013}}%
}]{%
Vasko13:pop}
\APACinsertmetastar {%
Vasko13:pop}%
\begin{APACrefauthors}%
{Vasko}, I\BPBI Y.%
, {Artemyev}, A\BPBI V.%
, {Popov}, V\BPBI Y.%
\BCBL {}\ \BBA {} {Malova}, H\BPBI V.%
\end{APACrefauthors}%
\unskip\
\newblock
\APACrefYearMonthDay{2013}{{\APACmonth{02}}}{}.
\newblock
{\BBOQ}\APACrefatitle {{Kinetic models of two-dimensional plane and axially
  symmetric current sheets: Group theory approach}} {{Kinetic models of
  two-dimensional plane and axially symmetric current sheets: Group theory
  approach}}.{\BBCQ}
\newblock
\APACjournalVolNumPages{Physics of Plasmas}{20}{2}{022110}.
\newblock
\begin{APACrefDOI} \doi{10.1063/1.4792263} \end{APACrefDOI}
\PrintBackRefs{\CurrentBib}

\bibitem [\protect \citeauthoryear {%
{Vasko}%
, {Petrukovich}%
, {Artemyev}%
, {Nakamura}%
\BCBL {}\ \BBA {} {Zelenyi}%
}{%
{Vasko}%
\ \protect \BOthers {.}}{%
{\protect \APACyear {2015}}%
}]{%
Vasko15:jgr:cs}
\APACinsertmetastar {%
Vasko15:jgr:cs}%
\begin{APACrefauthors}%
{Vasko}, I\BPBI Y.%
, {Petrukovich}, A\BPBI A.%
, {Artemyev}, A\BPBI V.%
, {Nakamura}, R.%
\BCBL {}\ \BBA {} {Zelenyi}, L\BPBI M.%
\end{APACrefauthors}%
\unskip\
\newblock
\APACrefYearMonthDay{2015}{{\APACmonth{10}}}{}.
\newblock
{\BBOQ}\APACrefatitle {{Earth's distant magnetotail current sheet near and
  beyond lunar orbit}} {{Earth's distant magnetotail current sheet near and
  beyond lunar orbit}}.{\BBCQ}
\newblock
\APACjournalVolNumPages{\jgr}{120}{}{8663-8680}.
\newblock
\begin{APACrefDOI} \doi{10.1002/2015JA021633} \end{APACrefDOI}
\PrintBackRefs{\CurrentBib}

\bibitem [\protect \citeauthoryear {%
{Verneta}%
\ \BBA {} {Somov}%
}{%
{Verneta}%
\ \BBA {} {Somov}%
}{%
{\protect \APACyear {1987}}%
}]{%
Verneta&Somov87}
\APACinsertmetastar {%
Verneta&Somov87}%
\begin{APACrefauthors}%
{Verneta}, A\BPBI I.%
\BCBT {}\ \BBA {} {Somov}, B\BPBI V.%
\end{APACrefauthors}%
\unskip\
\newblock
\APACrefYearMonthDay{1987}{{\APACmonth{07}}}{}.
\newblock
{\BBOQ}\APACrefatitle {{Stabilizing Effect of a Transverse Magnetic Field in
  Current Sheets of Solar Flares}} {{Stabilizing Effect of a Transverse
  Magnetic Field in Current Sheets of Solar Flares}}.{\BBCQ}
\newblock
\APACjournalVolNumPages{Soviet Astronomy Letters}{13}{}{302}.
\PrintBackRefs{\CurrentBib}

\bibitem [\protect \citeauthoryear {%
Virtanen%
\ \protect \BOthers {.}}{%
Virtanen%
\ \protect \BOthers {.}}{%
{\protect \APACyear {2020}}%
}]{%
Virtanenetal2020Scipy}
\APACinsertmetastar {%
Virtanenetal2020Scipy}%
\begin{APACrefauthors}%
Virtanen, P.%
, Gommers, R.%
, Oliphant, T\BPBI E.%
, Haberland, M.%
, Reddy, T.%
, Cournapeau, D.%
\BDBL {}others%
\end{APACrefauthors}%
\unskip\
\newblock
\APACrefYearMonthDay{2020}{}{}.
\newblock
{\BBOQ}\APACrefatitle {SciPy 1.0: fundamental algorithms for scientific
  computing in Python} {Scipy 1.0: fundamental algorithms for scientific
  computing in python}.{\BBCQ}
\newblock
\APACjournalVolNumPages{Nature Methods}{}{}{1--12}.
\PrintBackRefs{\CurrentBib}

\bibitem [\protect \citeauthoryear {%
{Walsh}%
, {Owen}%
, {Fazakerley}%
, {Forsyth}%
\BCBL {}\ \BBA {} {Dandouras}%
}{%
{Walsh}%
\ \protect \BOthers {.}}{%
{\protect \APACyear {2011}}%
}]{%
Walsh11}
\APACinsertmetastar {%
Walsh11}%
\begin{APACrefauthors}%
{Walsh}, A\BPBI P.%
, {Owen}, C\BPBI J.%
, {Fazakerley}, A\BPBI N.%
, {Forsyth}, C.%
\BCBL {}\ \BBA {} {Dandouras}, I.%
\end{APACrefauthors}%
\unskip\
\newblock
\APACrefYearMonthDay{2011}{{\APACmonth{03}}}{}.
\newblock
{\BBOQ}\APACrefatitle {{Average magnetotail electron and proton pitch angle
  distributions from Cluster PEACE and CIS observations}} {{Average magnetotail
  electron and proton pitch angle distributions from Cluster PEACE and CIS
  observations}}.{\BBCQ}
\newblock
\APACjournalVolNumPages{\grl}{38}{}{6103}.
\newblock
\begin{APACrefDOI} \doi{10.1029/2011GL046770} \end{APACrefDOI}
\PrintBackRefs{\CurrentBib}

\bibitem [\protect \citeauthoryear {%
{Wang}%
, {Zaharia}%
, {Lyons}%
\BCBL {}\ \BBA {} {Angelopoulos}%
}{%
{Wang}%
\ \protect \BOthers {.}}{%
{\protect \APACyear {2013}}%
}]{%
Wang13:ions}
\APACinsertmetastar {%
Wang13:ions}%
\begin{APACrefauthors}%
{Wang}, C\BHBI P.%
, {Zaharia}, S\BPBI G.%
, {Lyons}, L\BPBI R.%
\BCBL {}\ \BBA {} {Angelopoulos}, V.%
\end{APACrefauthors}%
\unskip\
\newblock
\APACrefYearMonthDay{2013}{{\APACmonth{01}}}{}.
\newblock
{\BBOQ}\APACrefatitle {{Spatial distributions of ion pitch angle anisotropy in
  the near-Earth magnetosphere and tail plasma sheet}} {{Spatial distributions
  of ion pitch angle anisotropy in the near-Earth magnetosphere and tail plasma
  sheet}}.{\BBCQ}
\newblock
\APACjournalVolNumPages{\jgr}{118}{}{244-255}.
\newblock
\begin{APACrefDOI} \doi{10.1029/2012JA018275} \end{APACrefDOI}
\PrintBackRefs{\CurrentBib}

\bibitem [\protect \citeauthoryear {%
Wang%
, Lee%
\BCBL {}\ \BBA {} Wei%
}{%
Wang%
\ \protect \BOthers {.}}{%
{\protect \APACyear {1988}}%
}]{%
wang1988streaming}
\APACinsertmetastar {%
wang1988streaming}%
\begin{APACrefauthors}%
Wang, S.%
, Lee, L.%
\BCBL {}\ \BBA {} Wei, C.%
\end{APACrefauthors}%
\unskip\
\newblock
\APACrefYearMonthDay{1988}{}{}.
\newblock
{\BBOQ}\APACrefatitle {Streaming tearing instability in the current sheet with
  a super-Alfvenic flow} {Streaming tearing instability in the current sheet
  with a super-alfvenic flow}.{\BBCQ}
\newblock
\APACjournalVolNumPages{The Physics of fluids}{31}{6}{1544--1548}.
\PrintBackRefs{\CurrentBib}

\bibitem [\protect \citeauthoryear {%
{Yamada}%
, {Kulsrud}%
\BCBL {}\ \BBA {} {Ji}%
}{%
{Yamada}%
\ \protect \BOthers {.}}{%
{\protect \APACyear {2010}}%
}]{%
Yamada10:reconnection}
\APACinsertmetastar {%
Yamada10:reconnection}%
\begin{APACrefauthors}%
{Yamada}, M.%
, {Kulsrud}, R.%
\BCBL {}\ \BBA {} {Ji}, H.%
\end{APACrefauthors}%
\unskip\
\newblock
\APACrefYearMonthDay{2010}{{\APACmonth{01}}}{}.
\newblock
{\BBOQ}\APACrefatitle {{Magnetic reconnection}} {{Magnetic
  reconnection}}.{\BBCQ}
\newblock
\APACjournalVolNumPages{Reviews of Modern Physics}{82}{}{603-664}.
\newblock
\begin{APACrefDOI} \doi{10.1103/RevModPhys.82.603} \end{APACrefDOI}
\PrintBackRefs{\CurrentBib}

\bibitem [\protect \citeauthoryear {%
{Yoon}%
\ \BBA {} {Lui}%
}{%
{Yoon}%
\ \BBA {} {Lui}%
}{%
{\protect \APACyear {2005}}%
}]{%
YL05}
\APACinsertmetastar {%
YL05}%
\begin{APACrefauthors}%
{Yoon}, P\BPBI H.%
\BCBT {}\ \BBA {} {Lui}, A\BPBI T\BPBI Y.%
\end{APACrefauthors}%
\unskip\
\newblock
\APACrefYearMonthDay{2005}{{\APACmonth{01}}}{}.
\newblock
{\BBOQ}\APACrefatitle {{A class of exact two-dimensional kinetic current sheet
  equilibria}} {{A class of exact two-dimensional kinetic current sheet
  equilibria}}.{\BBCQ}
\newblock
\APACjournalVolNumPages{J. Geophys. Res.}{110}{}{1202}.
\newblock
\begin{APACrefDOI} \doi{10.1029/2003JA010308} \end{APACrefDOI}
\PrintBackRefs{\CurrentBib}

\bibitem [\protect \citeauthoryear {%
{Zelenyi}%
, {Artemyev}%
, {Malova}%
\BCBL {}\ \BBA {} {Popov}%
}{%
{Zelenyi}%
\ \protect \BOthers {.}}{%
{\protect \APACyear {2008}}%
}]{%
Zelenyi08JASTP}
\APACinsertmetastar {%
Zelenyi08JASTP}%
\begin{APACrefauthors}%
{Zelenyi}, L\BPBI M.%
, {Artemyev}, A\BPBI V.%
, {Malova}, H\BPBI V.%
\BCBL {}\ \BBA {} {Popov}, V\BPBI Y.%
\end{APACrefauthors}%
\unskip\
\newblock
\APACrefYearMonthDay{2008}{{\APACmonth{02}}}{}.
\newblock
{\BBOQ}\APACrefatitle {{Marginal stability of thin current sheets in the
  Earth's magnetotail}} {{Marginal stability of thin current sheets in the
  Earth's magnetotail}}.{\BBCQ}
\newblock
\APACjournalVolNumPages{Journal of Atmospheric and Solar-Terrestrial
  Physics}{70}{}{325-333}.
\newblock
\begin{APACrefDOI} \doi{10.1016/j.jastp.2007.08.019} \end{APACrefDOI}
\PrintBackRefs{\CurrentBib}

\bibitem [\protect \citeauthoryear {%
{Zelenyi}%
, {Dolgonosov}%
, {Peroomian}%
\BCBL {}\ \BBA {} {Ashour-Abdalla}%
}{%
{Zelenyi}%
, {Dolgonosov}%
\BCBL {}\ \protect \BOthers {.}}{%
{\protect \APACyear {2006}}%
}]{%
Zelenyi06:beamlet}
\APACinsertmetastar {%
Zelenyi06:beamlet}%
\begin{APACrefauthors}%
{Zelenyi}, L\BPBI M.%
, {Dolgonosov}, M\BPBI S.%
, {Peroomian}, V.%
\BCBL {}\ \BBA {} {Ashour-Abdalla}, M.%
\end{APACrefauthors}%
\unskip\
\newblock
\APACrefYearMonthDay{2006}{{\APACmonth{09}}}{}.
\newblock
{\BBOQ}\APACrefatitle {{Effects of nonlinearity on the structure of PSBL
  beamlets}} {{Effects of nonlinearity on the structure of PSBL
  beamlets}}.{\BBCQ}
\newblock
\APACjournalVolNumPages{\grl}{33}{}{L18103}.
\newblock
\begin{APACrefDOI} \doi{10.1029/2006GL026176} \end{APACrefDOI}
\PrintBackRefs{\CurrentBib}

\bibitem [\protect \citeauthoryear {%
{Zelenyi}%
, {Malova}%
, {Artemyev}%
, {Popov}%
\BCBL {}\ \BBA {} {Petrukovich}%
}{%
{Zelenyi}%
\ \protect \BOthers {.}}{%
{\protect \APACyear {2011}}%
}]{%
Zelenyi11PPR}
\APACinsertmetastar {%
Zelenyi11PPR}%
\begin{APACrefauthors}%
{Zelenyi}, L\BPBI M.%
, {Malova}, H\BPBI V.%
, {Artemyev}, A\BPBI V.%
, {Popov}, V\BPBI Y.%
\BCBL {}\ \BBA {} {Petrukovich}, A\BPBI A.%
\end{APACrefauthors}%
\unskip\
\newblock
\APACrefYearMonthDay{2011}{{\APACmonth{02}}}{}.
\newblock
{\BBOQ}\APACrefatitle {{Thin current sheets in collisionless plasma:
  Equilibrium structure, plasma instabilities, and particle acceleration}}
  {{Thin current sheets in collisionless plasma: Equilibrium structure, plasma
  instabilities, and particle acceleration}}.{\BBCQ}
\newblock
\APACjournalVolNumPages{Plasma Physics Reports}{37}{}{118-160}.
\newblock
\begin{APACrefDOI} \doi{10.1134/S1063780X1102005X} \end{APACrefDOI}
\PrintBackRefs{\CurrentBib}

\bibitem [\protect \citeauthoryear {%
{Zelenyi}%
, {Malova}%
\BCBL {}\ \protect \BOthers {.}}{%
{Zelenyi}%
, {Malova}%
\BCBL {}\ \protect \BOthers {.}}{%
{\protect \APACyear {2006}}%
}]{%
Zelenyi06}
\APACinsertmetastar {%
Zelenyi06}%
\begin{APACrefauthors}%
{Zelenyi}, L\BPBI M.%
, {Malova}, H\BPBI V.%
, {Popov}, V\BPBI Y.%
, {Delcourt}, D\BPBI C.%
, {Ganushkina}, N\BPBI Y.%
\BCBL {}\ \BBA {} {Sharma}, A\BPBI S.%
\end{APACrefauthors}%
\unskip\
\newblock
\APACrefYearMonthDay{2006}{{\APACmonth{03}}}{}.
\newblock
{\BBOQ}\APACrefatitle {{``Matreshka'' model of multilayered current sheet}}
  {{``Matreshka'' model of multilayered current sheet}}.{\BBCQ}
\newblock
\APACjournalVolNumPages{Geophys. Res. Lett.}{33}{}{5105}.
\newblock
\begin{APACrefDOI} \doi{10.1029/2005GL025117} \end{APACrefDOI}
\PrintBackRefs{\CurrentBib}

\bibitem [\protect \citeauthoryear {%
{Zhou}%
\ \protect \BOthers {.}}{%
{Zhou}%
\ \protect \BOthers {.}}{%
{\protect \APACyear {2009}}%
}]{%
Zhou09}
\APACinsertmetastar {%
Zhou09}%
\begin{APACrefauthors}%
{Zhou}, X.%
, {Angelopoulos}, V.%
, {Runov}, A.%
, {Sitnov}, M\BPBI I.%
, {Coroniti}, F.%
, {Pritchett}, P.%
\BDBL {}{Glassmeier}, K.%
\end{APACrefauthors}%
\unskip\
\newblock
\APACrefYearMonthDay{2009}{{\APACmonth{03}}}{}.
\newblock
{\BBOQ}\APACrefatitle {{Thin current sheet in the substorm late growth phase:
  Modeling of THEMIS observations}} {{Thin current sheet in the substorm late
  growth phase: Modeling of THEMIS observations}}.{\BBCQ}
\newblock
\APACjournalVolNumPages{J. Geophys. Res.}{114}{}{3223}.
\newblock
\begin{APACrefDOI} \doi{10.1029/2008JA013777} \end{APACrefDOI}
\PrintBackRefs{\CurrentBib}

\end{thebibliography}

\end{document}